\documentclass{aa}  

\usepackage{hyperref}
\hyperlink{}
\hypersetup{
    colorlinks=true,
    citecolor=blue,   
    linkcolor=blue,   
    urlcolor=blue     
}

\usepackage{graphicx}
\usepackage{xcolor}
\usepackage{pdflscape}
\usepackage{booktabs}
\usepackage{array}
\usepackage{orcidlink}
\usepackage{academicons}
\usepackage{afterpage}
\usepackage{tabularx}
\usepackage[switch]{lineno}
\nolinenumbers
\usepackage{siunitx} 
\usepackage{textcomp}
\usepackage{multirow}
\usepackage{multicol}
\usepackage[utf8]{inputenc}
\DeclareUnicodeCharacter{2212}{-}
\usepackage{txfonts}
\newcommand{\gsim}{\ \raise -2.truept\hbox{\rlap{\hbox{$\sim$}}\raise 5.truept\hbox{$>$}\ }}

\definecolor{emerald}{rgb}{0.4,0.66,0.31}
\newcommand{\orcid}[1]{\href{https://orcid.org/#1}{\textcolor{black}{\aiOrcid}}}

\definecolor{gr}{rgb}{0.2, 0.20, 0.80}

\titlerunning{VEGAS-SSS: Intra-Group GCs in NGC\,5018}
\begin{document} 

   \title{VEGAS-SSS: An intra-group component in the globular cluster system of NGC 5018 group of galaxies using VST data}

   \subtitle{}
 
\author{Pratik Lonare\inst{1,2}\orcidlink{ https://orcid.org/0009-0000-0028-0493}
          \and
          Michele Cantiello\inst{1}\orcidlink{https://orcid.org/0000-0003-2072-384X}
          \and
          Marco Mirabile\inst{1,3}\orcidlink{https://orcid.org/0009-0007-6055-3933}
          \and
          Marilena Spavone\inst{5}\orcidlink{https://orcid.org/0000-0002-6427-7039}
          \and
          Marina Rejkuba\inst{6}\orcidlink{https://orcid.org/0000-0002-6577-2787}
          \and 
          Michael Hilker\inst{6}\orcidlink{https://orcid.org/0000-0002-2363-5522}
          \and
          Rebecca Habas\inst{1}\orcidlink{https://orcid.org/0000-0002-4033-3841}
          \and
          Enrichetta Iodice\inst{5}\orcidlink{https://orcid.org/0000-0003-4291-0005}
          \and
          Nandini Hazra\inst{1,3,4}\orcidlink{https://orcid.org/0000-0002-3870-1537}
          \and
          Gabriele Riccio\inst{1}\orcidlink{https://orcid.org/0000-0002-6399-2129}
          }

\institute{INAF-Osservatorio Astronomico d’Abruzzo, Via Maggini, 64100 Teramo, Italy\\
              \email{pratik.lonare@inaf.it}
         \and
             University of Rome Tor Vergata, Via della Ricerca Scientifica 1, I-00133 Rome, Italy
         \and
             Gran Sasso Science Institute (GSSI), Viale Francesco Crispi 7, 67100 L'Aquila, Italy
          \and
             INFN Laboratori Nazionale del Gran Sasso, Via Giovanni Acitelli 22, 67100 L'Aquila, Italy
          \and
             INAF-Osservatorio Astronomico di Capodimonte, Salita Moiariello 16, I-80131 Naples, Italy
          \and
             European Southern Observatory, Karl-Schwarzschild-Strasse 2, 85748 Garching bei München, Germany
}

  \abstract
   {Globular clusters (GCs) represent a valuable tool as a fossil tracer of the formation and evolution of galaxies and their environment. As such, studying the properties of these stellar systems provides crucial insights into the past formation and interaction events of the galaxies, especially in galaxy group and cluster environments.}
   {We study the properties of globular cluster (GC) candidates in an area of 1.25 $\times$ 1.03 sq. degrees centred on the NGC\,5018 group of galaxies using the deep, wide field and multi-passband ($ugr$) observations obtained with the VLT Survey Telescope (VST) as part of the VST Elliptical GAlaxy Survey (VEGAS). With a focus on studying small stellar systems (SSS) associated with bright galaxies, this paper is a continuation of the VEGAS-SSS series to investigate the GCs in the NGC\,5018 group.}
   {We derived photometric catalogues of compact and extended sources in the area and identified GC candidates using a set of photometric and morphometric selection parameters. A GC candidates catalogue is provided and is inspected using a statistical background decontamination technique, benefiting from the wide area coverage of the data.}
   {The 2D distribution map of GC candidates reveals an overdensity of sources on the brightest member of the group, NGC\,5018. No significant GC overdensities are observed in the other bright galaxies of the group. We report the discovery of a candidate local nucleated LSB dwarf galaxy that is possibly in tidal interaction with NGC\,5018. The 2D map also reveals an intra-group GC population aligning with the bright galaxies and along the intra-group light (IGL) component of the group. The radial density profile of GC candidates in NGC\,5018 follows the galaxy surface brightness profile. The ($g-r$) colour profile of GC candidates centred on this galaxy shows no evidence of the well-known colour bimodality, which is instead observed in the intra-group population. From the GC luminosity function (GCLF) analysis, we find a low specific frequency $S_{\!\rm N}=0.59 \pm 0.27$ for NGC\,5018, consistent with previous results based on less deep optical data over a smaller area. This relatively low $S_{\!\rm N}$ and the lack of colour bimodality might be due to a combination of observational data limitations and the post-merger status of NGC\,5018, which might host a population of relatively young GCs. For the intra-group GC population, we obtain a lower limit of $S_{\!\rm {N,gr}}\sim0.6$. Using the GCLF as a distance indicator, we estimate that NGC 5018 is located $38.0 \pm 7.9$ Mpc away, consistent with values in the literature.}  
   {}   
   

\maketitle
%

\section{Introduction}
\label{sec:intro}

Globular clusters (GCs) are found in galaxies of all morphologies and sizes spanning a wide range of masses and magnitudes \citep[e.g.][]{harris91,georgiev10,harris13}. Being primarily old stellar systems, studying the spatial distributions of GCs and their stellar populations through their colours and phase-space distribution (if radial velocities are available) can provide crucial insights into the interaction events of the host galaxy and, more broadly, the galaxy group or cluster environment \citep{brodie06,georgiev09}. Observations of GCs are used to constrain the star formation and assembly history of galaxies and are valuable tools in theoretical and observational astronomy across a wide range of research topics in extragalactic astrophysics \citep[]{harris01,peng08,durrell14,forbes20}.

A GC is a class of dense stellar agglomerates characterised as bright (mean absolute magnitude $M_V$ = \textminus 7.5 mag), compact (mean effective radii $r_{\rm e}$ = 3 parsec)\footnote{Effective radius is defined as the radius within which half of the total luminosity of an extended source
is contained. This quantity is widely used to report the sizes of GCs ($r_{\rm e}$) and galaxies ($R_{\rm e}$).} and old (ages typically $\geq$ 10 Gyr) with a stellar mass usually ranging between $10^4$ and $10^6$$M_\odot$ \citep{harris01}. As a first approximation, GCs host a simple stellar population (SSP), that is they have a single age and single metallicity. However, studies of GCs in our Galaxy \citep[e.g.][]{piotto07,carretta2010} and the Magellanic Clouds \citep[e.g.][]{milone17,martocchia2019} have revealed the presence of multiple stellar populations with varying light element abundances. Despite this, GCs host stellar populations that are much simpler than those of galaxies, in terms of iron abundance and age distributions \citep{gratton19}. Therefore the intrinsic simplicity of GCs, together with their old age and high luminosity, makes them a powerful and robust tracer of a galaxy and its environment out to cosmological distances \citep{alamo13,lee2022,harris2023}. In the last few decades, and particularly after the Hubble Space Telescope (HST) began collecting data with superb image quality, the increased depth and quality of multi-passband data available for the detection of extragalactic GC systems have allowed astronomers to unveil numerous properties that represent a valuable tool to constrain the host galaxy properties. 
Because of the near-universal Gaussian shape of the GC luminosity function (GCLF) in the optical and near-IR passbands, that has a peak called the Turn-Over magnitude (TOM) at $M_{V}^{\rm TOM}$ $\sim$ \textminus 7.5 mag \citep[]{rejkuba12}, the use of old GCs has increased as a standard candle to act as a distance indicator \citep[]{hanes77,lee2018,ferrarese2020}.

Most GC systems exhibit a bimodal colour distribution with a blue and a red peak \citep{brodie06}. A combination of photometric and spectroscopic observations indicate that these GCs are quite homogeneous in terms of their old age, and so the colour differences mainly reflect their metallicity differences. \citet{brodie06} attributed this to the hierarchical formation which gives rise to two distinct GC sub-populations that have different peak metallicities. While the red GCs follow the radial trend of the stars in the bulge or the spheroid of the host galaxy and are more concentrated in their inner regions, the blue GCs on the other hand are observed to have a wider radial extension \citep[e.g.][]{harris09,dabrusco16,cantiello20,hazra22,dabrusco22}. This seems to indicate that the red GC population is formed in-situ, while the blue GCs could represent a population acquired through mergers, accretion and other tidal interactions \citep{brodie06,forbes11}.

\citet{harris81} introduced the parameter specific frequency ($S_{\!\rm N}$) as a measure of the richness of the GCs normalised to their host galaxy's luminosity. It is given by the following equation:
\begin{linenomath}
\begin{equation}
\label{eqn:spec_freq}
S_{\!\rm N} = N_{\rm GC} \times 10^{0.4(M_V+15)}
\end{equation}
\end{linenomath}
where $N_{\rm GC}$ is the total number of GCs and $M_{V}$ is the $V$-band absolute magnitude of the host galaxy. This quantity has since been widely used to assess galaxy formation mechanisms. 
Recent discussions on this topic can be found in \citet{georgiev10}, \citet{harris13}, \citet{mueller21} and \citet{marleau24}. In particular, \citet{harris13} studied the GCs of a large sample of galaxies of different morphologies situated in different environments and showed the behaviour of $S_{\!\rm N}$ with $M_{V}$. They found that the mid-range luminosity galaxies ($10^9-10^{\rm 10} L_\odot$) form a relatively uniform group with $S_{\!\rm N}$ $\approx$ 1, while the dwarfs and giant galaxies at the opposite ends show a higher mean $S_{\!\rm N}$ (although with a large scatter). 

Interactions between galaxies can lead to the stripping of field stars and GCs, forming tidal debris and contributing to the diffuse intra-group light (IGL) observed in galaxy systems \citep[e.g.][]{arnaboldi2022,spavone18}. The IGL serves as an effective tracer of stripped stellar components, providing valuable insights into the past interactions and history of galaxy systems \citep{arnaboldi2022,ahvazi2024,kluge2024}. Given that groups host at least 50\% of galaxies in the universe today \citep[]{munoz13}, studying GCs in these environments is key to understanding galaxy evolution.

Much remains to be understood about GCs. The origin of the observed colour bimodality is debated, with studies like \citet{yoon06} and \citet{cantiello07} suggesting it could result from a continuous metallicity distribution combined with non-linear colour-metallicity relations. We also lack clarity on the mechanisms responsible for stripping and dispersing GCs from their host galaxies, as well as their relative efficiencies \citep{forbes18}. Additionally, it is unclear what fraction of the original GC population has survived to the present day compared to the early formation stages of their host galaxies \citep{brodie06}. Another important parameter is the radial extent of the GC systems. In bright, massive galaxies,  half of the total GC population typically extends out to 3 to 5 times the $R_{\rm e}$ of the host galaxy \citep{forbes17}. However, GCs can also be found at much larger galactocentric distances, tracing galaxy formation and interactions \citep{lamers2017}. A comprehensive discussion on the open questions regarding GC formation and evolution within a cosmological framework can be found in \citet{forbes18}.


In this work, we use the deep, multi-passband, wide-field imaging of the NGC\,5018 galaxy group collected with the VLT Survey Telescope (VST) as part of the VST Elliptical GAlaxy Survey \citep[VEGAS; P.I. E. Iodice,][]{capaccioli15,iodice21} and conduct a systematic study of its GC system. The paper is organised as follows. Section \ref{sec:5018_literature} provides a literature review on the NGC\,5018 group. In Sect. \ref{sec:obs_and_data}, we provide the details of the data used. Section \ref{sec:data_analysis} describes the data analysis procedure. The process adopted for GC selection is explained in Sect. \ref{sec:gc_select}. We present the results of our analysis in Sect. \ref{sec:results}. The discussion of our results is provided in Sect. \ref{sec:discussion} and is summarised in Sect. \ref{sec:summary}.

\section{NGC\,5018 group from the literature}
\label{sec:5018_literature}

The NGC\,5018 galaxy group is named after its brightest member, NGC\,5018, which is a massive elliptical galaxy. The other bright galaxies in the group include the edge-on spiral NGC\,5022, the face-on spiral NGC\,5006 and two lenticulars MCG-03-34-013 and PGC\,140148  \citep{kourkchi17}. Table \ref{tab:gal_prop} lists the physical properties of these galaxies. For NGC\,5018, we adopt a group distance modulus of $32.8 \pm 0.1$ mag \citep{tully23}, derived from the weighted average of all the available measures from the literature, based on Type Ia Supernova, Fundamental Plane  and Tully-Fisher Relation. Using this value, we estimate the TOM of the GCLF in this galaxy at $\sim$ 26.7 mag, $\sim$ 25.3 mag and $\sim$ 24.7 mag in $u$-, $g$- and $r$-passbands, respectively (more details on TOM estimation in Sect. \ref{sec:phot_select}).

\begin{table*}[htb!]
    \centering
    \caption{Properties of the bright galaxies in the NGC\,5018 group.}
    \begin{tabular}{cccccccc}
    \hline
    \\[-2ex]
        & NGC\,5018 & NGC\,5022 & NGC\,5006 & MCG-03-34-013 & PGC\,140148\\
    \hline
    \\[-2ex]
      R.A. (J2000)$^a$ & \ \ \ 198.254305 & \ \ \ 198.378292 & \ \ \ 197.940708 & \ \ \ 198.078792 & \ \ \ 198.199456 \\ 
      
      Dec. (J2000)$^a$ & \ \ -19.518193 & \ \ -19.546639 & \ \ -19.261750 & \ \ -19.446028 & \ \ -19.371685 \\
      
      Morphology$^a$ & E3 & SBb pec & (R)SB0 & S0 & S0 \\

      $R_{\rm e}$ (arcmin)$^b$ & 0.54 & 0.29 & 0.33 & 0.22 & NA \\
      

      $D_{group}$ (Mpc)$^c$ & \multicolumn{4}{c}{36.0$\pm$1.7} \\
      
      $M_B$ (mag)$^d$ & $-21.7 \pm 0.2$ & $-20.6 \pm 0.3$ & $-20.0 \pm 0.5$ & $-18.5 \pm 0.4$ & $-17.9 \pm 0.6$ \\
      
      $M^*_{\rm total}~(M_\odot)^e$  & $2.9 \times 10^{11}$ & $1.4 \times 10^{10}$ & $6.8 \times 10^{10}$ & $3.3 \times 10^{9}$ & $1.5 \times 10^8$ \\
      
      $cz$ (km/s)$^f$ & $2687 \pm 47$ & $2960 \pm 45$ & $2751 \pm 45$ & $2748 \pm 45$ & $3110 \pm 0$ \\

      $E_{B-V}$ (mag)$^g$ & 0.082 & 0.083 & 0.077 & 0.082 & 0.083 \\
      \hline
    \end{tabular}
    \tablefoot{$a$) Coordinates and morphologies are taken from the NASA/IPAC Extragalactic Database (NED); $b$) Effective radii of galaxies are taken from \citet{spavone18}, except for NGC\,5006 which is from NED; $c$) Group distance is from the Extragalactic Distance Database \citep[EDD;][]{tully23}; $d$) Absolute $B$-passband magnitudes (corrected for extinction) are taken from the HyperLeda database \citep{makarov14}; $e$) Total stellar masses are from \citet{spavone18}, except for NGC\,5006 and PGC\,140148 which are derived in this work (Sect. \ref{sec:5018_literature}); $f$) Heliocentric recession velocities are from \citet{jones09}, except for PGC\,140148 which is from \citet{kourkchi17}  and $g$) Reddening values from NASA/IPAC InfraRed Science Archive \citep[IRSA;][]{schlafly11}.}
    \label{tab:gal_prop}
\end{table*}

NGC\,5018 displays several signs of tidal interactions like shells, ripples, a complex system of dust lanes in the inner regions, a tail on the north-west side and a bridge of HI gas towards its neighbouring galaxy NGC\,5022 \citep[]{buson04,kim12}. The evidence of a perturbed inner structure indicates the occurrence of a minor merging event in the history of this galaxy \citep{spavone18}. The galaxy also shows some peculiar features such as a much weaker $\rm Mg_{2}$ index compared to ellipticals of similar absolute magnitude \citep{schweizer90}, a low UV flux level \citep{rampazzo07} and a metallicity lower than giant ellipticals \citep{bertola93}. 

HI observations of NGC\,5018 have revealed important evidence about the evolution of this galaxy. \citet{kim88} estimated the HI mass associated with NGC\,5018 and NGC\,5022 to be $\sim$ $4\times10^8$ $M_\odot$ and $\sim$ $2\times10^9$ $M_\odot$, respectively. \citet{guhathakurta90} made follow-up observations and found that the HI bridge actually connects NGC\,5022 and MCG-03-34-013 passing NGC\,5018 along the way. These authors estimate that the interaction that led to the observed plume in the north-west of NGC\,5018 is as recent as $\sim$ 600 Myr and suggest this was the first direct observational evidence for the formation of a shell system through the merger of an elliptical galaxy with a cold disk system.

X-ray observations have shed further light on the evolution of NGC\,5018. \citet{ghosh05} studied this galaxy using the Chandra X-ray Observatory's Advanced Imaging Charge-Coupled Device (CCD) Spectrometer. They detect diffuse hot gas that could be the remnant of interactions of NGC\,5018 with its neighbouring galaxies. These authors suggest that although there is little current star formation in the galaxy, there is a large enough reservoir of gas to maintain a low but steady level of star formation explaining the observed diffuse X-ray emission. Indeed, there seems to be evidence for the presence of a relatively young ($\sim$ 3 Gyr) stellar population in the central regions of NGC\,5018 \citep[]{bertola93,hilker96,leonardi2000,buson04}.

The GC system of NGC\,5018 has been studied previously. \citet{hilker96}, using $6.4\arcmin \times 6.4\arcmin$ area observations in the Bessel $V$- and Gunn $i$-passbands (each with an exposure time of $\sim$ 1 hr) from the 1.54m Danish telescope at the European Southern Observatory (ESO) at La Silla in Chile, detected a poor GC system with $S_{\!\rm N}$ = $1.10 \pm 0.60$ (a higher limit). These authors find that the GCs they detected in NGC\,5018 can be divided into two sub-populations: a small population of young GCs (with age between several hundred Myr to 6 Gyr) and a second, larger population of older GCs. They suggest that the younger GCs must have formed during the last tidal interaction with one of the neighbouring galaxies and the older GCs could have formed in-situ representing the original population of the galaxy. \citet{hilker96} also report the presence of a dust lane across NGC\,5018, extending along south-east and north-west direction, which leads to a lack of blue GCs ($V-I$ < 0.8). 
\citet{humphrey09} studied GCs in NGC\,5018 as a part of their sample of 19 early-type galaxies using the Hubble Space Telescope (HST) Wide Field Planetary Camera 2 (WFPC2; exposure time of 13 and 30 minutes in the HST equivalent of $V$- and $I$-passbands, respectively). Similar to \citet{hilker96}, they find a poor GC system and estimate $S_{\!\rm N}$ = $0.46 \pm 0.22$ by inspecting an area of $2.5\arcmin \times 2.5\arcmin$ around the core of the galaxy. 
No known works exist in the literature on the GCs in the other four bright galaxies of the group.

Using the same dataset as the present work, \citet{spavone18} investigated the IGL in the NGC\,5018 galaxy group, focusing on the region containing NGC\,5018, NGC\,5022 and MCG-03-34-013. They found that about 41\% of the $g$-band luminosity of the group ($L_g = 1.7 \times 10^{11}$ $L_\odot$) is attributed to IGL. The estimated colour of the IGL component ($g-r = 0.78 \pm 0.35$ mag) is consistent with the halo of NGC\,5018, aligning with findings in similar groups \citep{white2003,darocha2008}. Their analysis suggests that stellar stripping within the group is ongoing, with the total accreted mass fraction ranging from 78\% to 92\%. Additionally, they provided stellar mass estimates for the three galaxies involved (see Table \ref{tab:gal_prop}). In addition to the three galaxies studied in detail by \citet{spavone18}, in Table \ref{tab:gal_prop}, we also report the stellar masses for NGC\,5006 and PGC\,140148. 
For this,  we use the updated theoretical colour-$M_*/L$ relations in the optical from \citet{into13}. These authors provide the colour-$M_*/L$ relations for exponential, disk and dusty galaxy models for all the optical and NIR bands. We use the ($B$-$K$) colour and $K$-passband magnitude from the Hyperleda database \citep{makarov14} to get the stellar mass estimate for NGC\,5006 using the disk model. Due to a lack of a reliable $K$-passband magnitude for PGC\,140148, we use the ($B$-$I$) colour and $I$-passband magnitude to get its stellar mass estimate using the disk model.

\section{Observations and Data Reduction}
\label{sec:obs_and_data}

The observational dataset used in this work is a part of the VEGAS survey \citep[P.I. E. Iodice;][]{capaccioli15,iodice21}\footnote{Project webpage: \url{http://old.na.astro.it/vegas/VEGAS/Welcome.html}}. It is a deep, multi-passband ($u, g, r, i$) imaging survey carried out using the VST telescope, an Istituto Nazionale di Astrofisica (INAF) facility\footnote{Facility website: \url{https://vst.inaf.it}}, located at the Paranal Observatory in Chile. This survey was approved as a Guaranteed Time Observation project for the period 2016-2021 with a total allocated observing time of $\sim$ 500 hrs. VST is an optical telescope with an aperture size of 2.6m and a field of view (FoV) of 1 sq. degree offered by the OmegaCAM camera that has a resolution of 0.21 arcsec/pixel \citep{kuijken11}. 
A detailed description of the survey, the selected targets and the main scientific aims can be found in \citet{capaccioli15}.

Thanks to the wide FoV available and long exposure times, the VEGAS survey has acquired deep optical data which has enabled us to:
\begin{enumerate}

    \item Study galaxy outskirts out to 8--10 $R_{\rm e}$ and detect IGL and Low Surface Brightness (LSB) features in the intra-cluster/group space \citep[]{spavone18,iodice21,ragusa22};  
    
    \item Estimate the mass assembly of galaxies by deriving the accreted mass fraction in the stellar halos \citep[][and references therein]{spavone17,spavone17aa}; 
    
    \item Trace the properties of GCs out to large galactocentric distances (up to 20 $R_{\rm e}$ or more) in different galaxy environments \citep[]{cantiello15,dabrusco16};
    
    \item Make a census and provide a large catalogue of dwarf galaxies and Ultra Diffuse Galaxies (UDGs) in several clusters like Fornax \citep{venhola18}, Hydra I \citep{lamarca2022} and galaxy groups like NGC\,3640 \citep{mirabile2024}.
    
\end{enumerate}

\begin{figure*}[htb!]
   \centering   \includegraphics[width=18.5cm]{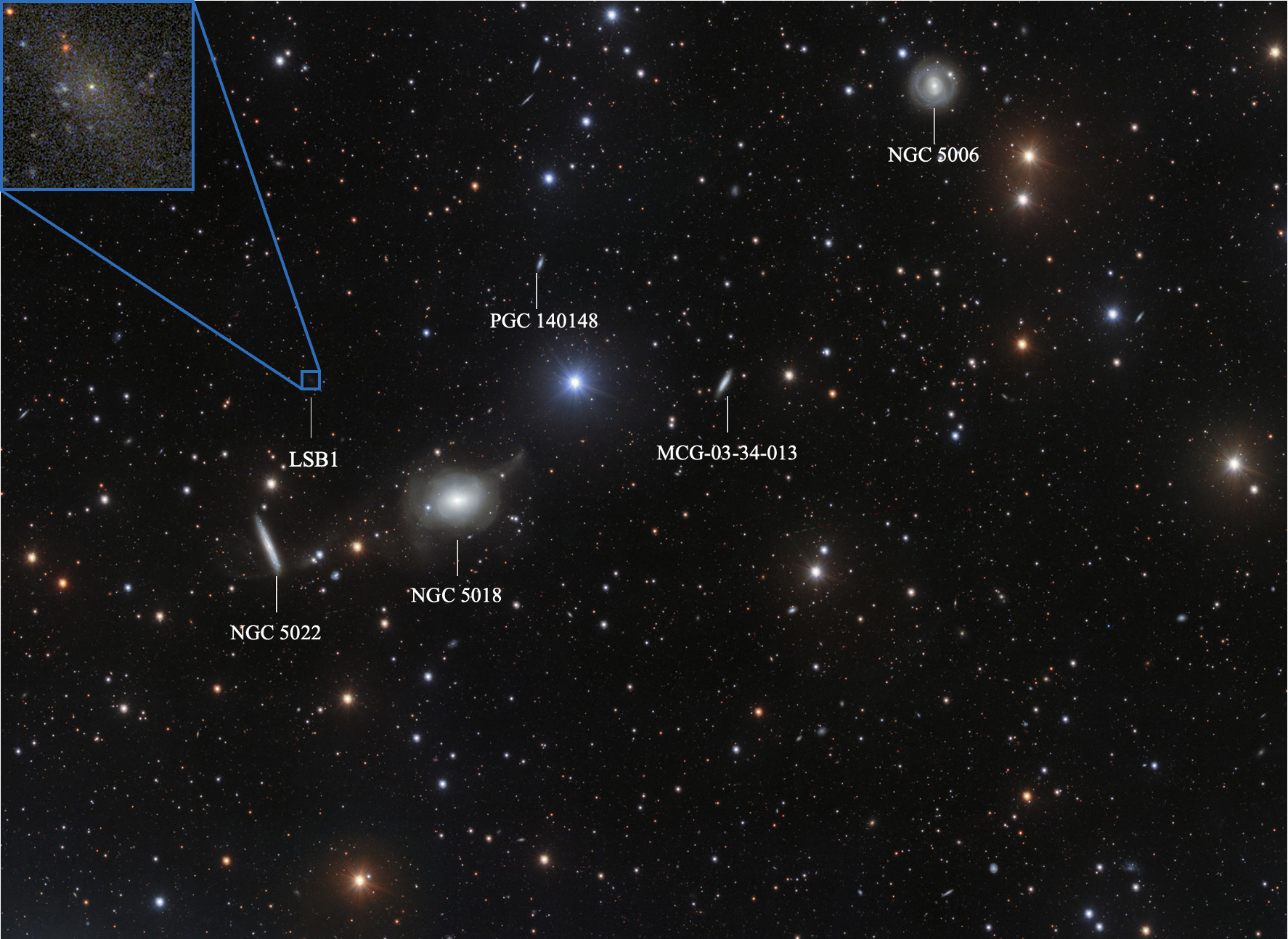}
   \caption{A colour composite image of the bright galaxies in the NGC\,5018 group. Numerous foreground stars from our Milky Way and some background galaxies are visible in this image. The image size is 48\arcmin $\times$ 60\arcmin; north is up and east is to the left. The inset (0.65\arcmin $\times$ 0.65\arcmin) shows an LSB dwarf galaxy candidate (NGC\,5018-LSB1) that we report on in this work \citep[Credit: ESO/][]{spavone18}.
   }
    \label{fig:5018_group}
\end{figure*}

\begin{table*}[htb!]
    \centering
    \caption{VST/OmegaCAM data and image properties.}
    \begin{tabular}{cccccc}
    \hline
    \\[-2ex]
    Band & Exposure time (hrs) & FWHM (arcsec)$^a$ & $m_{80\%} \ (\rm mag)^b$ & $\rm a.c. \ (mag)^c$ & $\Delta mag \ (\rm mag)^d$ \\
    \\[-2ex]
    \hline
    \\[-2ex]
      $u$ & 4.08 & 1.2 & 22.8 & $0.54 \pm 0.02$ & $-0.036 \pm 0.085$\\
      $g$ & 3.46 & 0.8 & 24.0 & $0.34 \pm 0.01$ & $-0.046 \pm 0.028$ \\
      $r$ & 3.33 & 0.9 & 23.2 & $0.46 \pm 0.02$ & \ \ $0.043 \pm 0.046$ \\
      \\[-2ex]
    \hline
    \end{tabular}
    \tablefoot{$a$) Median FWHM value for point sources in the field; $b$) 80\% completeness limit on NGC\,5018; $c$) Aperture correction values (median and $RMS_{\rm MAD}$); $d$) Photometric comparison with APASS for point sources in the field (median and $RMS_{\rm MAD}$ values).}
    \label{tab:obs_image_prop}
\end{table*}

The $u$-, $g$-, and $r$-band data used in this work (no $i$-band observations were taken for this field) were obtained in visitor mode and processed with the VST-Tube pipeline \citep{grado12}. The procedure is described in detail in Appendix A of \citet{capaccioli15}. A normalisation of Zero Point (ZP) to 30.0 mag was applied to the final co-added images. The calibration was done using a Point Spread Function (PSF) magnitude within 4.0\arcsec \ circular aperture using standard stars from the Sloan Digital Sky Survey \citep[SDSS;][]{blanton17}. Figure \ref{fig:5018_group} shows the ESO press release colour composite image of the NGC\,5018 group\footnote{Press release: \url{https://www.eso.org/public/news/eso1827/}}. Table \ref{tab:obs_image_prop} lists the VST exposures used in this work. Our work is focused on the detection and characterisation of GCs in the 1.25 $\times$ 1.03 sq. degrees area centred on this galaxy group.

For the NGC\,5018 field, images were acquired using a step-dither observing strategy, involving a cycle of short exposures on both the target and an adjacent field taken close in time and space. This method allows for a larger field of view around galaxies with large angular sizes, though it requires more telescope time compared to standard dithering. The technique enhances sky background estimation around bright, extended galaxies. More details on the data reduction process are available in \citet{spavone17aa}.

\section{Data analysis}
\label{sec:data_analysis}

\subsection{Galaxy modelling and subtraction}
\label{sec:gal_model_and_resid}

Detecting faint sources in NGC\,5018 is particularly challenging due to its peculiar morphology and the presence of dust in its inner regions. To enhance GC detection, a useful technique is to model and subtract the galaxy  light profile from the image. For this work, we employed the strategy developed by \citet{hazra22}, which is briefly outlined below.

To model the light distribution of the galaxy, we use the Python Elliptical Isophote Analysis package \citep{bradley24}\footnote{\url{https://photutils.readthedocs.io/en/stable/isophote.html}}. As a first step, we obtain a cutout frame of NGC\,5018 from the original full frame. The size of the cutout is 8.4\arcmin $\times$ 8.4\arcmin \ extending out to $\sim$ 8$R_{\rm e}$ on each side ($R_{\rm e}$ = 0.54\arcmin \ for NGC\,5018 in $g$-passband; see Table \ref{tab:gal_prop}). Selecting a wide region makes sure that we cover the entire light distribution of the galaxy to get a good model while also avoiding getting too close to its neighbouring galaxies: NGC\,5022 and MCG-03-34-013. To obtain a reliable galaxy model, we mask all the bright sources in and around NGC\,5018 within the cutout frame. We then use the ellipse task based on the algorithm by \citet{Jedrzejewski87} to fit the isophotes taking as input our initial guesses for the position angle, ellipticity and the centre of the galaxy. In the fitting process, all geometric and photometric parameters are allowed to vary. Then, using the "build\_ellipse\_model" task from the Isophote package, we create a model of the galaxy, which is subtracted from the cutout frame to obtain the residual image. This procedure (masking, modelling, and subtraction) is repeated until the residuals are satisfactory, that is, not dominated by modelling artefacts but instead by potential merger signatures. Finally, we pad the residual frame to the original full frame. This residual full frame now contains our data with the light distribution of NGC 5018 subtracted for more efficient detection of faint, point-like sources.

\begin{figure*}[htb!]
    \centering    
    \begin{minipage}{0.33\textwidth}
        \centering        \includegraphics[width=0.997\textwidth]{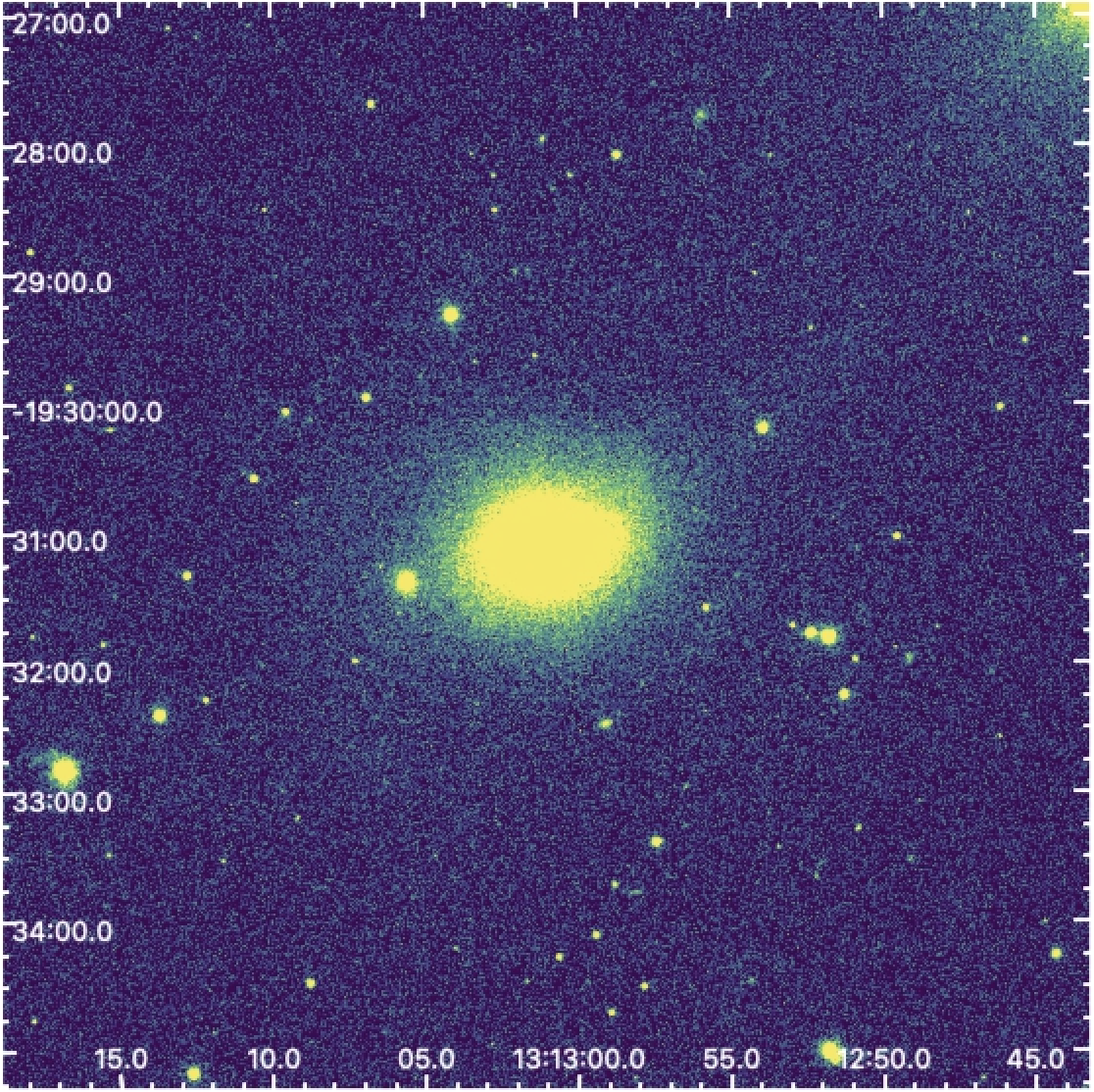}
    \end{minipage}
    \begin{minipage}{0.33\textwidth}
        \centering        \includegraphics[width=0.997\textwidth]{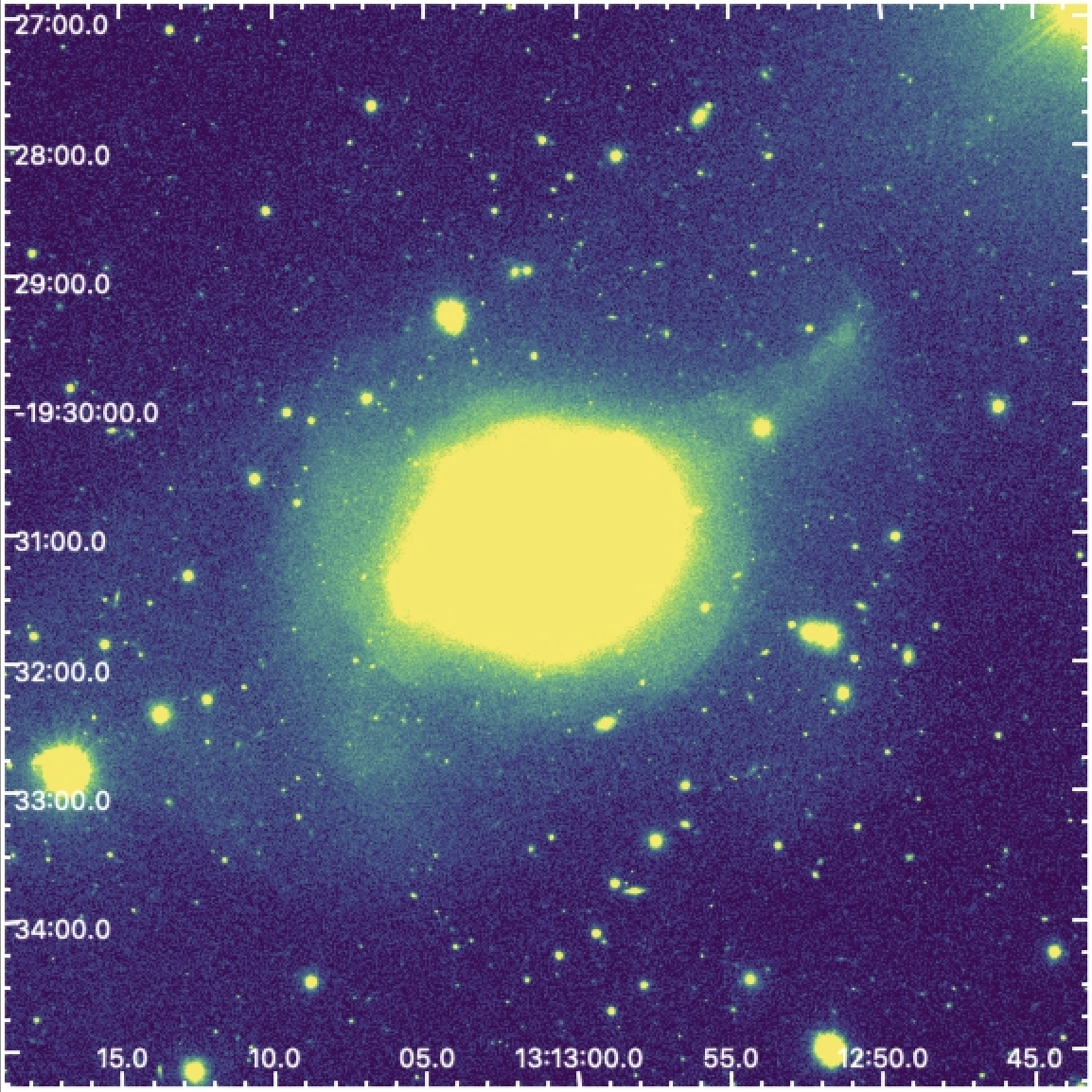}
    \end{minipage}
    \begin{minipage}{0.33\textwidth}
        \centering        \includegraphics[width=0.997\textwidth]{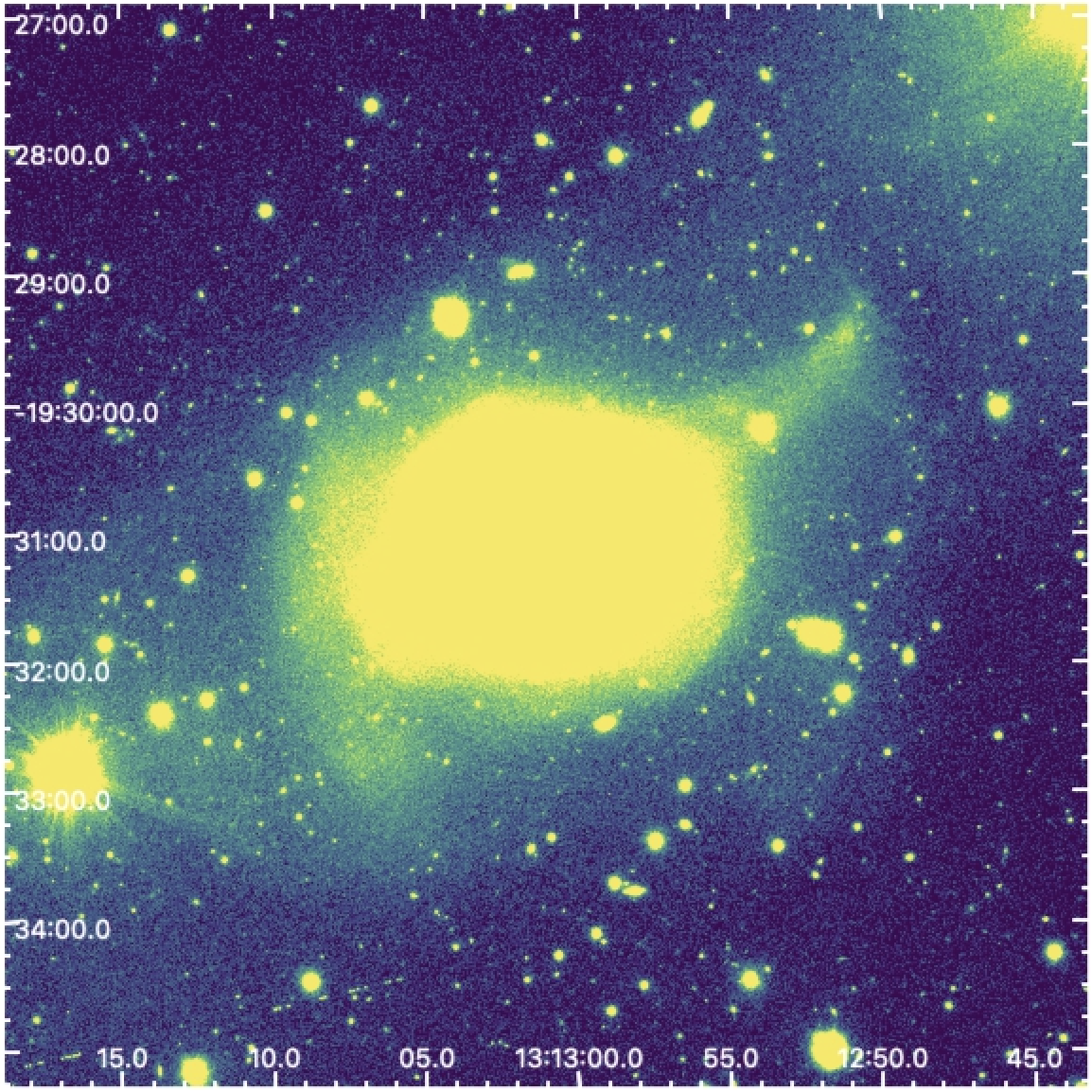}
    \end{minipage}

    \centering    
    \begin{minipage}{0.33\textwidth}
        \centering        \includegraphics[width=0.997\textwidth]{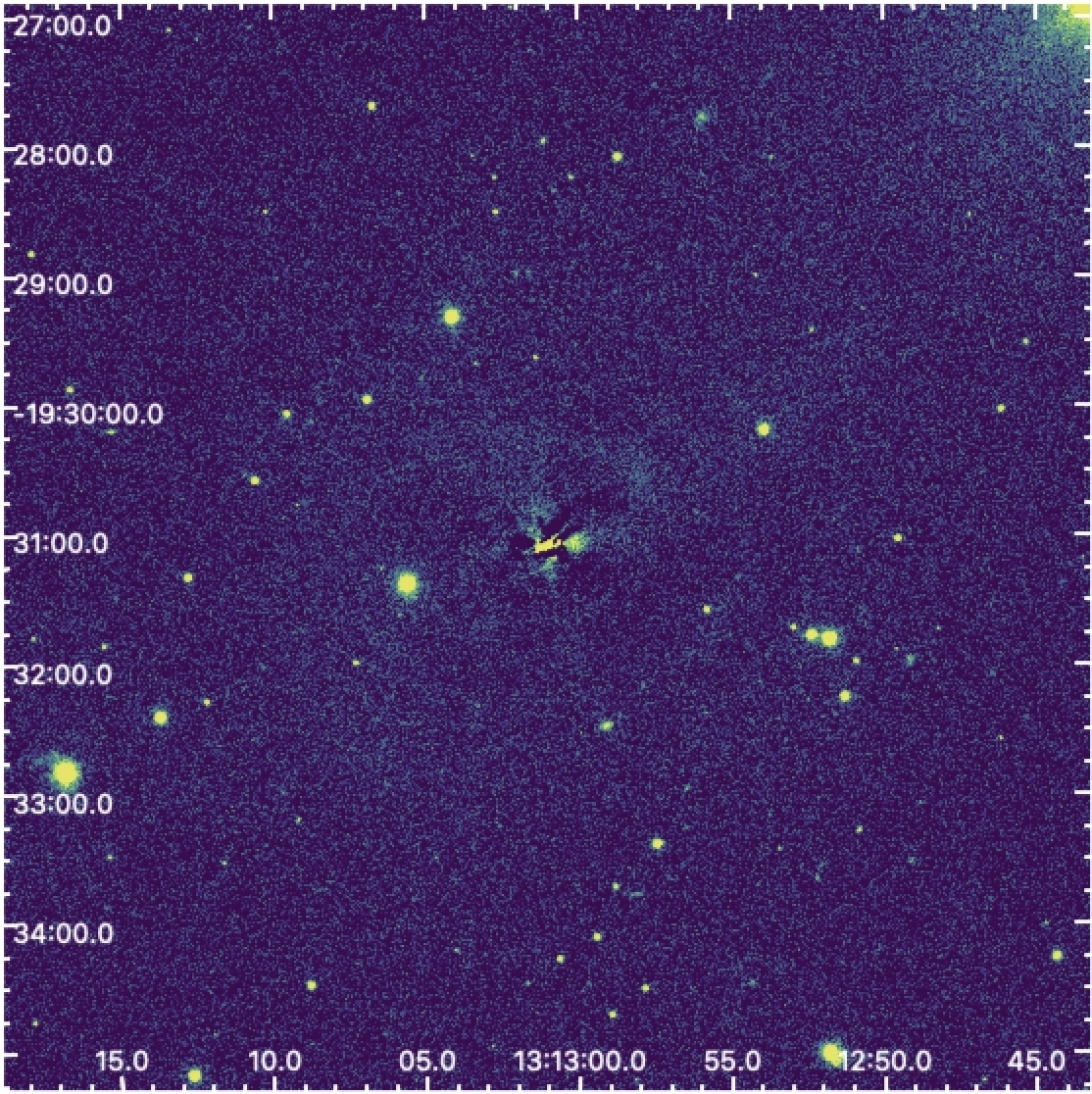}
    \end{minipage}
    \begin{minipage}{0.33\textwidth}
        \centering        \includegraphics[width=0.997\textwidth]{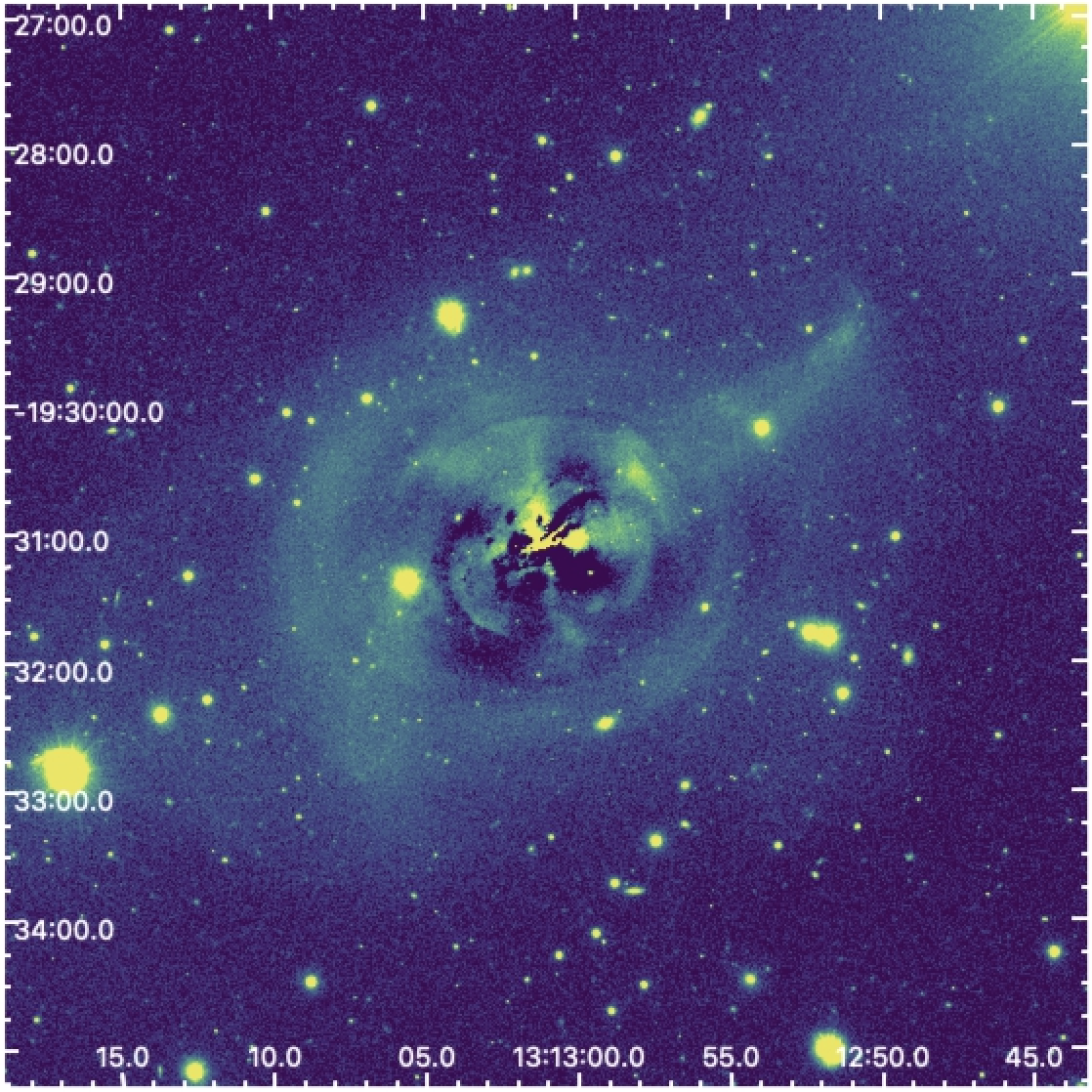}
    \end{minipage}
    \begin{minipage}{0.33\textwidth}
        \centering        \includegraphics[width=0.997\textwidth]{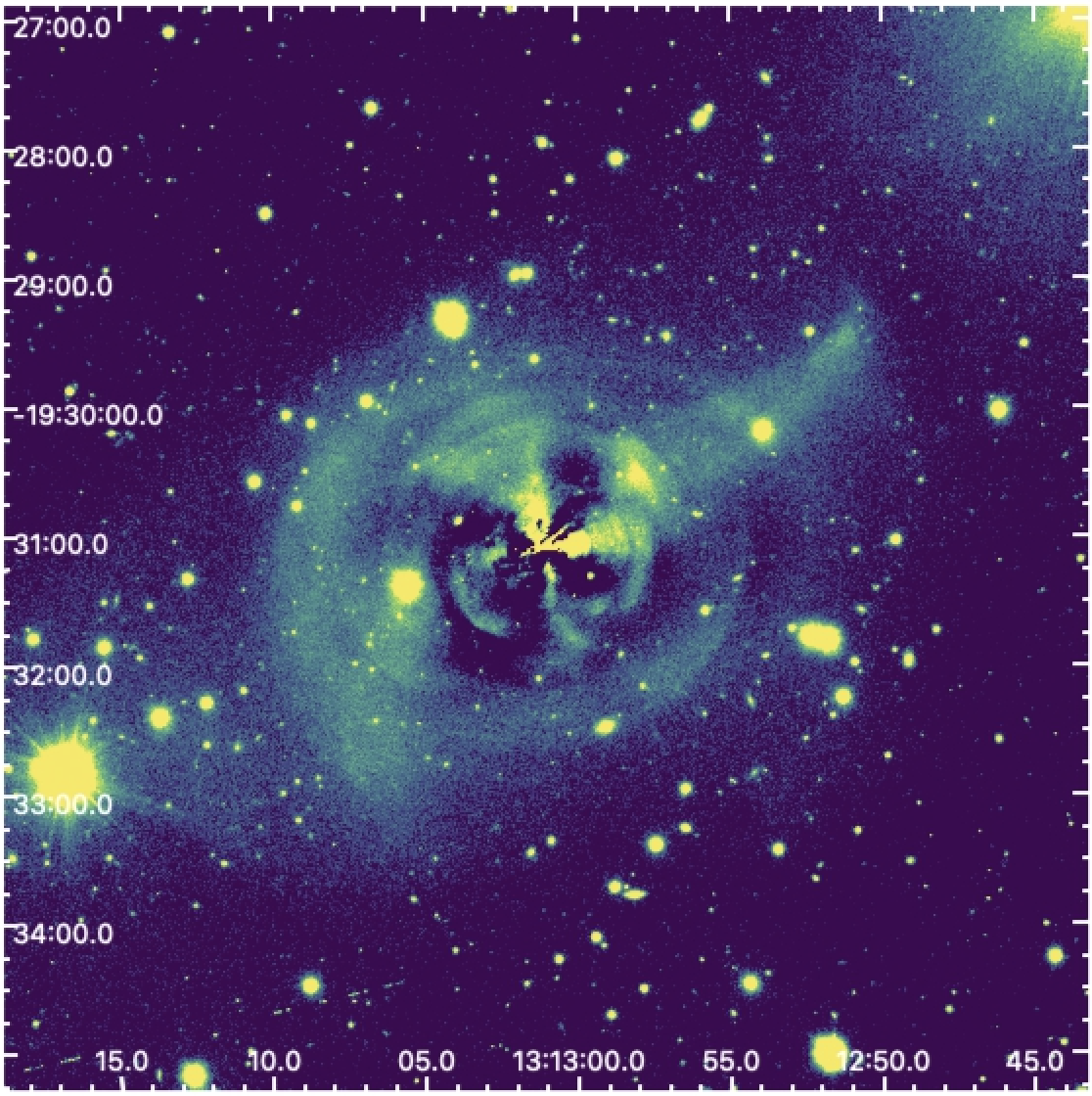}
    \end{minipage}
    \caption{Cutout frames centred on NGC\,5018 (top panels) and residuals obtained after model subtraction (bottom panels) in $u$- (left), $g$- (middle) and $r$- (right) bands. Each frame is of the size 8.4\arcmin $\times$ 8.4\arcmin; north is up and east is left.}
    \label{fig:resid}
\end{figure*}

Figure \ref{fig:resid} illustrates the results of our procedure where we show the cutout frames of NGC\,5018 (top panels) and the residuals obtained after model subtraction (bottom panels) for all the available passbands. Many faint sources now become visible in the residual frames. Some diffuse LSB features are also visible. These are mostly shells, ripples and filaments discussed in Sect. \ref{sec:5018_literature}. 

\subsection{Source detection and photometry}
\label{sec:source_det_and_phot}

To obtain the photometry of sources in our field, which is tailored to detect compact sources: stars, GCs and unresolved galaxies, we use the photometry and source detection tool SExtractor \citep{bertin96}. To optimise the detection of faint sources close to the bright galaxies, we adopt the following strategy: 
\begin{enumerate}

    \item Run SExtractor on the residual full frame with initial guesses for the input parameters (e.g FWHM, detection threshold, analysis threshold) using the weight maps from the VST-Tube pipeline;
    
    \item Visually inspect the image generated by SExtractor where all the detected sources are subtracted from the residual full frame ('\textminus OBJECT' frame). This is done to check if all the sources in the frame are detected, especially in the central regions of NGC\,5018;
    
    \item Check if the FWHM parameter we set initially is correct by comparing it with FWHM measured on the bright, compact and isolated sources after the initial SExtractor run;
    
    \item Run SExtractor again on the residual full frame with the updated input parameters to improve the detection, adopting a more consistent FWHM.
    
\end{enumerate}

Individual catalogues of sources in the observed field were obtained in the $u$-, $g$- and $r$-passbands using the above strategy and were matched using a 1\arcsec \ matching radius. The catalogue matched in all the three passbands ($ugr$ matched catalogue hereafter) contains $\sim$ 12000 sources, whereas the catalogue matched in $g$- and $r$-passbands ($gr$ matched catalogue hereafter) contains $\sim$ 100000 sources. The significant difference between the two matched catalogues is due to the highly incomplete $u$-band photometry, which will be discussed in later sections.

The following colour corrections, provided by the VST-Tube pipeline \citep{grado12}, were applied to the magnitudes obtained from the SExtractor run:
\begin{linenomath}
\begin{equation}
\label{eqn:col_corr}
\begin{aligned}
m_{u_{corr}}& = m_u + 0.0255\times(m_u{-}m_g) \\
m_{g_{corr}}& = m_g + 0.0288\times[1.3336\times(m_g{-}m_r)+0.0444] \\
m_{r_{corr}}& = m_r + 0.0406\times[0.3853\times(m_g{-}m_r)+0.0137] \\
\end{aligned}
\end{equation}
\end{linenomath}

As in other works on GCs using the VEGAS survey data \citep[e.g.][]{dabrusco16,cantiello18a,cantiello20}, we use the magnitude within an 8 pixel circular aperture as our reference magnitude. This means that the flux of a source is measured within this finite aperture and, thus, there is a need to correct for the flux outside the 8 pixel radius. To derive this aperture correction (a.c.), we measure the difference in magnitude 
at our selected aperture size and a larger aperture size of 19.05 pixels. This larger aperture size was selected based on the fact that the data is reduced by the VST-Tube pipeline which uses a calibration radius of 19.05 pixels \citep[i.e. 4$\arcsec$,][]{grado12}. The a.c. values for the three passbands are listed in Table \ref{tab:obs_image_prop}. To obtain these values, we select only the bright and isolated stars from the individual photometric catalogues from SExtractor.

To verify the quality of our photometry, we compare it with the AAVSO Photometric All-Sky Survey (APASS). First, we obtain a catalogue of sources in the observed field from APASS using the VizieR catalogue query tool\footnote{\url{https://vizier.cds.unistra.fr/viz-bin/VizieR?-source=II/336}}. As APASS lacks $u$-passband, we adopt the strategy described in \citet{cantiello20}. We transform the $B$-passband magnitude from APASS to $u$-passband using the  transformation equations available at the SDSS web pages\footnote{\url{http://www.sdss3.org/dr8/algorithms/sdssUBVRITransform.php}}. In particular, we use the following equation:
\begin{linenomath}
\begin{equation}
u = B_{\rm APASS} + 0.8116\times(u-g)_{\rm fit} \ {-} \ 0.1313
\end{equation}
\end{linenomath}
where $(u-g)_{\rm fit}$ colour index is derived from the APASS ($g-i$) and ($g-r$) colour indices using a second-degree polynomial fit derived from the SDSS data over the M87 region. We select M87 particularly because it is situated at a high galactic latitude and thus has very small extinction. Therefore, by using the $u$-passband magnitude of sources in APASS derived as a function of the $B, g, r$ and $i$ photometry, we proceed with the photometric comparison. For this, we select only the bright and isolated stars from the $ugr$ and APASS matched catalogue (matched using a 1\arcsec \ radius). For this work, we adopt AB mag photometric system.

\begin{figure*}[htb!]
    \centering    
    \begin{minipage}{0.33\textwidth}
        \centering        \includegraphics[width=0.997\textwidth]{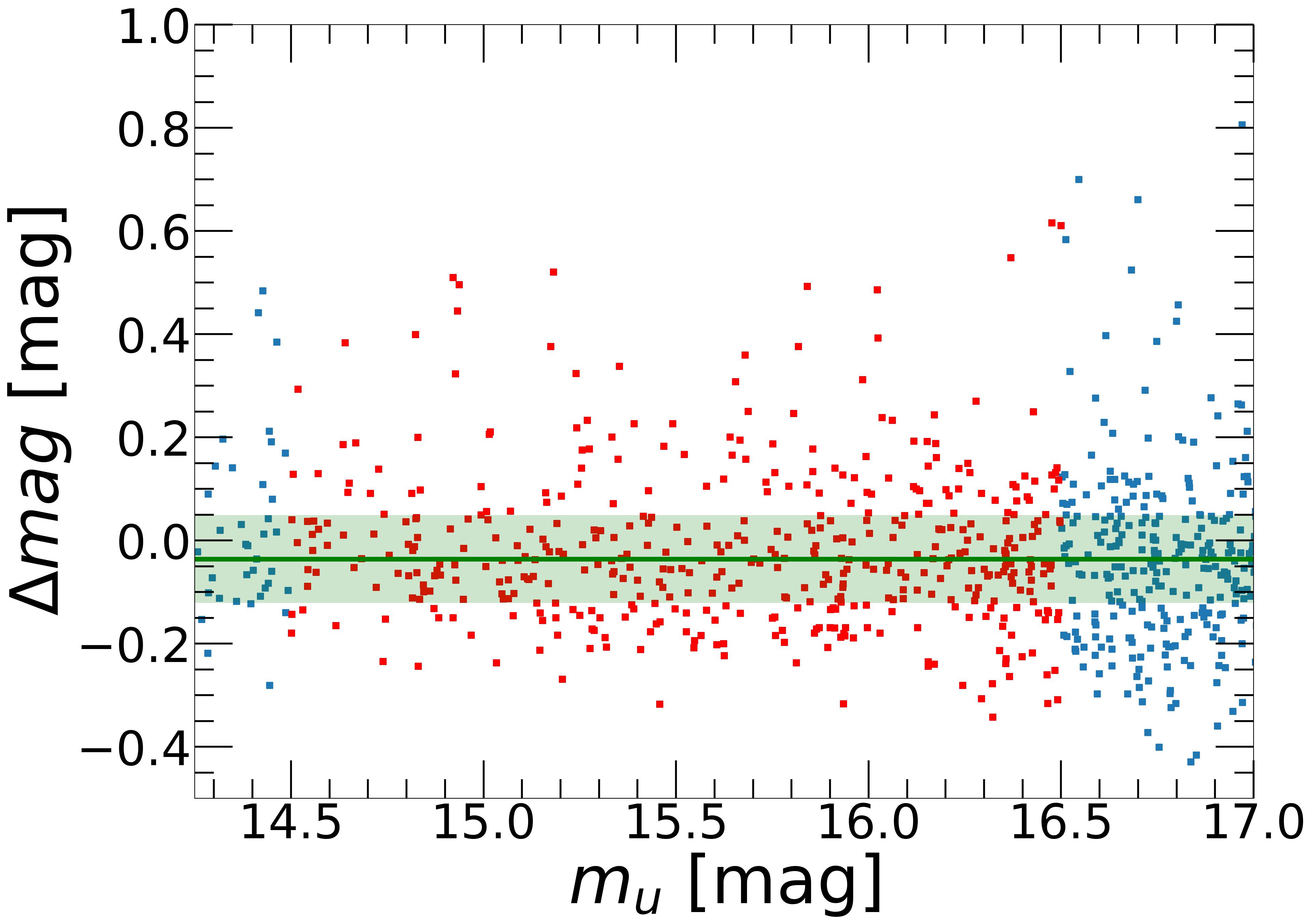}
    \end{minipage}
    \begin{minipage}{0.33\textwidth}
        \centering        \includegraphics[width=0.997\textwidth]{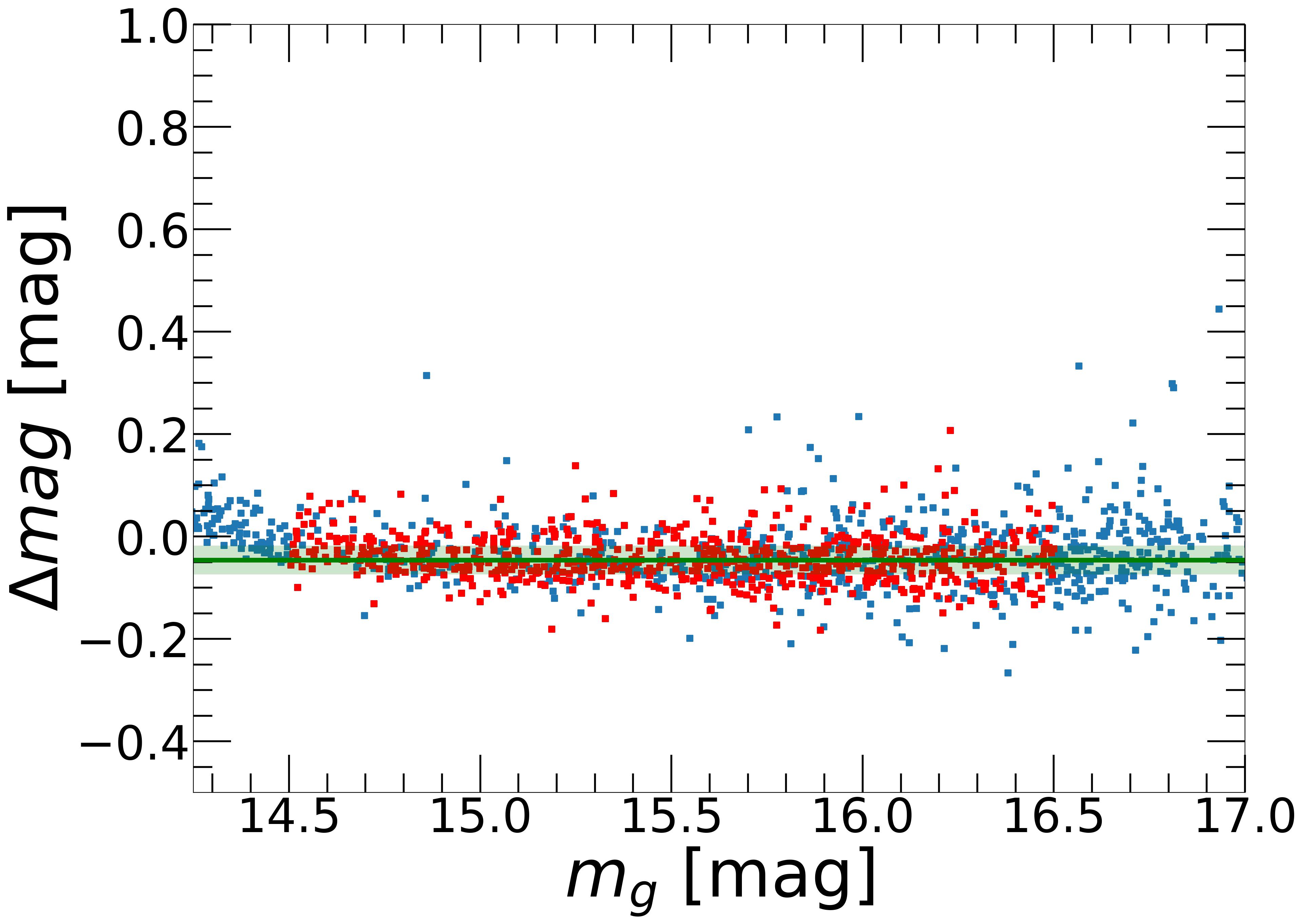}
    \end{minipage}
    \begin{minipage}{0.33\textwidth}
        \centering        \includegraphics[width=0.997\textwidth]{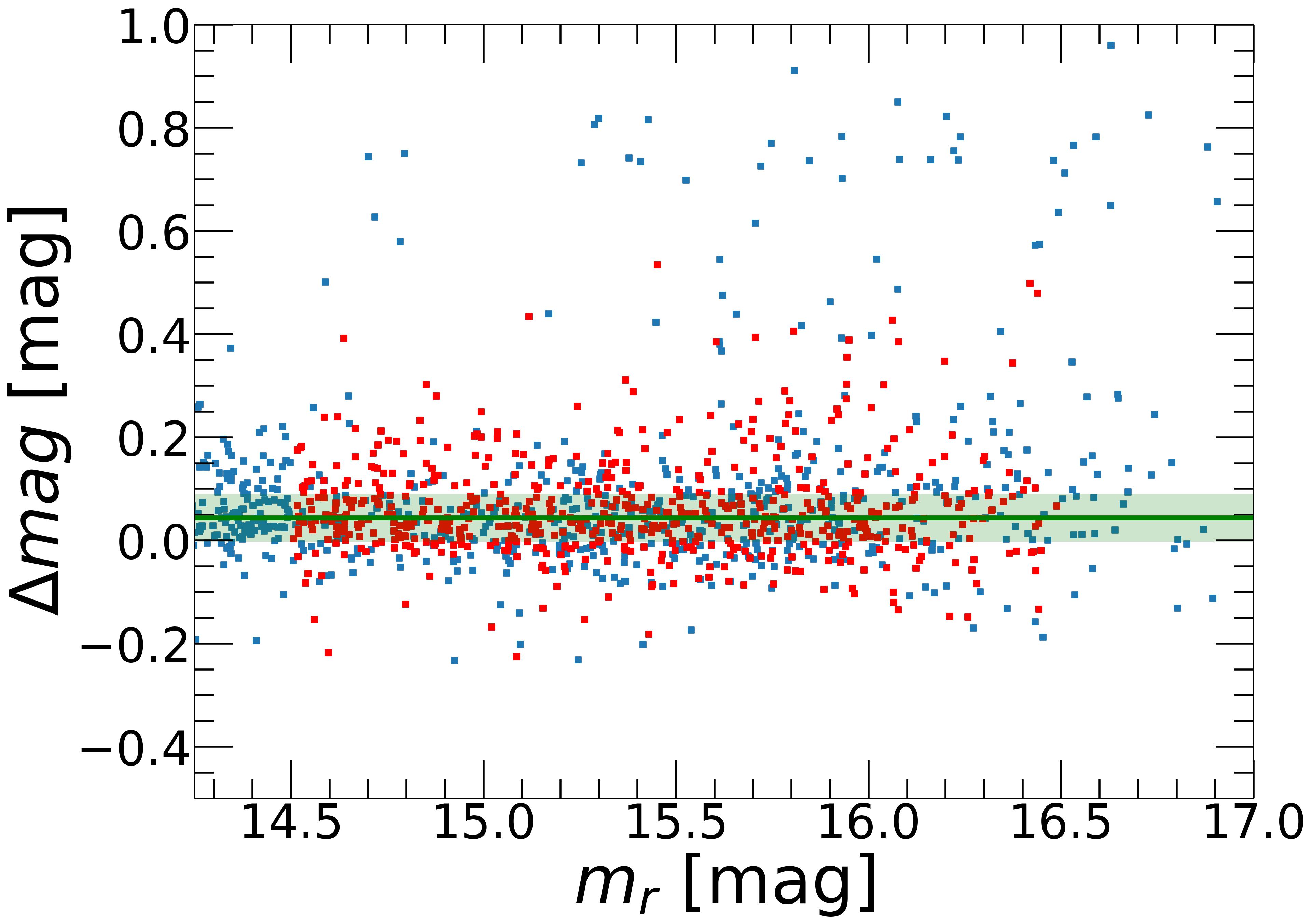}
    \end{minipage}
    \caption{Comparison between VST and APASS magnitudes ($\Delta mag$ = $m_{\rm VST} - m_{\rm APASS}$) in $u$- (left), $g$- (middle) and $r$- (right) passbands. Blue dots represent all the sources in $ugr$ matched catalogue which is matched with APASS catalogue. Red dots represent the bright, unsaturated and point-like sources that we select for photometric comparison. Median and $RMS_{\!\rm MAD}$ values of $\Delta mag$ in the three passbands are shown by green horizontal line and shaded regions, respectively and the values are reported in Table \ref{tab:obs_image_prop}.}
    \label{fig:phot_comp}
\end{figure*}

Figure \ref{fig:phot_comp} shows the comparison between VST and APASS photometry in all three passbands. In Table \ref{tab:obs_image_prop}, we report the median (green horizontal line in Fig. \ref{fig:phot_comp}) and the $RMS$  (green shaded region; derived from the median absolute deviation $RMS_{\!MAD}$) values of $\Delta mag$ (= $m_{\rm VST} - m_{\rm APASS}$). The small $\Delta mag$ values in all the passbands indicate that the VST photometry is in good agreement with that of APASS. The larger scatter in the $r$-band (right panel in Fig. \ref{fig:phot_comp}) is due to its slightly worse image quality compared to the $g$-band which is visible in Table \ref{tab:obs_image_prop} (FWHM and a.c.) and in Figs. \ref{fig:comp_reg_funcs} and \ref{fig:gr_gcsel} (e.g. from the concentration index, CI, panel). The outliers in the right panel in Fig. \ref{fig:phot_comp} are mostly due to poor matching related to such lower image quality. Finally, to correct for galactic extinction, we obtain the reddening ($E_{B-V}$) values from the IRSA dust query module of Astropy\footnote{\url{https://astroquery.readthedocs.io/en/latest/ipac/irsa/irsa.html}} at the position of sources in the observed field. These $E_{B-V}$ values are then multiplied by extinction factors of 4.239, 3.303 and 2.285 for $u$-, $g$- and $r$-passbands, respectively, from \citet{schlafly11} to obtain the extinction correction. The $E_{B-V}$ variation across the field is of the order of $\sim0.03$ mag.

\subsection{Completeness}
\label{sec:completeness}

A further ingredient that will be useful in our forthcoming analysis is the completeness function in the observed field. This function describes how complete our photometric catalogue is in a given passband for point sources as a function of magnitude. The function is derived from simulations by injecting artificial sources into a given frame and then detecting them using the same strategy described in the previous section. The fraction of the number of detected sources over the number of injected sources ($f = N_{\rm d}$/$N_{\rm i}$) as a function of magnitude provides the completeness function. This fraction is close to one for bright sources and as we move to fainter magnitudes, an increasing number of sources are undetected, so the fraction approaches zero. The procedure we adopt to estimate the completeness function is described in more details in \citet{mirabile2024}. Here, we briefly summarise the main steps.

\begin{figure*}[htb!]
\centering  
\includegraphics[width=0.69\textwidth]{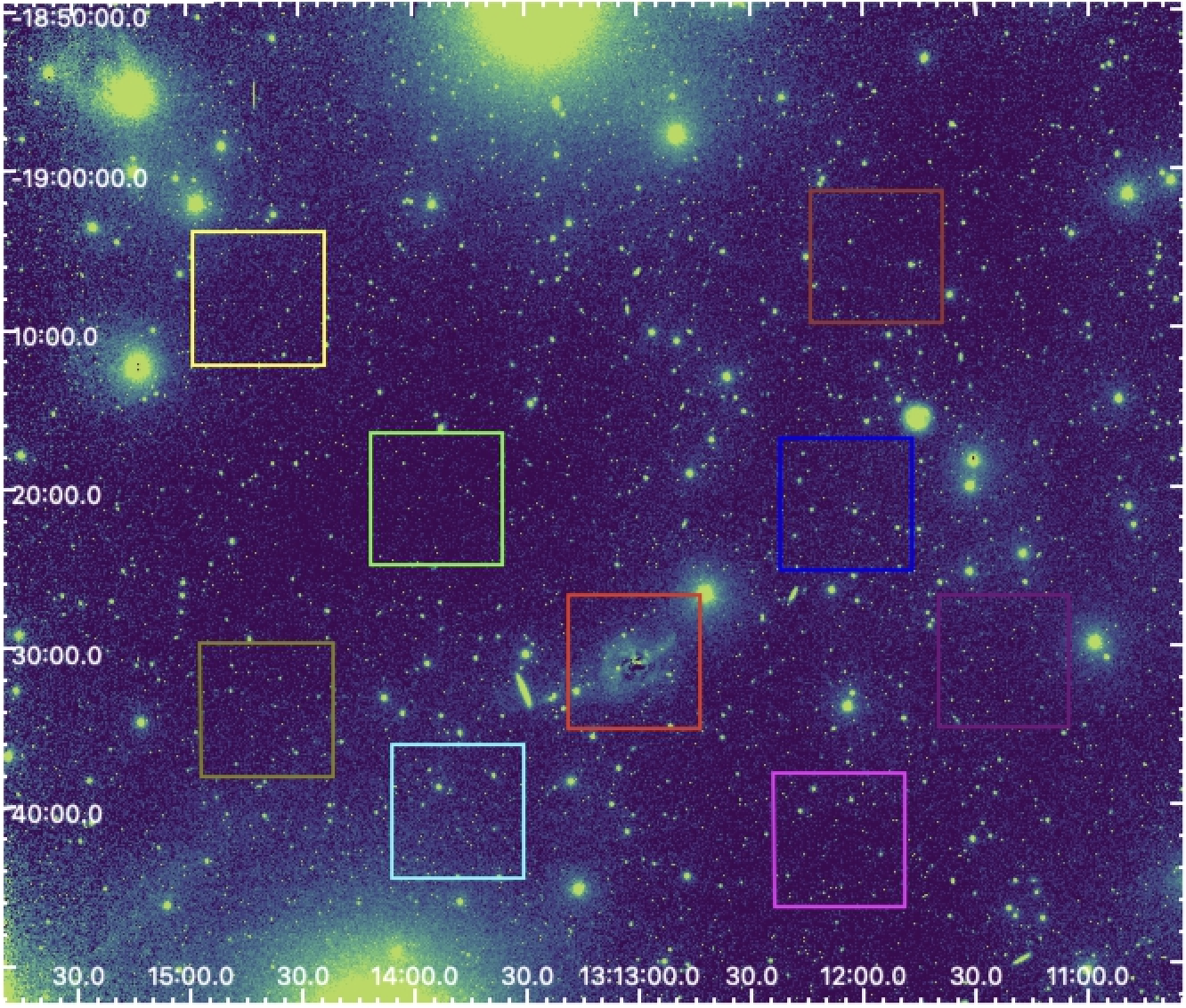} 
\vskip .3cm
\includegraphics[width=0.33\textwidth]{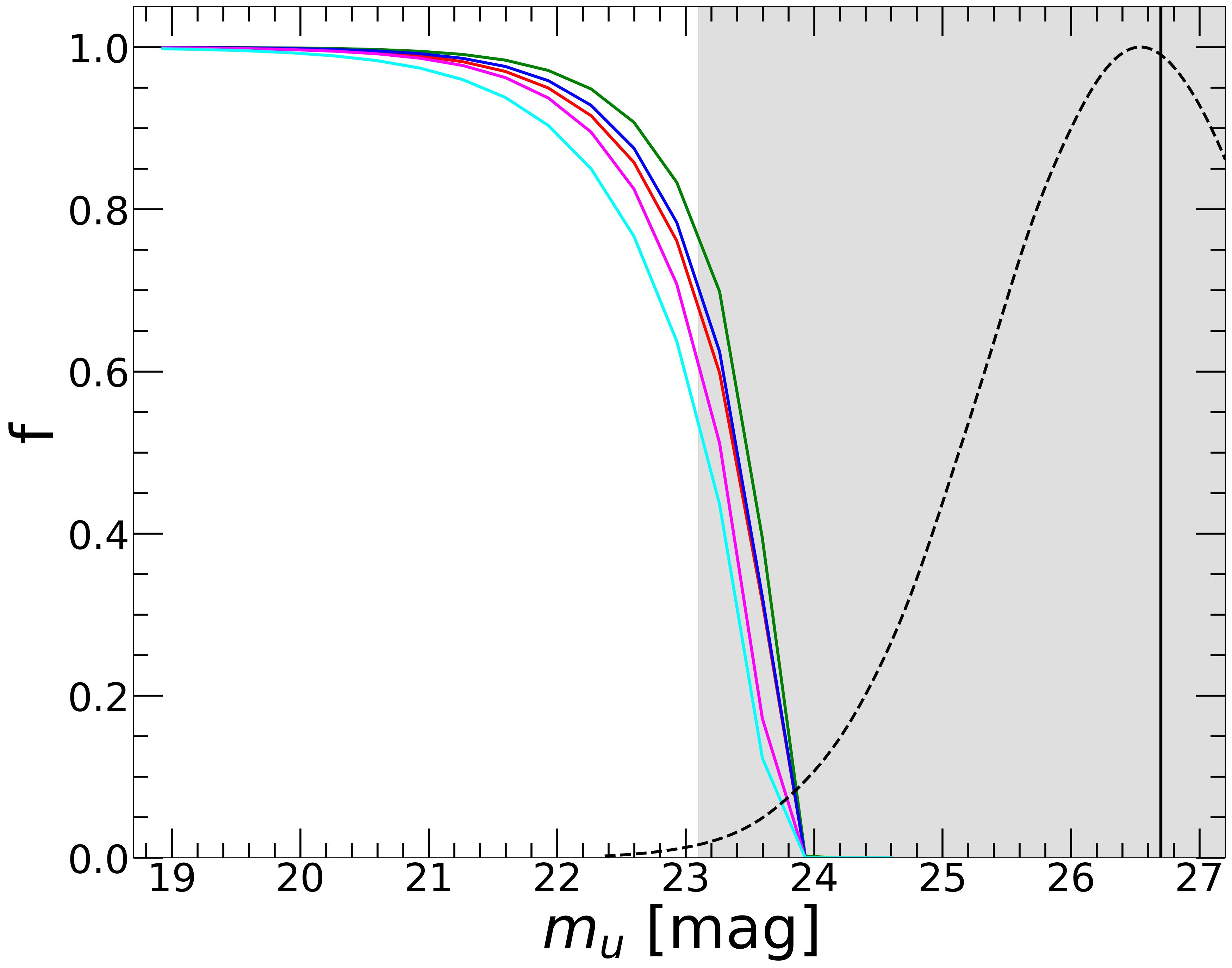}
\includegraphics[width=0.33\textwidth]{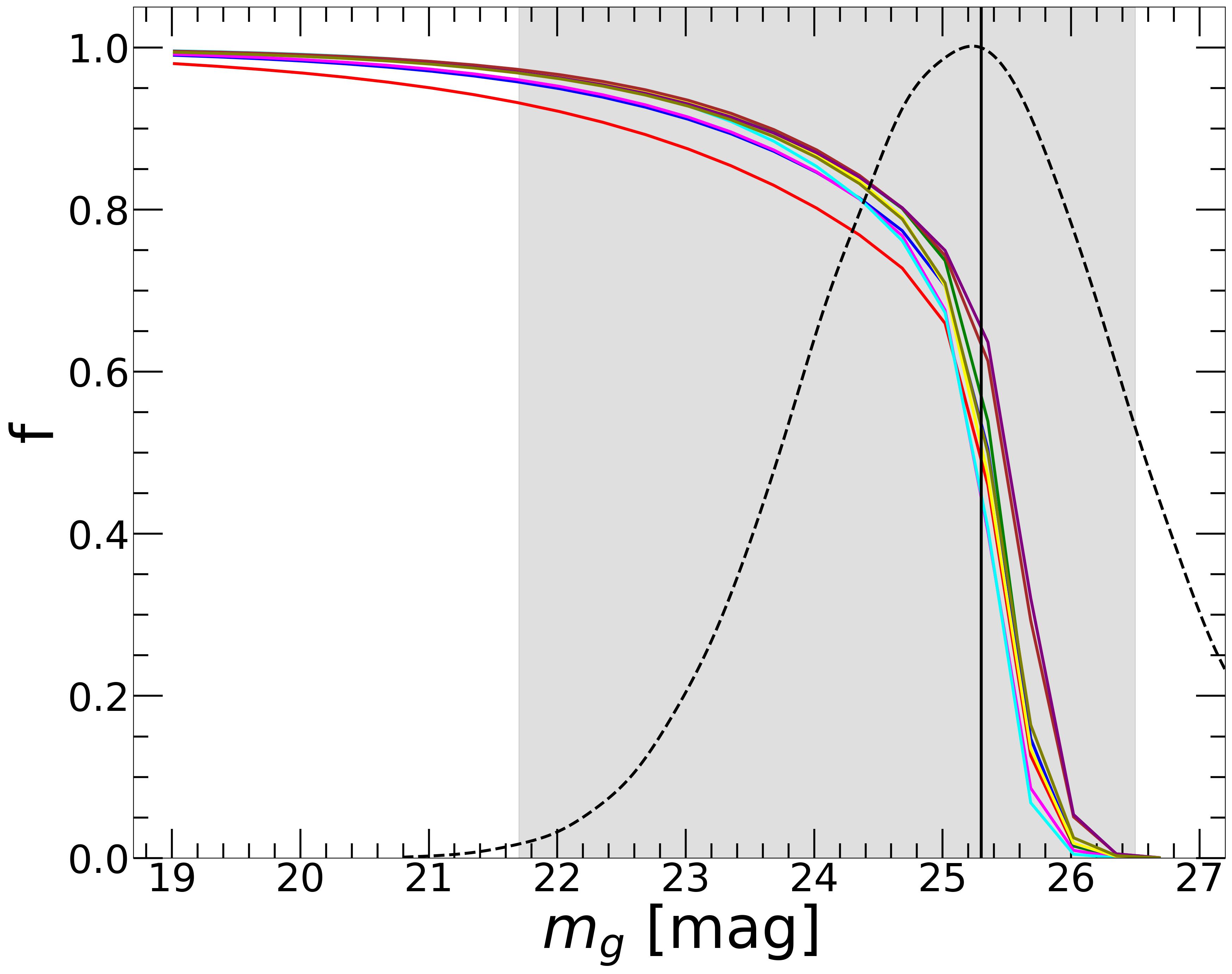}
\includegraphics[width=0.33\textwidth]{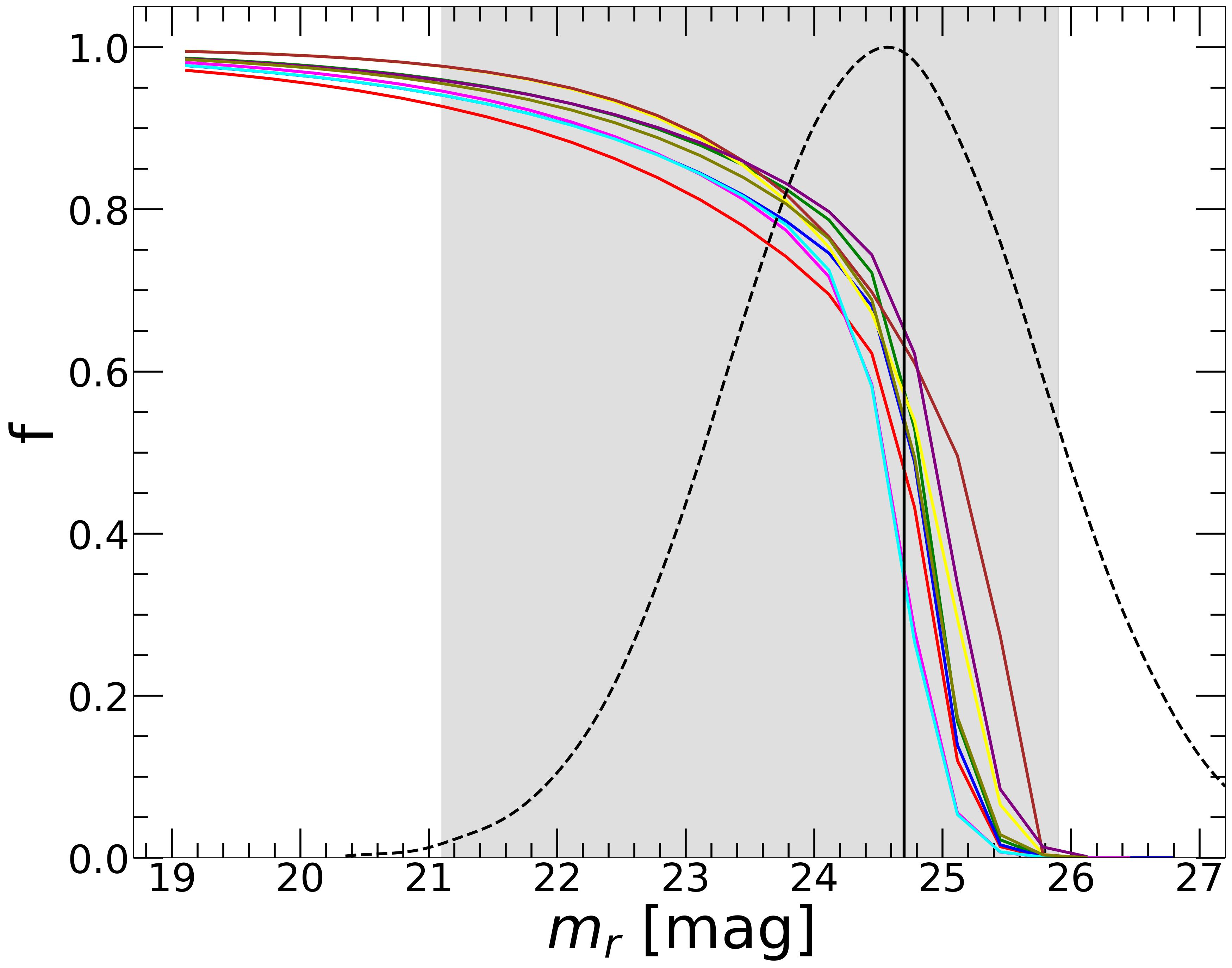}
    \caption{Top panel: $r$-passband image highlighting the regions selected for completeness analysis in the observed field. The area of the entire frame inspected for GCs is 1.25 $\times$ 1.03 sq. degrees; north is up, and east is left. The region highlighted in red is centred on NGC\,5018. The eight off-galaxy regions surrounding the galaxy are highlighted with green, blue, magenta, cyan, olive, yellow, brown, and purple boxes. Each region is 8.4\arcmin\ $\times$ 8.4\arcmin\ in size. Bottom panels: Completeness functions simulated and fitted using the modified Fermi function (Eq. \ref{eqn:fermi_func}) for the $u$- (left), $g$- (middle), and $r$- passbands (right). The solid curves are colour-coded according to the boxed regions in the top panel. The vertical solid black line marks the expected TOM of the GCLF in the respective passbands. The grey-shaded region indicates the adopted magnitude range for GC selection in the three passbands (see Sect. \ref{sec:phot_select}). The dashed black curves represent the expected GCLF in the respective passbands (see Sect. \ref{sec:completeness} for details).}
   \label{fig:comp_reg_funcs}
\end{figure*}

In a case like this where we have a large area format observations, it is ideal to characterise the homogeneity in terms of depth and image quality over the observed area. 
To study the stability of the completeness function with our data, we select nine regions (shown in the top left panel in Fig. \ref{fig:comp_reg_funcs}) in the field, including one on NGC\,5018. 
The region on NGC\,5018 is of the same size as the cutout used in Sect. \ref{sec:gal_model_and_resid} (8.4\arcmin $\times$ 8.4\arcmin). For consistency, all the eight off-galaxy regions are of the same size.

To run the completeness tests, we select the bright, compact and isolated sources in the selected regions to generate models of PSF using the EPSFBuilder\footnote{\url{https://photutils.readthedocs.io/en/stable/epsf.html}} routine of the Photutils package. Then we simulate the magnitude of sources that are to be artificially injected in the image by generating a random sample of magnitudes using numpy.random.choice\footnote{A routine of the NumPy package that generates a random sample from a given 1-D array (\url{https://numpy.org/doc/stable/reference/random/generated/numpy.random.choice.html})} in the shape of the luminosity function of all the real sources detected in those regions. We inject $\sim$ 1700 artificial sources in the regions, using the PSF model obtained from the corresponding region, along an equispaced grid whose position is varied over 75 iterations. This means that during each iteration, the magnitude and the position of all the injected artificial sources in the regions are randomly changed, and they do not increase local density of sources.

As a test of the reliability of the PSF models, we compare the a.c. values obtained from the artificially injected sources with those derived from the real sources and we find a good agreement between the two, within 0.01 mag. Adopting the same input parameters of SExtractor that were used for the detection of real sources in the residual full frame, we now obtain a catalogue of sources detected on each of the simulated frames containing artificially injected sources. We match this catalogue (using a 1\arcsec \ matching radius) to the catalogue of injected sources and clean for spurious sources. For this, we inspect the broadening of difference between injected and detected magnitude versus the detected magnitude for each source. We use an iterative sigma clipping approach that is magnitude dependent. 
Finally, the ratio of the number of sources retrieved vs injected ($f$) in each magnitude bin gives us the completeness fraction of the data in all the three passbands.


We fit the data points (not shown in Fig. \ref{fig:comp_reg_funcs} to avoid their over-plotting on the curves) using the following modified Fermi function \citep{alamo13}:
\begin{linenomath}
\begin{equation}
\label{eqn:fermi_func}
    f \rm (\textit{m})=\frac{1 + \textit{C}\cdot exp[\rm \textit{b}(\textit{m} - \textit{m}_{50})]}{1 + exp[\rm \textit{a}(\textit{m} - \textit{m}_{50})]}
\end{equation}
\end{linenomath}
where $m_{\rm 50}$ is the 50\% magnitude completeness limit, \lq \textit{a}\rq \ is a parameter that regulates the steepness of the cut-off, parameter \lq \textit{b}\rq \ (which must be < \textit{a}) influences the point at which the deviation above unity starts and parameter \lq \textit{C}\rq \ determines the amplitude of the deviation. The 80\% completeness limits in the three passbands for the frame centred on NGC\,5018 are reported in Table \ref{tab:obs_image_prop}. 

Figure \ref{fig:comp_reg_funcs} shows the completeness functions (coloured curves) for the selected regions in all three passbands, along with the expected GCLFs (dashed black curves; see Sect. \ref{sec:phot_select}). Each curve assumes a reference $S_{\!\rm N}=1$ \citep{hilker96} and our estimate of $M_V$ = \textminus 22.28 $\pm$ 0.15 mag for NGC\,5018 (see Sect. \ref{sec:gclf}), which were used in Eq. \ref{eqn:spec_freq}
 to estimate the total number of expected GCs. Completeness functions in the $u$-passband were derived only for the five regions nearest the galaxy. This is because, given the completeness tests and the bright magnitude limit of the $u$-data (Sect. \ref{sec:source_det_and_phot}), our results did not rely on catalogues based on $u$-photometry, so no $u$-band completeness correction was required.

By inspecting the completeness functions shown in Fig. \ref{fig:comp_reg_funcs}, we make the following observations:
\begin{enumerate}

    \item In the case of $u$-passband (bottom left panel), we do not reach the estimated TOM of the GCLF (26.7 mag; Sect. \ref{sec:phot_select} explains the estimation procedure of TOM in all the three passbands);
    
    \item We have $\sim$ 50\% completeness at the TOM in the $g$- (25.3 mag) and $r$- (24.7 mag) passbands in the region centred on NGC\,5018 (red curves in the bottom middle and right panels);
    
    \item Relative to the respective TOM, the $r$- and the $g$-passband have very similar completeness, while in absolute terms the completeness drops to zero at $\sim$ 25.8 mag in $r$- compared to $\sim$ 26.2 mag in the $g$-band;
    
    \item The completeness varies by a larger fraction in the $r$-passband compared to the $g$-passband. As an example, around the TOM, the value of $f$ from the different completeness experiments has an $rms$ of 2.5\% in $g$-passband and 4.7\% in $r$-passband. This is also expected, as the $g$ image has slightly better image quality parameters compared to the $r$ image, as evidenced by the larger scatter in the comparison with APASS photometry (Fig. \ref{fig:phot_comp}) and the quality parameters reported in Table \ref{tab:obs_image_prop}. This behaviour will motivate our choice of selection of $g$-passband to characterise the GCLF (see Sect. \ref{sec:gclf}).
    
\end{enumerate}

Finally, we note that, because of the peculiar observing strategy adopted in VEGAS (Sect. \ref{sec:obs_and_data}), the completeness drops towards the image edges \citep[see also][]{mirabile2024}.

\section{GCs selection}
\label{sec:gc_select}

The identification of GCs in distant galaxies is challenging, as it suffers from contamination from foreground Milky Way stars and background galaxies. Following the same approach successfully adopted in the other works on GCs using VEGAS data \citep[e.g.][]{cantiello15,cantiello18b,cantiello20,dabrusco22,ragusa22}, we use a set of photometric (magnitude and colour) and morphometric (shape and compactness) properties to identify GCs described as follows and summarised in Tables \ref{tab:gr_gcsel_crit} and \ref{tab:ugr_gcsel_crit}.

\subsection{Photometric properties}
\label{sec:phot_select}

In order to clean our catalogue from non-GC sources, we first apply a magnitude cut taking advantage of the expected magnitude range of GCs in NGC\,5018. The GCLF exhibits a universal profile in the form of a Gaussian \citep{harris01} and has a peak at an absolute magnitude of $M_g^{\rm TOM}$ = \ $- 7.5 \pm 0.2$ mag \citep{villegas10}. It has a width $\sigma^{\rm GCLF}$ which scales with the luminosity of its host galaxy \citep[]{harris01,villegas10}.

Using Eq. 4 from \citet{villegas10} and adopting a total galaxy magnitude $M_z$ = $-23.2 \pm 0.1$ mag for NGC\,5018,
we estimate $\sigma_g^{\rm GCLF}$ = $1.2 \pm 0.2$ mag for the GCLF of NGC\,5018. At our adopted distance modulus for this galaxy (Table \ref{tab:gal_prop}), we estimate the TOM of its GCLF at $m_g^{\rm TOM}$\,= $25.3 \pm 0.3$ mag. Given the proximity of the $g$-passband to the $r$-passband, we assume $\sigma_r^{\rm GCLF}=\sigma_g^{\rm GCLF}$. Then we take advantage of the GC median colour ($g-r$) = 0.6 mag obtained from spectroscopically confirmed GCs in the Fornax cluster using the Fornax Deep Survey data from VST \citep[][FDS data hereafter]{cantiello20} to obtain the TOM in the $r$-passband ($m_r^{\rm TOM}$ = $24.7 \pm 0.3$ mag). Finally, for selection of GCs in the $g$-passband, we use the range from $-3\sigma_g^{\rm GCLF}$ for the brighter part to $+1\sigma_g^{\rm GCLF}$ for the fainter part around the TOM\footnote{As will be discussed in later sections, we tested the robustness of the results presented in this work against several of the selection criteria adopted, most notably the faint magnitude cut. Since the completeness --and thus the photometric signal-to-noise ratio (S/N)-- drops beyond the TOM in both the $g$- and $r$-bands, a cut at the TOM limits the analysis to the most reliable sources, at the expense of GC completeness.(Sect. \ref{sec:results} and \ref{sec:appendix}).}. We use a similar magnitude range from $-3\sigma_r^{\rm GCLF}$ to $+1\sigma_r^{\rm GCLF}$ in the $r$-passband. Figure \ref{fig:comp_reg_funcs} (bottom panels) visually motivates our choice for the faint magnitude limit of $+1\sigma^{\rm GCLF}$ in both $g$- and $r$-passbands as the completeness fractions reach zero at these magnitude levels.

For the $u$-passband, adopting the GC median colour ($u-r$) = 2.0 mag from the FDS data, we estimate $m_u^{\rm TOM}$ = $26.7 \pm 0.3$ mag and assume $\sigma_u^{\rm GCLF}$ = $1.2 \pm 0.2$ mag, due to its proximity to the $g$-passband. As discussed in Sect. \ref{sec:source_det_and_phot}, from Fig. \ref{fig:comp_reg_funcs} (bottom left panel) it is evident that the $u$-passband lacks the depth to reach the estimated TOM, and hence we will use the $gr$ matched catalogue as our reference, for which we present our results in the next section. The $ugr$ catalogue will still be analysed, but primarily as a sanity check for the results obtained with the two-band catalogue. Assuming a common distance modulus for the galaxies in the group, we also expect to observe the GCs of four other bright galaxies within these magnitude ranges across the three passbands.

\begin{figure}[htb!]
   \centering
   \includegraphics[width=9cm]{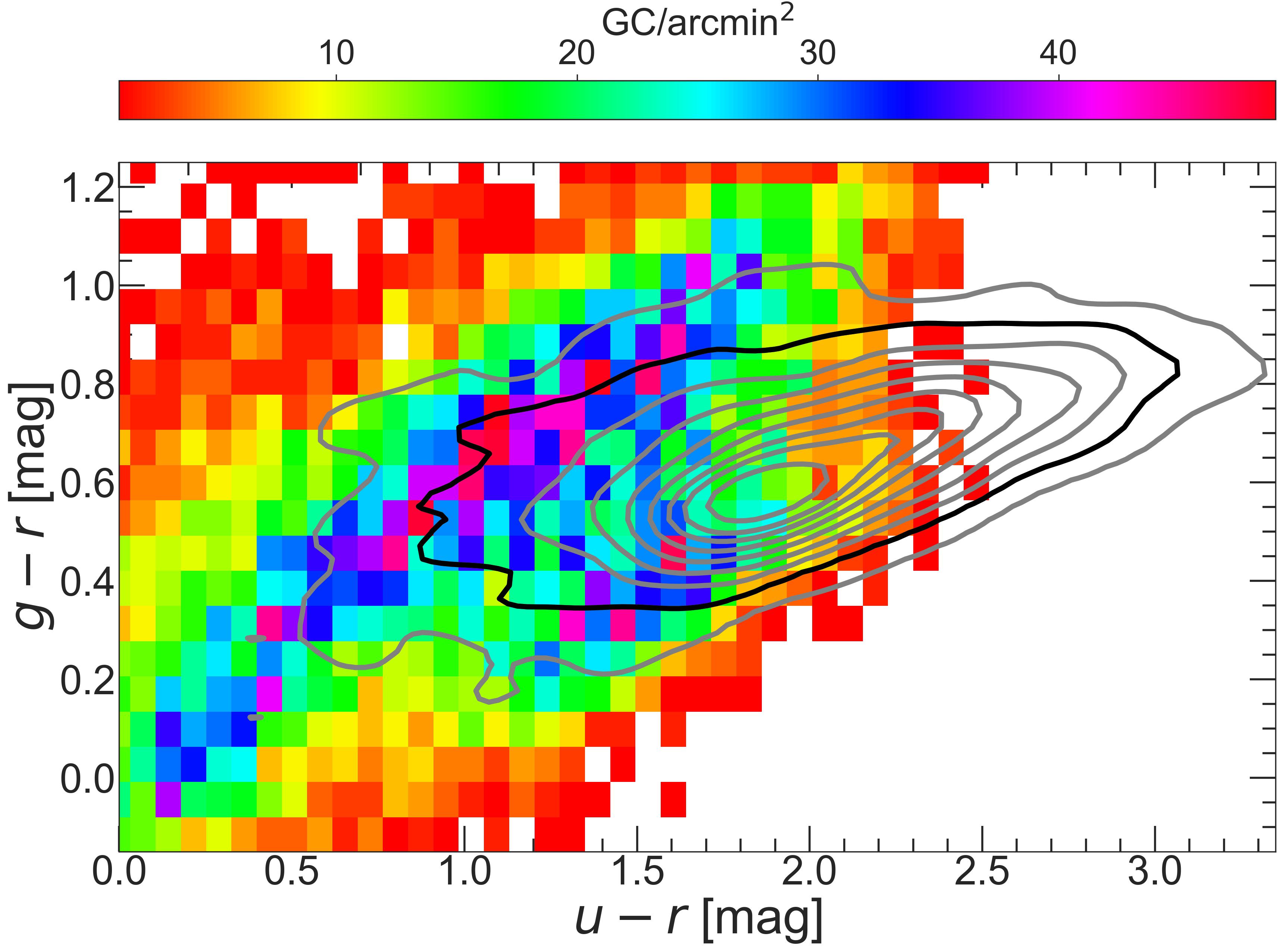}
   \caption{2D histogram of pre-selected GC candidates (based on morphometric and magnitude parameters only) in the ($u-r$) vs. ($g-r$) plane, showing sources from the $ugr$ matched catalogue. The overdense regions (red/pink) represent sequences of contaminants, mostly foreground stars and background galaxies. The isodensity contours, computed using spectroscopically confirmed GCs in the Fornax cluster from FDS data \citep{cantiello20}, are over-plotted. The contours range from 10\% (innermost) to 90\% (outermost) in steps of 10\%. Only the sources within the 80\% contour (highlighted with a black thick contour) are selected as GC candidates from the $ugr$ matched catalogue.}
   \label{fig:ugr_gcsel}
\end{figure}

Using colour for GC selection helps reduce contamination from fore- and background compact sources. To define a colour range in the $gr$ matched catalogue, we use the range 0.3 $\leq (g-r) \leq$ 1.1 mag from the FDS data. In the case of the $ugr$ matched catalogue, we perform a more refined selection in the ($u-r$) vs. ($g-r$) plane (see Fig. \ref{fig:ugr_gcsel}). Adopting the FDS data, we derive the isodensity contours and overlay them on the ($u-r$) vs. ($g-r$) colour-colour plot of GC candidates, pre-selected based on the morphometric parameters from our $ugr$ matched catalogue. The isodensity contours represent fractions of the total GC population in the Fornax cluster, increasing outwards. In the figure, the red side of the plot ($u-r$ > 2.5 mag) is nearly empty due to the shallow $u$-passband data. The overdense regions visible in Fig. \ref{fig:ugr_gcsel} are sequences of foreground stars and background galaxies, that is, contaminants that passed our preliminary selection criteria in the $ugr$ matched catalogue. Selecting a narrow contour implies lower contamination but also lower completeness, while a wide contour increases both contamination and completeness. We choose the 80\% contour level as optimal for GC selection (bold in the figure) and select all sources within this contour as our final GC candidates from the $ugr$ matched catalogue. Although this choice allows some contaminants to enter (as seen in Fig. \ref{fig:ugr_gcsel}), these are effectively addressed by the background decontamination technique described in Sect. \ref{sec:rad_and_col_prof}.

\subsection{Morphometric properties}
\label{sec:morpho_select}

Assuming a mean $R_{\rm e}$ of 3 pc for GCs (Sect. \ref{sec:intro}) and using our adopted distance of 36 Mpc for NGC\,5018 group (Table \ref{tab:gal_prop}), we estimate the angular size of typical GCs in this group to be $\sim$ 0.02 arcsec. Therefore, the GCs at the distance of the NGC\,5018 group are unresolved with our data. However, as has been shown in previous GC works using VEGAS \citep[e.g.][]{dabrusco16,cantiello18a,cantiello20}, we can still take advantage of morphometric properties to identify GCs through their shape.

A first selection is made using the Concentration Index (CI) parameter which is defined as the difference between magnitudes at two aperture sizes. The CI is a good indicator of compactness of the source \citep{peng11}. For this measurement, we adopted apertures with 4 and 8 pixel radii. These values were chosen as a good compromise between CI flatness and small $rms$ after testing different radii. We select as GC candidates the sources with CI in the range 0.5 to 1.6 mag in both $g$- and $r$-passbands from the $gr$ matched catalogue and, additionally, between 0.5 and 2.0 mag in the $u$-passband from the $ugr$ matched catalogue. A slightly wider range is adopted in $u$-passband because of its worse image quality compared to $g$- and $r$-passbands (see Table \ref{tab:obs_image_prop}).

Further cleaning of the sample is done using the 'ELONGATION' output parameter from SExtractor (elongation hereafter), which describes the shape of the detected source in terms of the ratio between its major and minor axes. We  select all the sources with elongation $\leq$ 3 in both the $gr$ and $ugr$ matched catalogues. 
The limit we apply is larger than what is typically adopted for GCs in the previous papers focussing on the analysis of small stellar systems (SSS) using VEGAS data \citep[VEGAS-SSS; e.g.][]{dabrusco16,cantiello15,cantiello18b,cantiello20}. We adopted such a wide cut mainly because, when matching our GC candidates in NGC\,5018 with the catalogue of \citet[][mainly composed of candidates close to the galaxy centre]{humphrey09}, we found that some of their GCs had large elongation. This is due to the highly variable background in the innermost galaxy regions, caused by the combined effects of dust and the steep surface brightness gradient and the possibility of blending of multiple sources in the ground-based images.

\begin{figure*}[htb!]
   \centering
   \includegraphics[width=18.5cm]{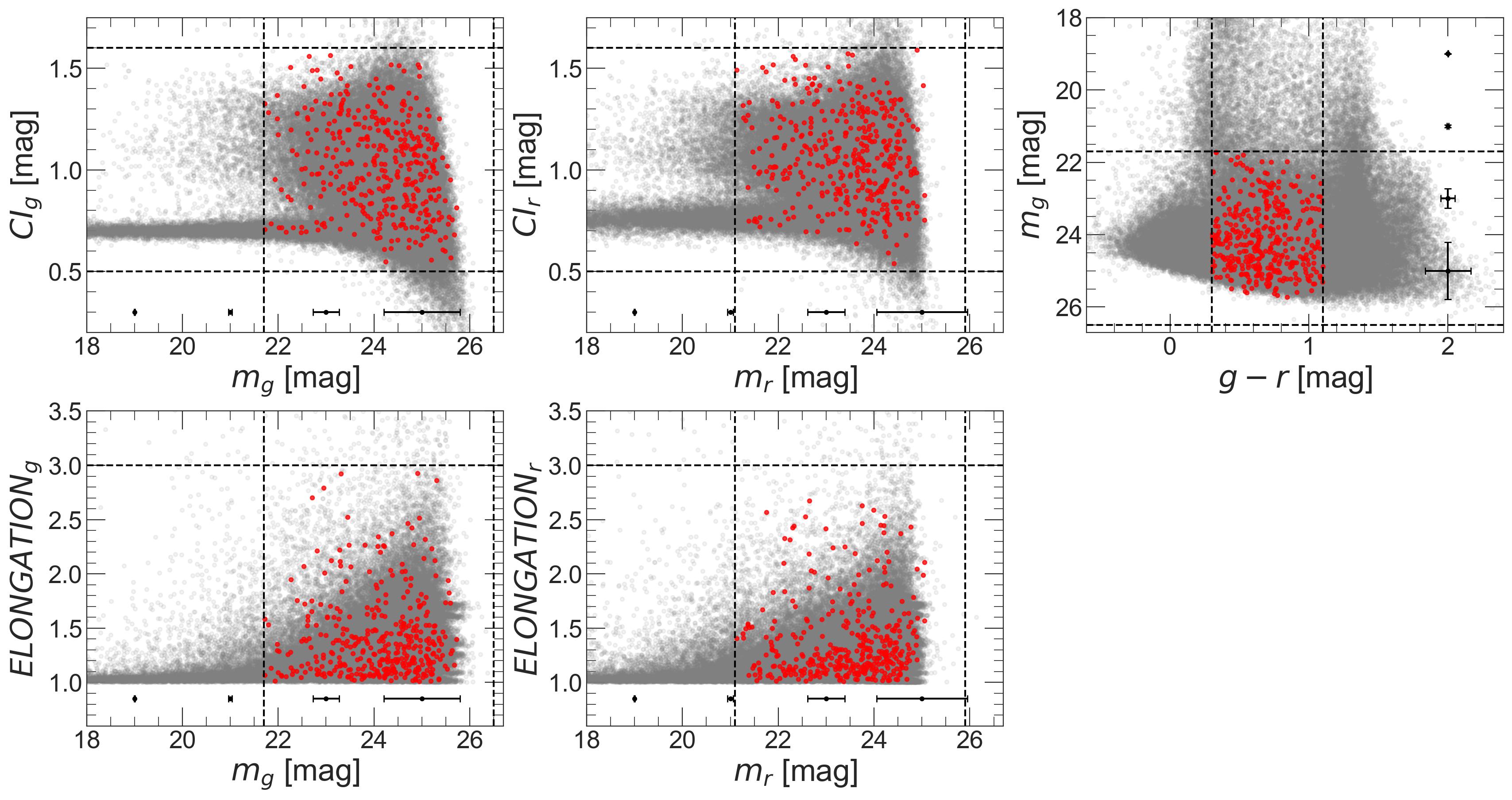}
   \caption{Morphometric parameters used for the selection of GC candidates: CI, elongation, and ($g-r$) colour are shown in the top left/middle panels, bottom left/middle panels, and top right panel, respectively. Grey dots represent all sources in the $gr$ matched catalogue, while red dots indicate GC candidates within $R_{\rm e,GC}$ of NGC\,5018 (see Sect. \ref{sec:morpho_select}). The dashed vertical and horizontal black lines indicate the adopted range for GC selection (see Table \ref{tab:gr_gcsel_crit}). The black error bars in all the panels denote the median error on magnitude (multiplied by a factor of 7 for visibility purpose) with a bin size of 2.0 mag, and on ($g-r$) colour (top right panel) within the given magnitude bin size.}
    \label{fig:gr_gcsel}
\end{figure*}

However, we verified that using an elongation cut of $\leq$ 2.0 in both passbands does not change the results at large galactocentric distances (see right panel in Fig. \ref{fig:2d_maps_gr_TOM_elong_gr_2.0_0.14}), and the results remain consistent with the general trends we observe, albeit with some offset due to the impact on background count levels when using different GC selection criteria. Applying such a lower cut on elongation came at the expense of losing many sources in the core region of NGC\,5018. Since we wanted to include information from the internal galaxy regions, we retained the larger cut on elongation. Moreover, we highlight that, with all the other cuts made, the number of GCs with elongation between 2 and 3 is $\sim10\%$ of the selected sample, or less (see Fig. \ref{fig:gr_gcsel}, middle lower panel). Finally, assuming, as we always do, that the background contamination is constant, even with such a large cut, the interlopers are properly accounted for with the background characterisation and decontamination procedure we adopt (see Sect. \ref{sec:rad_and_col_prof}).

Figure \ref{fig:gr_gcsel} shows the plots for the parameters that we adopt for GC selection in the case of $gr$ matched catalogue. The GC candidates within 5$R_{\rm e}$ of NGC\,5018 are shown (red dots) along with all the sources (grey dots) in this catalogue. The choice of 5$R_{\rm e}$ comes from the observation that half the total GC population is observed to lie within $\sim$ 5$R_{\rm e}$ of the host galaxy in the case of bright, massive ellipticals \citep{forbes18}. In the following sections, we will identify this 5$R_{\rm e}$ radius of NGC\,5018 as its GC half-number radius: $R_{\rm e,GC}$. Tables \ref{tab:gr_gcsel_crit} and \ref{tab:ugr_gcsel_crit} list the criteria that we adopt for the GC selection in the $gr$ and $ugr$ matched catalogues, respectively.

\begin{table}[htb!]
    \centering
    \caption{Parameters adopted for the selection of GCs in the $gr$ matched catalogue.}
    \begin{tabular}{ccc}
    \hline
    \\[-2ex]
    Parameter & $g$-band & $r$-band \\
    \\[-2ex]
    \hline
    \\[-2ex]
      Mag. & $\geq 21.7, \leq$ 26.5 & $\geq 21.1, \leq$ 25.9 \\ 
      ELONG. & $\leq 3$ & $\leq 3$ \\
      CI & $\geq 0.5, \leq 1.6$ & $\geq 0.5, \leq 1.6$ \\
     \hline
     \noalign{\smallskip}
     Colour & \multicolumn{2}{c}{$0.3 \leq g-r \leq 1.1$} \\ 
     \\[-2ex]
    \hline
    \end{tabular}
    \label{tab:gr_gcsel_crit}
\end{table}

\begin{table}[htb!]
    \centering
    \caption{Parameters adopted for the selection of GCs in the $ugr$ matched catalogue. For colour selection in this case, see Fig. \ref{fig:ugr_gcsel}.}
    \begin{tabular}{cccc}
    \hline
    \\[-2ex]
    Parameter & $u$-band & $g$-band & $r$-band \\
    \\[-2ex]
    \hline
    \\[-2ex]
      Mag. & $\geq 23.1, \leq 27.9$ & $\geq 21.7, \leq 26.5$ & $\geq 21.1, \leq 25.9$ \\ 
      ELONG. & $\leq 3$ & $\leq 3$ & $\leq 3$ \\
      CI & $\geq 0.5, \leq 2.0$ & $\geq 0.5, \leq 1.6$ & $\geq 0.5, \leq 1.6$ \\
      \\[-2ex]
      \hline
      \noalign{\smallskip}
    \end{tabular}
    \label{tab:ugr_gcsel_crit}
\end{table}

SExtractor has other output parameters such as 'FWHM\_IMAGE', 'FLUX\_RADIUS' and 'CLASS\_STAR' that describe the morphometry of detected sources. These parameters can be exploited for selection of GCs \citep[e.g.][]{cantiello18a,cantiello20,hazra22}. 
However, after testing the GC candidates selected using these parameters, we opted not to use them as no improvement over our existing technique was found.

Using the selection criteria described in Table \ref{tab:gr_gcsel_crit}, we obtain a catalogue of GC candidates from the $gr$ matched catalogue which is inspected in the next section to analyse the various properties of the GC system. Table \ref{tab:gc_catalog} contains an extract of  the final list of GC candidates with the position of the sources and their measured photometric and morphometric properties, including the $u$-passband data when available.

In conclusion, since the $u$-passband is shallower compared to the other passbands, by adopting the $ugr$ sample, we expect to be potentially dominated by the fraction of blue GCs and, therefore, possibly contaminated by young massive clusters, as well as blue Milky Way stars and background compact galaxies. With the $gr$ matched catalogue, on the other hand, combined with our adopted criteria for GC selection, we expect to observe a dominant population of older GCs, although there may still be significant residual contamination (which depends on the distance from the bright galaxies). Consequently, we might miss intermediate- and young-age GCs that could exist in the NGC 5018 group (Sect. \ref{sec:5018_literature}).

\section{Results}
\label{sec:results}

As described in previous works based on VEGAS survey data \citep[e.g.][]{iodice16, cantiello18a, lamarca2022}, one of the substantial advantages of the VEGAS images is the availability of a large area, which is useful for constraining the background properties in the field around the bright galaxies. This provides valuable information for studying the overdensity of sources on the galaxies in our system which, under the selection criteria we adopt, are primarily the GCs hosted in the galaxy group. In the following sections, we examine the observational properties of the GC system in the NGC\,5018 group, using the GC candidates from Table \ref{tab:gc_catalog}, derived from the $gr$ matched catalogue as our reference.

\subsection{2D distribution of GCs}
\label{sec:2d_maps}

To generate the 2D distribution maps, we employ a kernel density estimator (KDE) from the Seaborn package \citep{waskom21}\footnote{\url{https://seaborn.pydata.org/generated/seaborn.kdeplot.html}}. The plots created with the KDE tool visualise the 2D distribution of the GC candidates using a continuous probability density curve. The bandwidth of this curve is determined by two parameters: \lq bw\_method\rq, which controls the smoothing using a Gaussian kernel, and \lq bw\_adjust\rq, a factor that multiplicatively scales the value of \lq bw\_method\rq. Increasing the \lq bw\_adjust\rq \ parameter will make the density curve smoother.

\begin{figure*}[htb!]
    \centering    
    \begin{minipage}{0.495\textwidth}
        \centering        \includegraphics[width=0.995\textwidth]{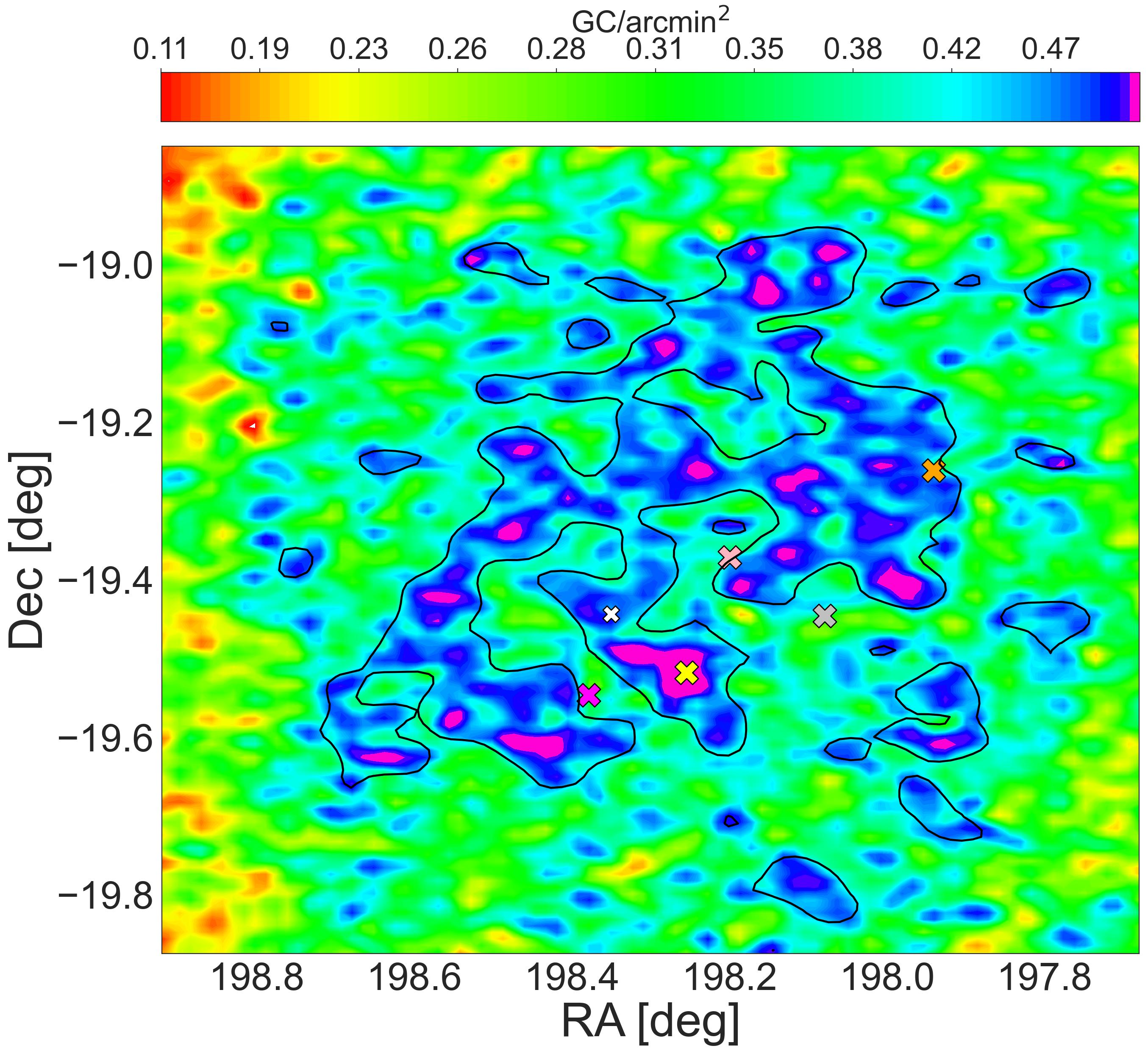}
    \end{minipage}
    \begin{minipage}{0.495\textwidth}
        \centering        \includegraphics[width=0.995\textwidth]{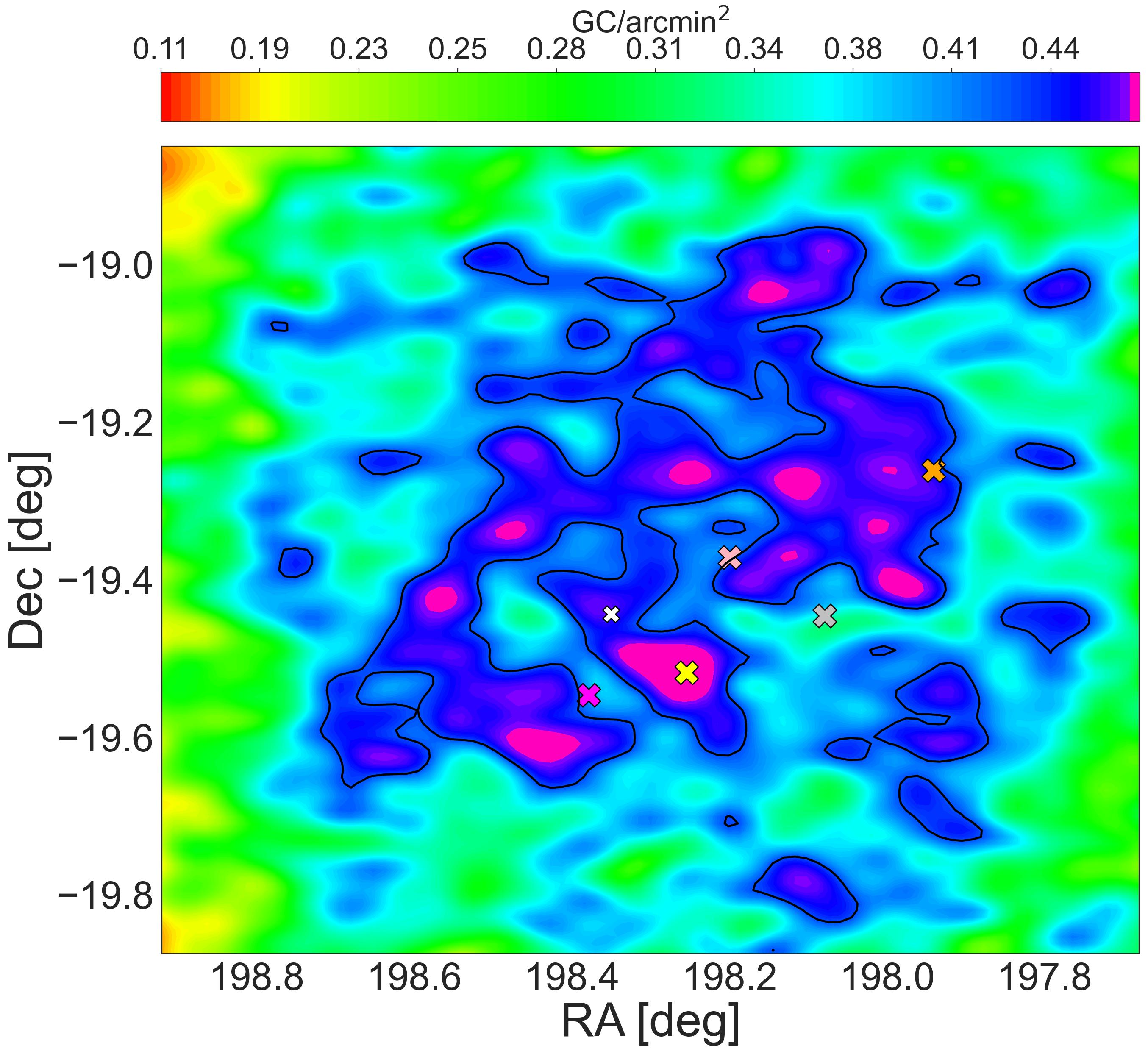}
    \end{minipage}
    \caption{Left panel: 2D distribution map of GC candidates in the observed field from the $gr$ matched catalogue obtained using  $bw\_adjust= 0.14$ and  $bw\_method$ parameter set to {\it Scott} (see Sect. \ref{sec:2d_maps} for the definitions). The magenta and blue coloured areas represent an overdensity of sources whereas the red and yellow-green coloured areas show the underdense regions. The red coloured areas, mostly situated around the edges of the frame, are underdense due to the low S/N because of the observing strategy. The position of the galaxies are indicated with crosses of colour yellow (NGC\,5018), magenta (NGC\,5022), cyan (MCG-03-34-013), orange (NGC\,5006), light pink (PGC\,140148). The white cross is the LSB dwarf galaxy candidate (NGC\,5018-LSB1) reported in Sect. \ref{sec:lsb_dwarf}. Right panel: As in left panel, but using $bw\_adjust = 0.28$. The black contours reported in both the panels indicate the regions higher than 5$\sigma$ level from the background $rms$, obtained using $bw\_adjust = 0.28$}. \label{fig:2d_map_gr_0.14_0.28}
\end{figure*}

Figure \ref{fig:2d_map_gr_0.14_0.28} displays the 2D distribution map for the GC candidates. We observe an overdensity of sources centred on NGC\,5018 while the other four bright galaxies in the group, particularly the two lenticulars, do not seem to exhibit significant overdensities of GC candidates. The plot reveals a radial trend, with higher density near the centre of the observed field and the group, decreasing towards the outskirts. The overdensities are spread in an elliptical shape extending diagonally along the direction of the five galaxies, including NGC\,5018, and around the group itself. Its centre (RA = 198.290 deg; Dec = -19.385 deg) is located $\sim$ 0.13 deg north of NGC\,5018.

To determine whether this structure is an artefact of our GC selection procedure or a genuine feature of the GCs in the field, we perform several tests, detailed in Appendix \ref{sec:ellipse_stab_tests}. These tests include examining the 2D distributions of: $i$) the GC catalogue obtained with narrower constraints compared to the reference selection criteria in Table \ref{tab:gr_gcsel_crit}, $ii$) two catalogues, one of very bright, saturated point sources representing foreground Milky Way stars and another of extended sources with non-GC colours resembling background galaxies, and $iii$) two GC catalogues, one with a magnitude cut up to the TOM and another with a narrower cut on elongation (as discussed in Sect. \ref{sec:morpho_select}). The results with such tests confirm that the elliptical-shaped overdensity is preserved in all cases, hence it is likely a real feature in the observed field, indicating the presence of an intra-group GC population, which will be analysed in the subsequent sections. To further support the reality of this feature, we anticipate that it shows a significant overlap with the IGL region studied by \citet{spavone18}, but extends to larger group-centric radii. The black contours in Fig. \ref{fig:2d_map_gr_0.14_0.28} represent the 5$\sigma$ level higher than the background, where the $\sigma$ is assumed to be the mean of the two $rms$ value of background densities described in the following Sect. \ref{sec:rad_and_col_prof} ($\sigma$ = 0.175 GC/arcmin$^2$).

A notable feature observed in our tests is a non-negligible overdensity of GC candidates north-west of NGC\,5018 around RA and Dec = [198.1 deg; -19.0 deg]. This region does not appear to contain any bright galaxies. However, it might represent an extension of the intra-group GC population resulting from interactions among group members or a region with higher foreground/background contamination. In the forthcoming discussion, we will focus on the central GC overdensity within the described elliptical region. Further studies on the detailed sub-structures in the GC population, including the north-west extension, will require deeper multi-passband observations, as expected from surveys like Legacy Survey of Space and Time (LSST) at the Vera C. Rubin Observatory.

\begin{figure}
   \centering   \includegraphics[width=9cm]{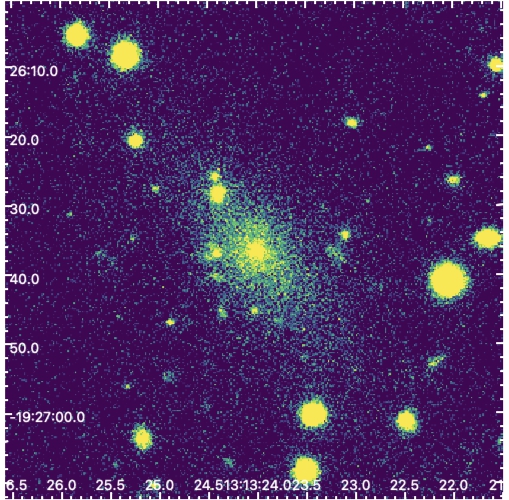}
   \caption{A cutout from $r$-passband data of the LSB dwarf galaxy candidate (NGC\,5018-LSB1) that we report in Sect. \ref{sec:2d_maps}. The image size is 1.2\arcmin \ $\times$ 1.2\arcmin; north is up and east is left. A zoomed-in colour version of this image is shown in the top left corner of Fig. \ref{fig:5018_group}.}
    \label{fig:lsb_dwarf_gal}
\end{figure}

One additional test we present involves studying the 2D map by further smoothing the GC candidate density, shown in the right panel of Fig. \ref{fig:2d_map_gr_0.14_0.28}. As expected, the small-scale structure of over/under-density changes due to the higher smoothing factor ($bw\_adjust=0.28$), causing the edges of some substructures to merge. Despite this, the overall appearance of the map, including the excess of candidates on NGC\,5018, the wider elliptical overdense area, the north-west extension, and the asymmetry towards a local dwarf galaxy (see below), is confirmed. Hence, for the remainder of our analysis we will focus on group-scale trends rather than small-scale overdensities. The only small-scale exception we highlight is a plume-like overdense region north-east of NGC\,5018. This feature is also preserved in the tests on 2D distribution maps (Figs. \ref{fig:2d_maps_narrow_sels_0.14} and \ref{fig:2d_maps_gr_TOM_elong_gr_2.0_0.14}). Although it does not appear to be connected to any distribution of diffuse light in the area, this plume is oriented towards what appears to be a local nucleated LSB dwarf galaxy of the group whose brightness profile stretches along the direction of NGC\,5018. 
Small cutouts centred on this LSB dwarf galaxy (NGC\,5018 - LSB1 hereafter) are shown in top left corner in Fig. \ref{fig:5018_group} (a zoomed-in colour image) and Fig. \ref{fig:lsb_dwarf_gal} ($r$-passband image). NGC\,5018 - LSB1 is further discussed in details in Sect. \ref{sec:lsb_dwarf}.

We also analysed the 2D distribution map of GC candidates from the $ugr$ matched catalogue. Due to the shallow depth of the $u$-passband data (see bottom left panel in Fig. \ref{fig:comp_reg_funcs}), only the brightest and bluest GCs in the population were detected. The 50\% completeness level in the $u$-passband is 23.4 mag, $\sim$3 mag brighter than the expected TOM in this band. Consequently, no significant overdensity of GC candidates around NGC\,5018 and the four other bright galaxies is observed in the $ugr$ matched catalogue. Furthermore, the few detected GCs are likely metal-poor, which in most galaxies are distributed at larger galactocentric distances than the red, metal-rich ones. This broader distribution dilutes their density over a larger area, making them more likely to be lost in statistical fluctuations of the background interlopers.

\subsection{Radial and colour distribution of GCs}
\label{sec:rad_and_col_prof}

To study the radial density and ($g-r$) colour distribution of the GC candidates, we take advantage of the large format of the VST imaging data to apply a background decontamination technique. This approach allows us to inspect the properties of the GC candidates in the observed field. We adopt a method previously used in similar works on GCs using VEGAS survey data \citep[e.g.][]{cantiello15,dabrusco16,cantiello20,mirabile2024}, which relies on the assumption that the fore- and background contaminants in the GC candidates catalogue are evenly distributed across the field. While there is cosmic variance among background galaxies, our approach of sampling different background areas helps to smooth this variance. By studying sources in the background fields, far from the target galaxies, we can constrain the properties of the contaminants. These constraints are then used to correct the GC properties through on-galaxy to off-galaxy density subtraction. We did not rely on GC counts derived from areas at larger galactocentric distances due to the incompleteness of the catalogue at those radii \citep[see Sect. 3 and 4.3, and][]{mirabile2024}.

A key ingredient of this analysis is the identification of suitable background regions to be used for decontamination. Considering the small variations in depth and image quality over the field in $g$- and $r$-passbands (see discussion in Sect. \ref{sec:completeness}), the observed geometry of the GC overdensity region (highlighted in Fig. \ref{fig:2d_map_gr_regions_0.14}) and taking advantage of the large observed area, we decide to use two different regions as background:
\begin{enumerate}

    \item We adopt a circular annulus centred on the observed field with radii between 26\arcmin and 29\arcmin (\textit{annular background} hereafter; large black annulus in Fig. \ref{fig:2d_map_gr_regions_0.14}) which has a density of $11.34 \pm 0.15$ GC/arcmin$^2$.
    
    \item We select ten random regions each with 2.7\arcmin \ radius and consider their mean ($10.60 \pm 0.20$ GC/arcmin$^2$) as a background (\textit{random background} hereafter; small black circles in Fig. \ref{fig:2d_map_gr_regions_0.14}). 
    
\end{enumerate}

\begin{figure*}[htb!]
   \centering   
   \includegraphics[width=\textwidth]{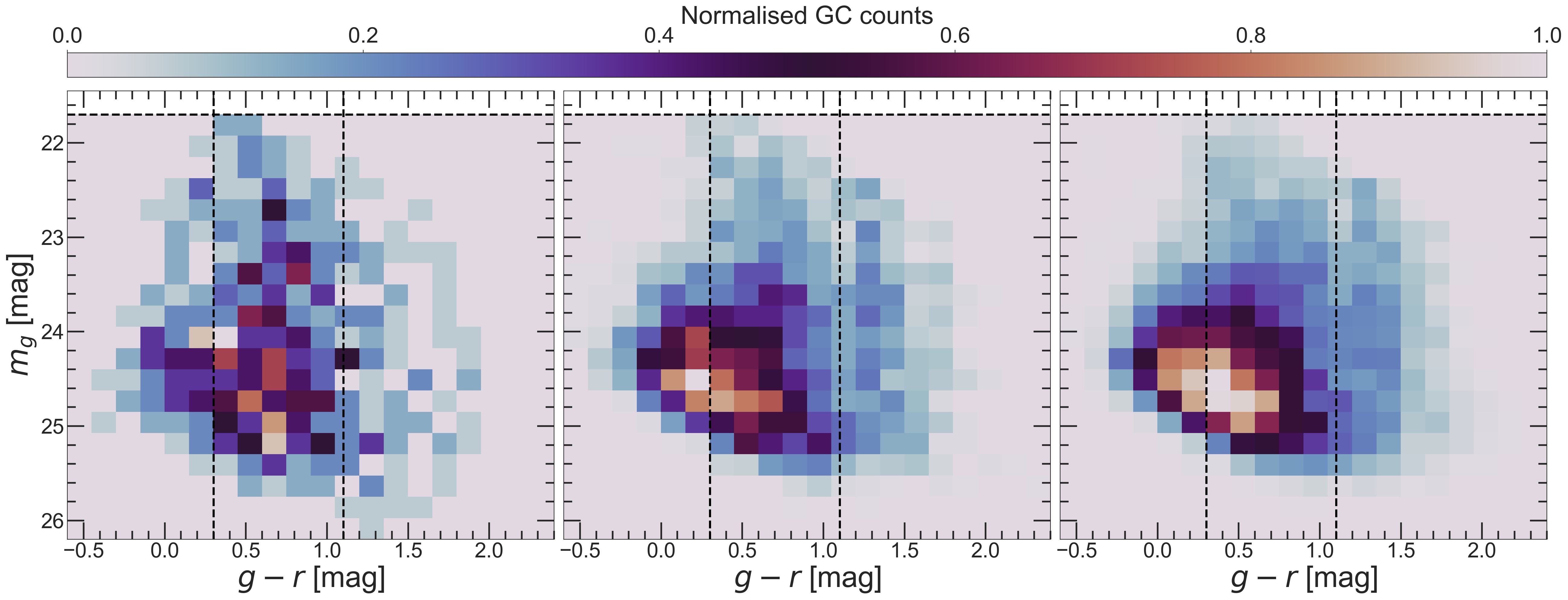}
   \caption{Normalised 2D histogram of colour-magnitude diagram of GC candidates selected using criteria reported in Table \ref{tab:gr_gcsel_crit}, except for ($g-r$) colour selection, around NGC\,5018 (within $R_{\rm e,GC}$; left panel), \textit{random background} (middle panel) and \textit{annular background} (right panel). The ($g-r$) colour selection interval, and the bright magnitude cut are also shown in the panels with dashed black lines.}   \label{fig:2d_hist_cmd_gc_wo_colour_5018_rb_all_an_normalised}
\end{figure*}

\begin{figure}[htb!]
   \centering   
   \includegraphics[width=9cm]{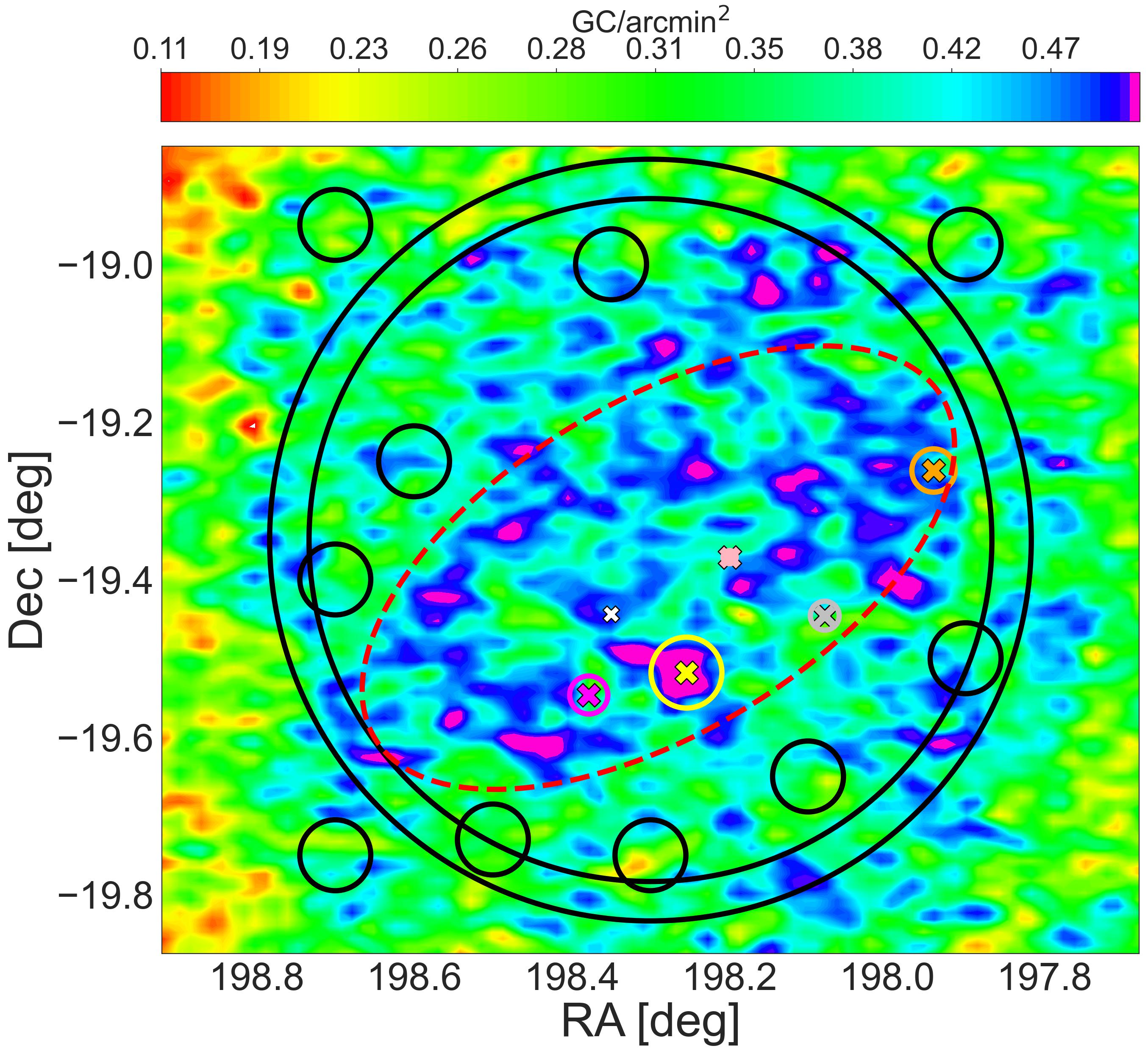}
   \caption{Same as Fig. \ref{fig:2d_map_gr_0.14_0.28}, with the different regions selected for background decontamination highlighted. The circles centred on the five bright galaxies represent their 5$R_{\rm e}$ radius ($R_{\rm e,GC}$), which corresponds to 2.7\arcmin \ in the case of NGC\,5018.
   The 10 small black circles, each with a radius of 2.7\arcmin, are the regions selected to collectively represent the \textit{random background}. The large black annulus (with inner and outer radii of 26\arcmin and 29\arcmin, respectively) represents the \textit{annular background}. The red dashed ellipse indicates the elliptical overdensity region, with major and minor axes of 50.4\arcmin and 25.2\arcmin, respectively. See text for details.}
   \label{fig:2d_map_gr_regions_0.14}
\end{figure}

The normalised 2D histograms of the colour-magnitude diagram for GC candidates in NGC\,5018 and the two background regions are shown in Fig. \ref{fig:2d_hist_cmd_gc_wo_colour_5018_rb_all_an_normalised}. The plot includes all pre-selected GC candidates within $R_{\rm e,GC}=2.7\arcmin$ \ of NGC\,5018 (left panel), in the \textit{random background} (middle panel), and in the \textit{annular background} (right panel), normalised to a peak density of one. The GC candidates in the figure were selected based on the criteria reported in Table \ref{tab:gr_gcsel_crit}, except for the ($g-r$) colour selection. The GC candidate overdensity in NGC\,5018 is visible in the left panel as a darker sequence (i.e. higher number counts) with ($g-r$) between $\sim$0.4 and $\sim$1.0 mag and $m_g$ between  $\sim$22.5 and $\sim$24.5 mag. The middle and right panels show the selected GC candidates from the control background regions, revealing a less prominent density at the expected GC colours and magnitudes.

To estimate the uncertainty in the GC density for the background, we used Poissonian statistics for the number of counts in the annular background, and the $rms$ between the GC counts in the selected circular areas for the random background, in both cases normalising to the area to obtain a density. Unlike in previous VEGAS-SSS works, we retain both background estimates in our analysis, as they are only broadly consistent with each other.
Furthermore, we do not select any region close to the edges of the frame because the S/N is low there due to the VEGAS survey observing strategy (see Sect. \ref{sec:obs_and_data}-\ref{sec:completeness}) and this affected our source detection. For selecting GC candidates within the elliptical overdensity region, we consider an ellipse with major and minor axes of length 50.4\arcmin \ and 25.2\arcmin, respectively (approximately 0.50 Mpc and 0.25 Mpc; see Fig. \ref{fig:2d_map_gr_regions_0.14}). This size is selected to make sure that the general outline of the feature observed in Fig. \ref{fig:2d_map_gr_0.14_0.28} is encompassed, including the galaxies in the group. For the radial density and ($g-r$) profiles of the GC candidates in this region, we performed multiple tests with varying size and the position angle of the ellipse and found consistent results with negligible differences.

\subsubsection{Radial density profile}
\label{sec:rad_prof}

To inspect the radial density profile of GC candidates on NGC\,5018, we derive the azimuthally averaged density profile obtained using concentric circular annuli centred on the galaxy. From the profile, we subtract the background level estimated in the two background regions.

\begin{figure}[htb!]
   \centering   
   \includegraphics[width=9 cm]{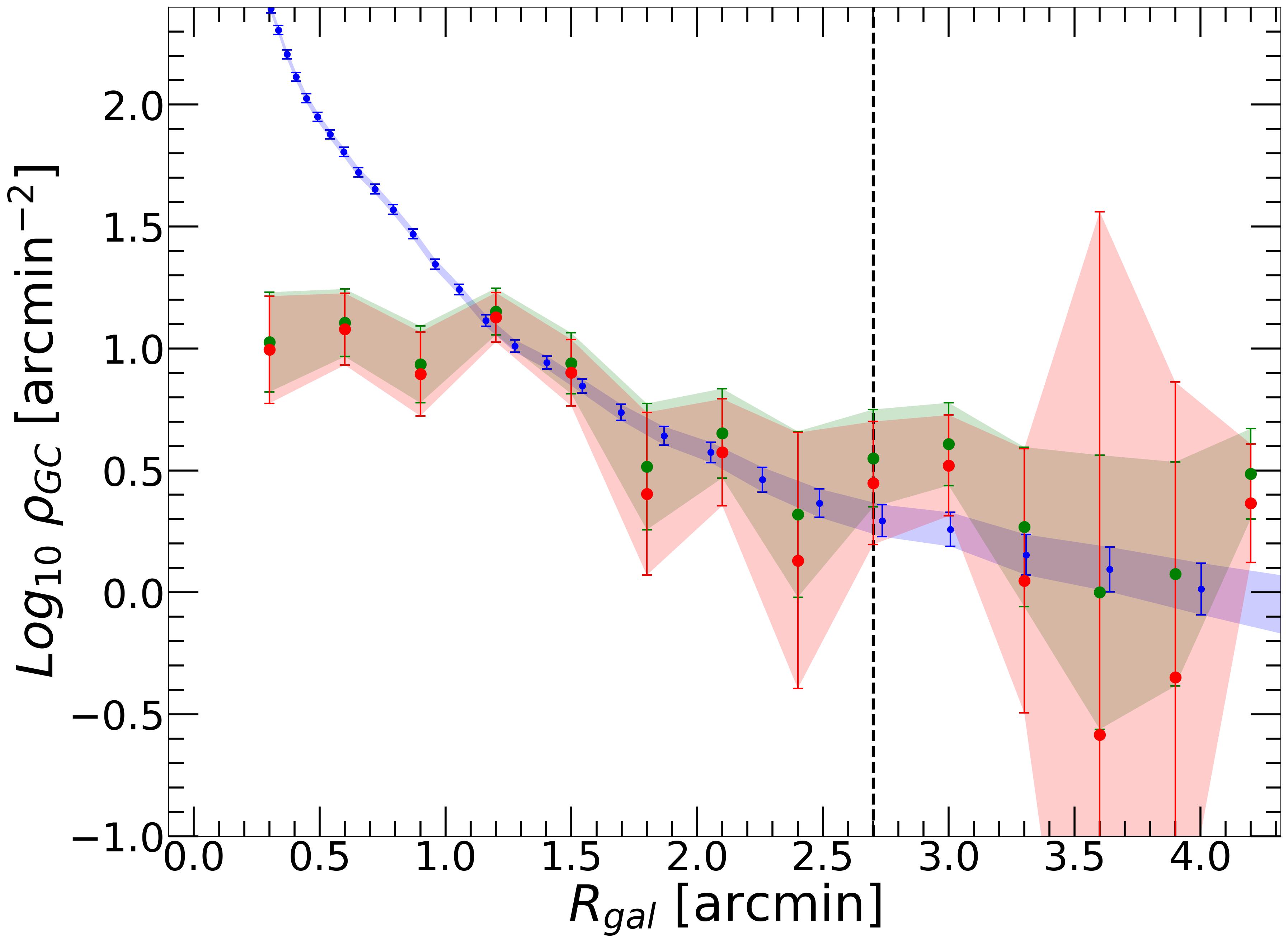}
   \caption{Radial density profile of GC candidates in NGC\,5018 after subtracting \textit{random background} (green points) and \textit{annular background} (red points). The blue points represent the light profile of the galaxy in the $g$-passband from \citet{spavone18} after an arbitrary vertical shift. The vertical black dashed line represents $R_{\rm e,GC}$ radius.}  \label{fig:rad_prof_5018_back_sub_log}
\end{figure}

Figure \ref{fig:rad_prof_5018_back_sub_log} shows our result where the profiles displays the expected larger GC density around NGC\,5018. The observed low density in the core (galactocentric radius, $R_{\rm gal}$ $\leq$ 1.2\arcmin) of the galaxy is expected because of the poor detection efficiency in this region and also due to the presence of dust (see Figs. \ref{fig:5018_group} and \ref{fig:resid}) which affects the observability of sources, especially at the faint magnitude levels. 

From the observed peak, a general decreasing trend is observed out to $R_{\rm gal}$ $\sim$ 3.6\arcmin \ which also marks the transition radius  between the two accreted components observed from the fit of the surface brightness profile of NGC\,5018 \citep[][Fig. 13]{spavone18}. Additionally,  a bump is observed at $R_{\rm gal}$ > 3.6\arcmin  which corresponds approximately to the plume overdensity region of GC candidates that we observe to the north-east of the galaxy extending towards NGC\,5018 - LSB1. The $g$-passband surface brightness profile of NGC\,5018 is plotted in Fig.\ref{fig:rad_prof_5018_back_sub_log} (with an arbitrary shift) from \citet{spavone18}. We observe that both the background corrected radial density profiles follow the galaxy light profile in the interval between 1.2\arcmin $\leq$ $R_{\rm gal}$ $\leq$ 3.6\arcmin.

\begin{figure*}[htb!]
   \centering   
   \includegraphics[width=\textwidth]{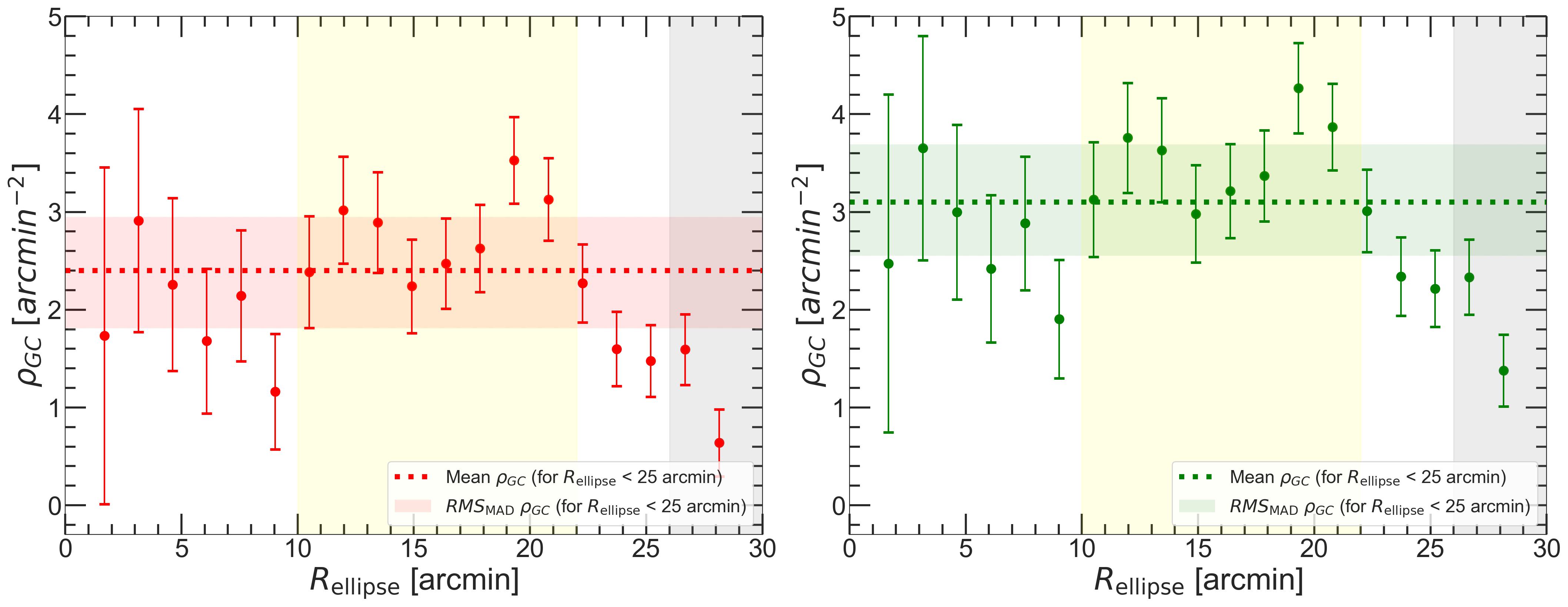}
   \caption{The density profile of GC candidates in the elliptical overdensity region obtained by subtracting \textit{annular background} (left panel) and \textit{random background} (right panel). The grey shaded region in both the panels beyond 26\arcmin \ represents the \textit{annular background} region. The red and green dotted horizontal lines and shaded regions in the left and right panels, respectively, represent the mean and $RMS_{\rm MAD}$ density values within elliptical radius $R_{\rm ellipse}$ < 25\arcmin. The yellow shaded regions in both the panels represent the region encompassing NGC\,5018 and other GC overdensities surrounding the group (see Sect. \ref{sec:2d_maps}).}  \label{fig:rad_gr_profs_ellipse_back_sub_linear}
\end{figure*}

Figure \ref{fig:rad_gr_profs_ellipse_back_sub_linear} shows the background subtracted density profiles within the elliptical overdensity region. The radial distributions are obtained by selecting GC candidates within concentric ellipses (with fixed ellipticity) of increasing semi-major axis reaching $28\arcmin$. 
The profiles appear to have a relatively complex shape. We observe the radial density as relatively flat out to the ellipse radius, $R_{\rm ellipse}$ $\sim$ 20\arcmin, with a small shift of 0.74 GC/arcmin$^{2}$ between the two different background profiles.

The grey shaded region in Fig. \ref{fig:rad_gr_profs_ellipse_back_sub_linear} represents the \textit{annular background} region, where the radial profiles show a sharp decrease. The observed distribution peaks between $R_{\rm ellipse}$ $\sim$ 10\arcmin \ and $R_{\rm ellipse}$ $\sim$ 22\arcmin \ (highlighted in yellow in Fig. \ref{fig:rad_gr_profs_ellipse_back_sub_linear}), corresponding to the region where NGC\,5018 and other observed GC overdensities are located. Moreover, the innermost regions of the distribution ($R_{\rm ellipse}$ < 10\arcmin) have a density lower than the average between $R_{\rm ellipse}$ = 10\arcmin \ and 22\arcmin. The median and the $RMS_{\rm MAD}$ for $R_{\rm ellipse}$ < 25\arcmin \ is 2.38 $\pm$ 0.57 GC/arcmin$^2$ for \textit{annular background} subtracted profile and 3.12 $\pm$ 0.57 GC/arcmin$^2$ for \textit{random background} subtracted profile. These values are shown in Fig. \ref{fig:rad_gr_profs_ellipse_back_sub_linear} by red and green dotted horizontal lines and shaded regions in the left and right panels, respectively. Although this inner decrease is consistent with the estimated uncertainties, if real, it could potentially be explained by the fact that the elliptical shape we have assumed is a simplification of the actual geometry of the GC candidates in the observed intra-group population.

We observe only a small GC overdensity around the other four bright galaxies in the group, which is expected since these galaxies are less massive compared to NGC\,5018 and host a smaller GC population, particularly the lenticulars (MCG-03-34-013 and PGC\,140148). Moreover, the two brighter members of the group, NGC\,5022 and NGC\,5006, are spiral galaxies, making their GC populations harder to identify due to confusion from dust, star-forming regions, and disk structures. However, for both spirals, we observe regions of GC overdensities at large galactocentric distances, possibly part of their extended GC systems transitioning into the intra-group GC population. For comparison, the Milky Way GCs AM\,1 and PAL\,4, located $\sim$125 kpc from the Milky Way centre,  at the distance of the NGC\,5018 group would correspond to a projected distance of $\sim 12\arcmin$, aligning with the overdense regions around these spirals.

\begin{figure}[htb!]
   \centering   
   \includegraphics[width=9 cm]{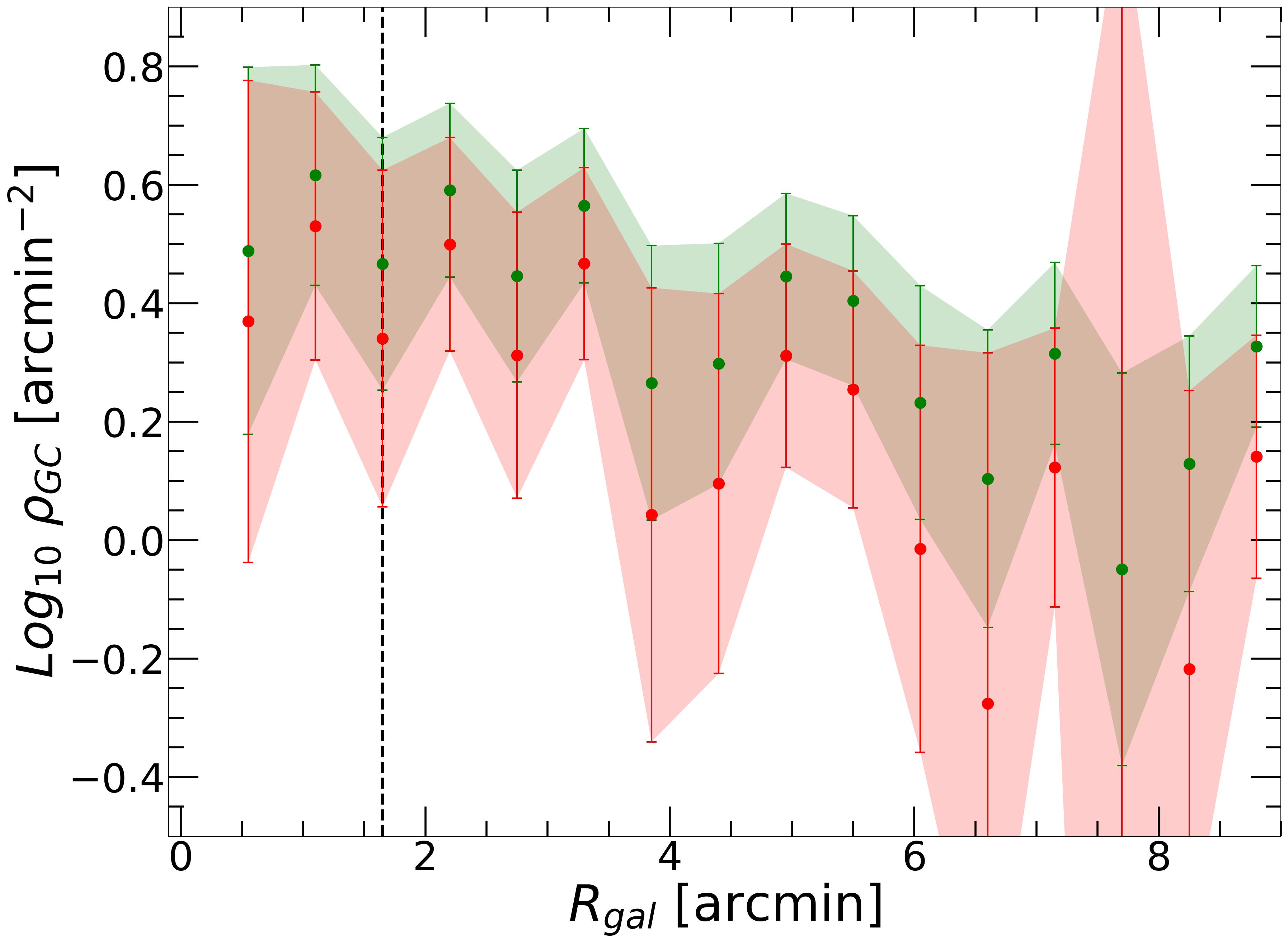}
   \caption{Radial density profile of GC candidates on NGC\,5006 after subtracting \textit{random background} (green points) and \textit{annular background} (red points). The vertical dashed black line represents 5 $R_{\rm e}$ radius.}  \label{fig:rad_gr_profs_5006_back_sub_log}
\end{figure}

Figure \ref{fig:rad_gr_profs_5006_back_sub_log} shows the background-subtracted radial density profile for NGC\,5006. Within 5 $R_{\rm e}$ (vertical dashed black line), we observe a small GC candidates overdensity. Beyond this, the GC population decreases, possibly transitioning into the intra-group component. For NGC\,5022 (magenta cross in Fig. \ref{fig:2d_map_gr_0.14_0.28}), we do not observe a local overdensity, but it is worth noting that it shows the strongest evidence of interaction with NGC 5018 \citep[see Fig. \ref{fig:resid} and][]{spavone18}, and lies between two GC overdense regions.

\subsubsection{Colour distribution}
\label{sec:col_prof}


As discussed in Sect. \ref{sec:intro}, the colour distribution of GC populations can provide insights into key properties of the host, in particular by allowing us to trace their metallicity distribution. Differences in integrated colours are a good proxy for differences in metallicities of old GCs, and this can help us to study the evolutionary history of the host galaxies.

\begin{figure*}[htb!]
    \centering    
    \begin{minipage}{0.495\textwidth}
        \centering        \includegraphics[width=0.999\textwidth]{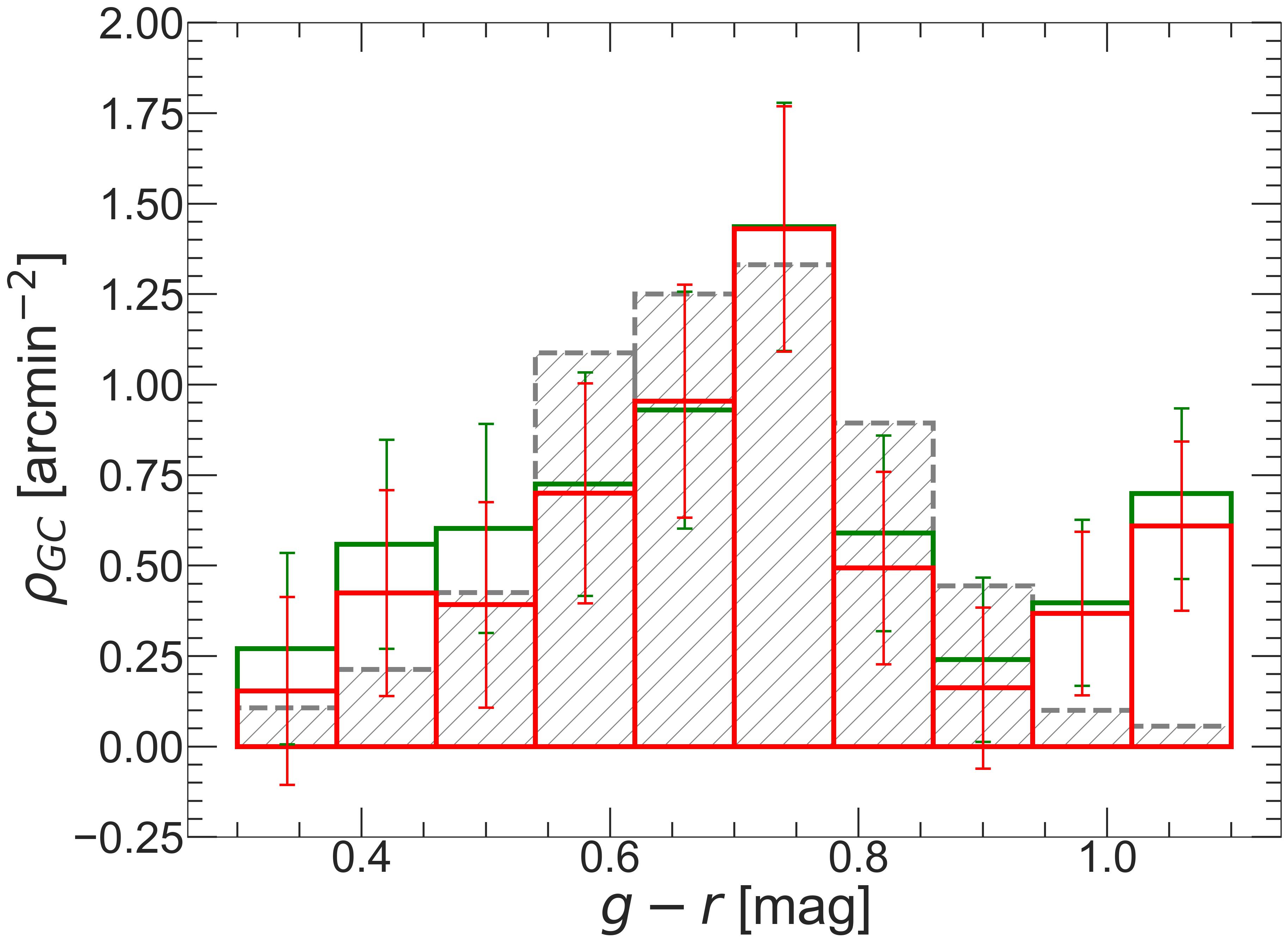}
    \end{minipage}
    \begin{minipage}{0.480\textwidth}
        \centering        \includegraphics[width=0.995\textwidth]{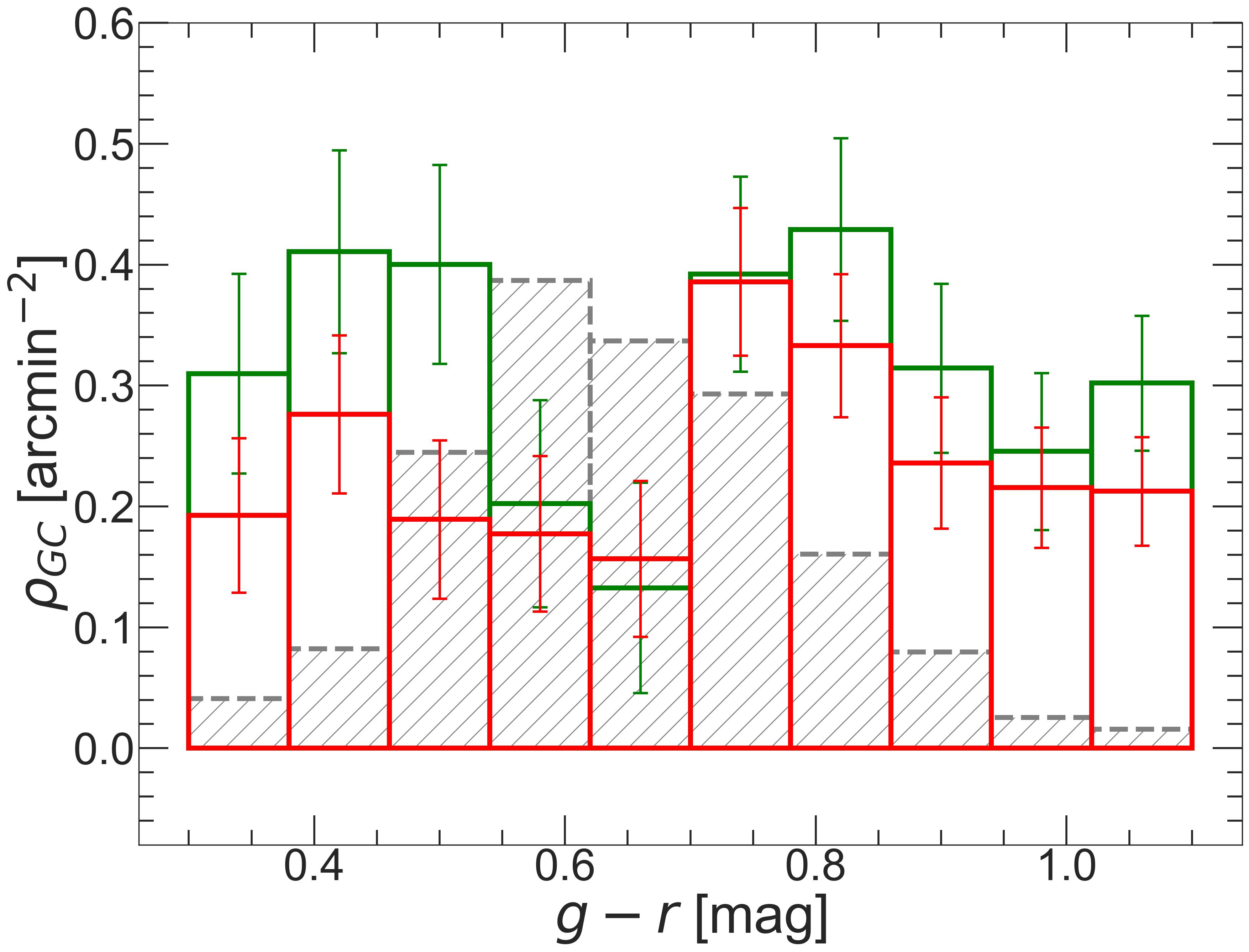}
    \end{minipage}
    \caption{$g-r$ colour distribution of GC candidates within $R_{\rm e,GC}$ of NGC\,5018 (left) and in the elliptical overdensity region (right) after \textit{annular} and \textit{random background} subtraction (shown in red and green histograms, respectively). The grey dashed histogram (with arbitrary normalisation, to get a similar scale) represents the spectroscopically confirmed GCs in NGC\,1399 (left) and in the entire Fornax cluster (right) from the FDS data \citep{cantiello20}.}    \label{fig:g_r_dist_5018_ell_back_sub}
\end{figure*}

We again take advantage of the large format of the VST images to apply the background decontamination technique for inspecting the ($g-r$) colour distribution of GC candidates. Figure \ref{fig:g_r_dist_5018_ell_back_sub} (left panel) shows the colour distribution of GC candidates density within $R_{\rm e,GC}$ of NGC\,5018. The red and green dashed histograms represent \textit{annular} and \textit{random background} subtracted colour density distributions, respectively, derived after normalizing to their respective areas. They reveal the presence of a relatively blue GC population with a peak at $(g{-}r) \sim$ 0.75 mag, along with a tail of red GC population with $(g{-}r) > 0.95$ mag. The colour distribution of GCs in NGC\,1399 (within an 8$\arcmin$  radius of the galaxy centre) from the FDS data is also shown as a grey hashed histogram. We observe that the GCs in NGC\,1399 have a peak at a very similar $(g{-}r)$ colour. We further examined the spatial distribution of the GC population with $(g{-}r) > 0.95$ mag around NGC\,5018 and found that a majority of these very red GC candidates lie on dust patches in its core region ($R_{\rm gal} \leq$ 1.2\arcmin). Their presence in dusty regions makes them appear redder than their actual colour. Hence, this red peak should be attributed to the presence of GCs reddened by dust.

The colour distribution of the GC candidates in NGC\,5018 shows no evidence of the well-known colour bimodality feature. This is likely due to the small wavelength separation between the $g$- and $r$-passbands combined with the relatively low S/N of most GC candidates at faint magnitudes, which increases photometric errors and smooths out any residual bimodality. Similarly, the GCs in NGC\,1399 do not show clear bimodality in ($g-r$) colour, though their distribution is bimodal in ($u{-}r$) and ($g{-}i$) colours, which involve wider wavelength separations \citep{cantiello20}. However, the absence of bimodality in NGC\,5018 might also result from some GC candidates being relatively young and blue \citep{hilker96}, but still passing our parametric selections, for example due to internal reddening from the dust in the galaxy core which makes blue GCs appear redder.

We also inspect the ($g-r$) colour distribution of the GC candidates over the elliptical overdensity region (right panel in Fig. \ref{fig:g_r_dist_5018_ell_back_sub}). The colour profiles reveal two peaks: a blue component with ($g-r$) $\sim$ 0.45 mag and a relatively redder component peaking at ($g-r$) $\sim$ 0.80 mag. For comparison, the entire GC sample from the FDS data is plotted as a grey hashed histogram. We observe a bimodal colour distribution for the GC candidates in the intra-group space of NGC\,5018. Unlike the Fornax cluster GCs, it appears that the bulk of the GC population in the elliptical overdensity region has two colour peaks. By inspecting the mean distance of blue ($g-r \leq$ 0.65 mag) and red ($g-r >$ 0.65 mag) GC populations from the centre of the elliptical overdensity, the two sub-populations appear indistinguishable, as the blue GCs have a mean group-centric radius of 13.2\arcmin, while the red GC population has a mean of 12.9\arcmin. When measured with respect to the centre of NGC\,5018, the mean radius is 16.3\arcmin \ for the blue and 13.0\arcmin \ for the red GCs. Based on this, we suggest that a certain fraction of the extended blue GC population in the intra-group space could represent the original blue old GCs system of NGC\,5018 that was possibly dispersed due to the tidal interactions with the neighbouring galaxies. The tidal interactions might also have resulted in some younger GCs forming in NGC\,5018, but it is unlikely that these young GCs are dispersed in the IGL, which would instead be more dominated by old and metal-poor ones.

\begin{figure*}
    \centering
    \includegraphics[width=0.995\textwidth]{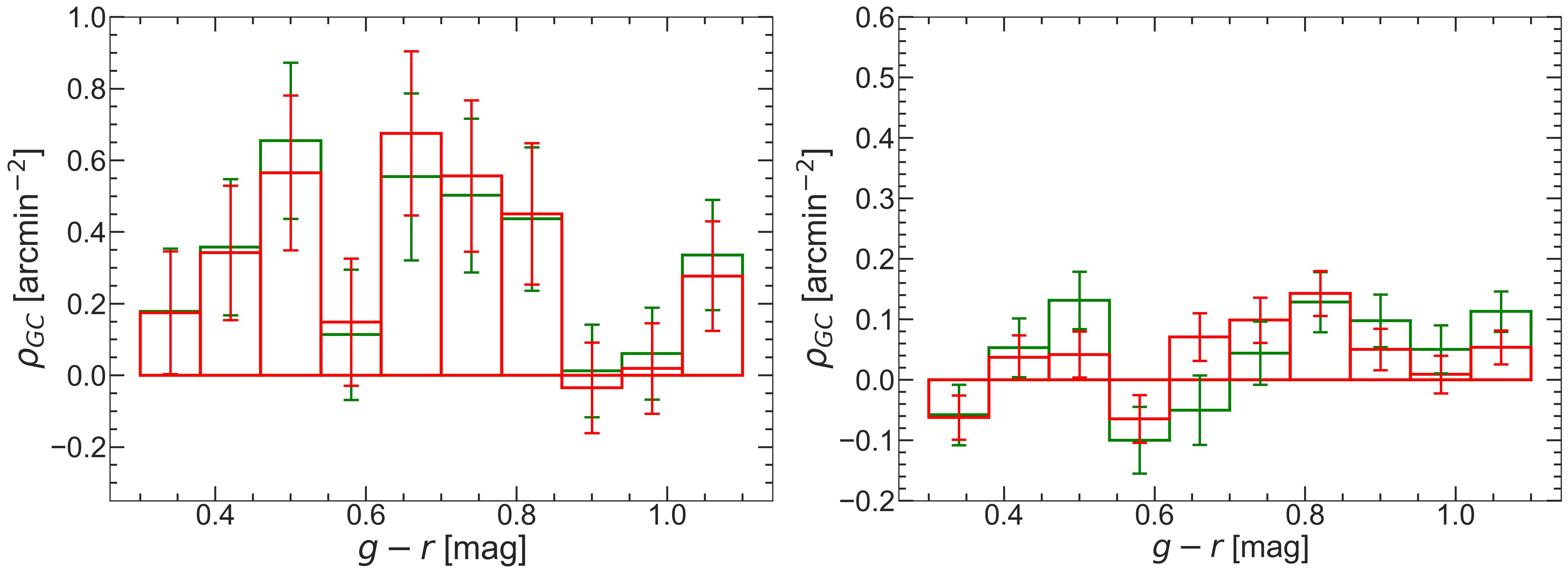}
    \includegraphics[width=0.995\textwidth]{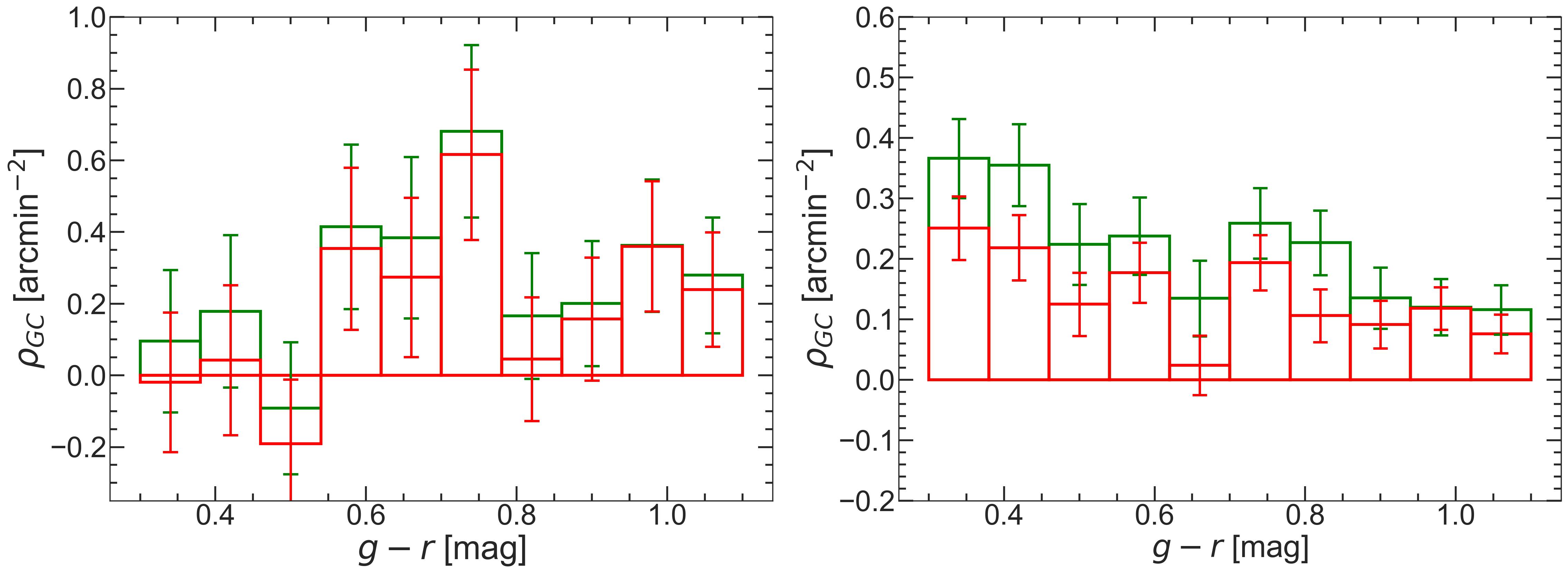}
    
    \caption{Top panel: Same as Fig. \ref{fig:g_r_dist_5018_ell_back_sub} but for GC candidates with 21.7 $\leq$ $m_g$ $\leq$ 24.1 mag. The change in magnitude is applied only in $g$-passband and the rest of the selection criteria is same as in Table \ref{tab:gr_gcsel_crit}.  Bottom panel: Similar to top panel, but for GC candidates with 24.1 $\leq$ $m_g$ $\leq$ 25.3 mag.}    \label{fig:g_r_dist_5018_ell_back_sub_bright_faint}
\end{figure*}

To further inspect the colour distribution, we tested the properties of the brightest GCs in the sample, which reduces scatter caused by sources with low S/N ratios. The results are shown in Fig. \ref{fig:g_r_dist_5018_ell_back_sub_bright_faint}, where we plot ($g-r$) colour histograms for GCs with: i) 21.7 mag $\leq$ $m_g$ $\leq$ 24.1 mag (brighter than 1$\sigma$ from the $m_g^{\rm TOM}$; bright sample, top panels), and; ii) 24.1 mag $\leq$ $m_g$ $\leq$ 25.3 mag (between 1$\sigma$ brighter than the $m_g^{\rm TOM}$ and up to $m_g^{\rm TOM}$; fainter sample, bottom panels). These histograms are presented for two regions: one within 5$R_e$ of NGC\,5018 (left panels), and the other for elliptical overdensity region (right panels). Magnitude selection was applied only in the $g$-band, while all other selection criteria remain the same reported in Table \ref{tab:gr_gcsel_crit}.

Similar to Fig. \ref{fig:g_r_dist_5018_ell_back_sub}, red and green histograms represent the \textit{annular} and \textit{random} background subtracted distributions in Fig. \ref{fig:g_r_dist_5018_ell_back_sub_bright_faint}. This test allows to draw the following considerations: i) Bright GCs on NGC\,5018 (top left panel): The background subtracted colour distribution shows a dip at ($g−r$) $\sim$ 0.6 mag, suggesting a bimodality that was not evident in the global sample (Fig. \ref{fig:g_r_dist_5018_ell_back_sub} left panel); ii) Fainter GCs on NGC\,5018 (bottom left panel): The peak at ($g−r$) $\sim$ 0.7 mag observed in Fig. \ref{fig:g_r_dist_5018_ell_back_sub} is preserved; iii) Bright intra-group GC candidates (top right panel): The ($g−r$) colour distribution shows two peaks, similar to the full sample in Fig. \ref{fig:g_r_dist_5018_ell_back_sub}. However, the sample size is insufficient for a robust identification, particularly of the blue peak; iv) Fainter intra-group GC candidates (bottom right panel): The distribution resembles that of the full sample in Fig. \ref{fig:g_r_dist_5018_ell_back_sub}, although with some differences in the shape. In conclusion, this test supports the presence of a broad colour distribution for intra-group GCs, with two colour peaks. For NGC\,5018, there are hints of potential bimodality among the brightest GC candidates. However, further insight would require deeper, multi-band data, such as those from the Legacy Survey of Space Time (LSST) survey.

\subsection{NGC\,5018-LSB1 dwarf galaxy}
\label{sec:lsb_dwarf}

In this section, we briefly discuss the properties of the dwarf galaxy identified in Sect. \ref{sec:2d_maps}. As previously noted, inspection of the $g$- and $r$-band images, along with the 2D GC candidates map, reveals a diffuse nucleated galaxy with a plume of GC overdensity extending towards it from NGC\,5018. The brightness profile of the galaxy also stretches in the direction of NGC\,5018, suggesting a possible tidal origin due to interactions between the two galaxies.

Using standard techniques for the morphological and photometric analysis of faint diffuse objects \citep[see][]{mirabile2024}, we estimated the main properties for the dwarf, summarised in Table \ref{tab:dwarf_prop}. Based on these characteristics and assuming the group distance (Table \ref{tab:gal_prop}), we classify NGC\,5018-LSB1 as an ultra-diffuse galaxy (UDG) candidate adopting the \citet{vandokkum15} definition. However, due to its faint magnitude and the presence of relatively bright point-like contaminants, further data are needed to confirm this result.

\begin{table}[htb!]
    \centering
    \caption{Properties of NGC\,5018-LSB1}
    \begin{tabular}{cc}
    \hline
    \\[-2ex]
    Parameter &  Value\\
    \\[-2ex]
    \hline
    \\[-2ex]
      R.A. (J2000) & 198.3500976 \\
      Dec (J2000) & -19.4434528 \\
      $m_g^{\rm tot}$ (mag) & 19.3 $\pm$ 0.5 \\ 
      $m_g^{\rm nucleus}$ (mag) & 22.9 $\pm$ 0.02 \\ 
      $m_r^{\rm nucleus}$ (mag) & 22.3 $\pm$ 0.02 \\ 
      $R_{\rm e}$ (arcsec) & 15 $\pm$ 5 \\
      $R_{\rm e}$ (kpc)$^{a}$ & $\sim2.4$ \\
      $\mu_{g,eff}$ (mag/arcsec$^2$) & 26.2 \\
     $g-r$ (mag) & $\sim0.8$ \\ 
     \\[-2ex]
    \hline
    \end{tabular}
    \tablefoot{$a$) Assuming the group distance reported in Table \ref{tab:gal_prop}.}
    \label{tab:dwarf_prop}
\end{table}

\subsection{Luminosity function (GCLF)}
\label{sec:gclf}

In this section, we focus on the GCLF analysis within $R_{\rm e,GC}$ of NGC\,5018. Using the $gr$ matched catalogue would require deriving combined $g$- and $r$-passband completeness functions. This approach would set our reference to the worst passband and require spatial and colour-dependent corrections, which, as anticipated, would be excessive for the present dataset. Therefore, we decided to proceed by inspecting the luminosity function using single-band data. Although single-passband catalogues exhibit higher contamination, they offer greater completeness. Due to poorer image quality and depth in the $r$-passband compared to the $g$-passband (see Figs. \ref{fig:phot_comp}, \ref{fig:comp_reg_funcs}, \ref{fig:gr_gcsel}, and Table \ref{tab:obs_image_prop}), we decided to analyse the GCLF using only the $g$-passband data. 

\begin{figure}
   \centering   
   \includegraphics[width=9cm]{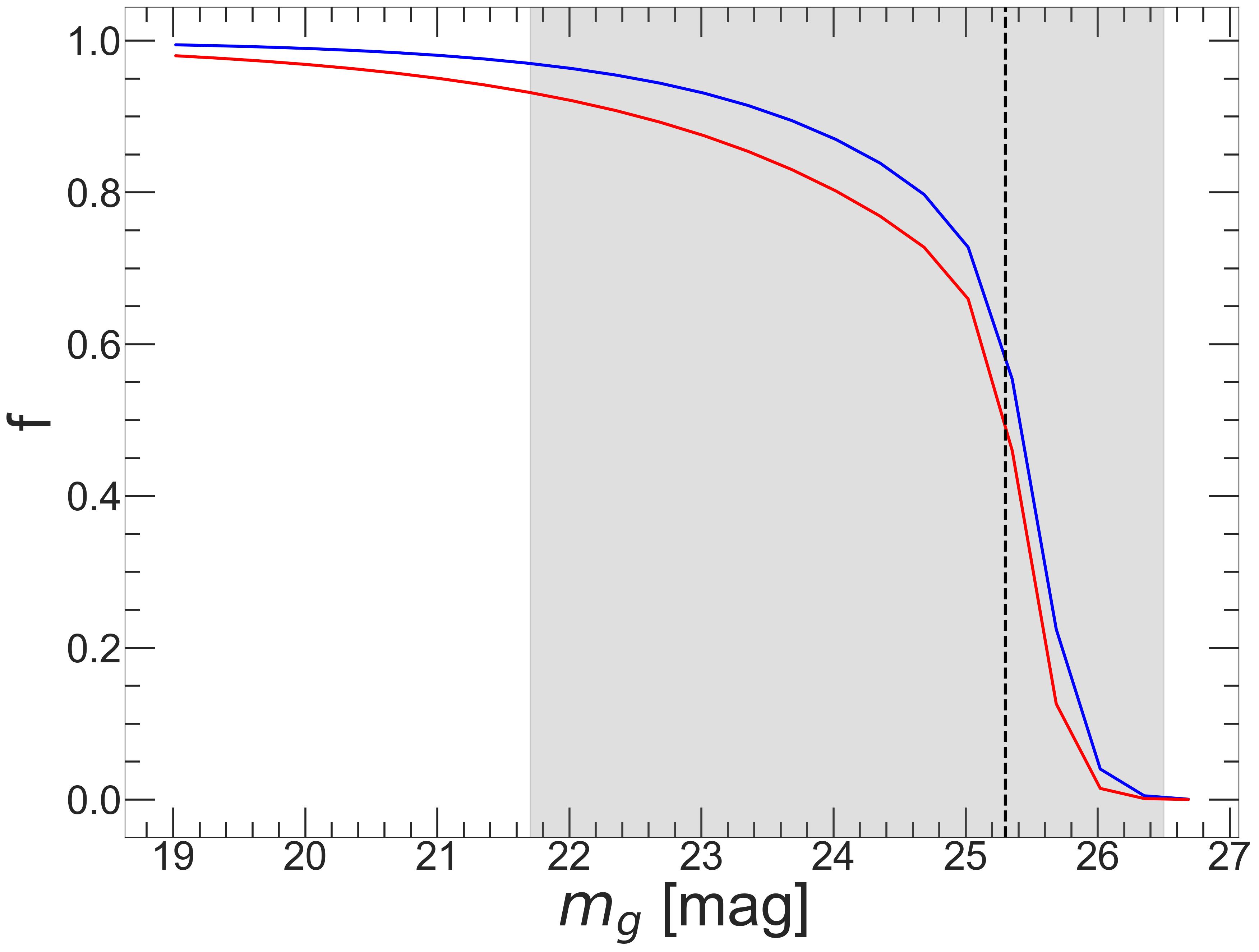}   \caption{Completeness functions derived for the GCLF analysis in the $g$-passband. The red curve represents the completeness function obtained from the region on NGC\,5018 (same as red curve in bottom left panel in Fig. \ref{fig:comp_reg_funcs}). The blue curve represents the mean completeness function derived for the two background regions (explained in Sect. \ref{sec:gclf}). The vertical dashed black line is the estimated TOM in this passband ($m_g^{TOM}$). The grey shaded region is the adopted GCLF range in this band (see Table \ref{tab:gr_gcsel_crit}).}    \label{fig:comp_funcs_5018_back_g}
\end{figure}

To obtain an accurate GCLF, we corrected it for the fraction of undetected sources. For the sources on NGC\,5018, we used the completeness function derived from the region containing this galaxy (red curve in Fig. \ref{fig:comp_funcs_5018_back_g}). For the background correction, we used the mean completeness derived from the four outermost off-galaxy regions (yellow, brown, purple, and olive-coloured boxes in the top left panel of Fig. \ref{fig:2d_map_gr_regions_0.14}) selected for the completeness analysis. These regions correspond to the locations of the two background areas used in this study. The derived completeness function for these backgrounds is shown in Fig. \ref{fig:comp_funcs_5018_back_g} (blue curve). As expected, Fig. \ref{fig:comp_funcs_5018_back_g}, similar to Fig. \ref{fig:comp_reg_funcs}, shows that the completeness on-galaxy is lower than that of the background (off-galaxy) regions. 

\begin{figure*}[htb!]
    \centering    
    \begin{minipage}{0.495\textwidth}
        \centering        \includegraphics[width=0.995\textwidth]{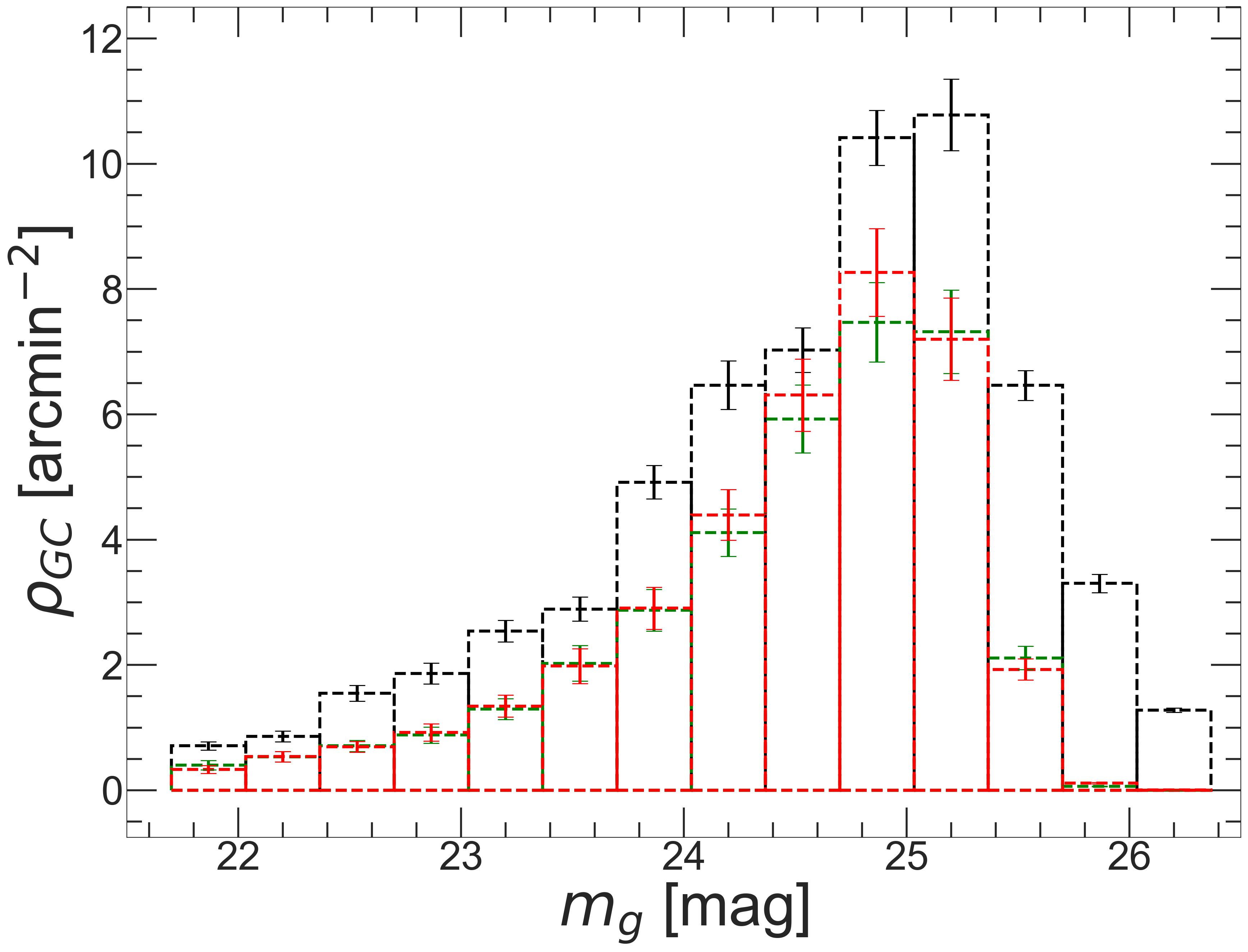}
    \end{minipage}
    \begin{minipage}{0.49\textwidth}
        \centering        \includegraphics[width=0.995\textwidth]{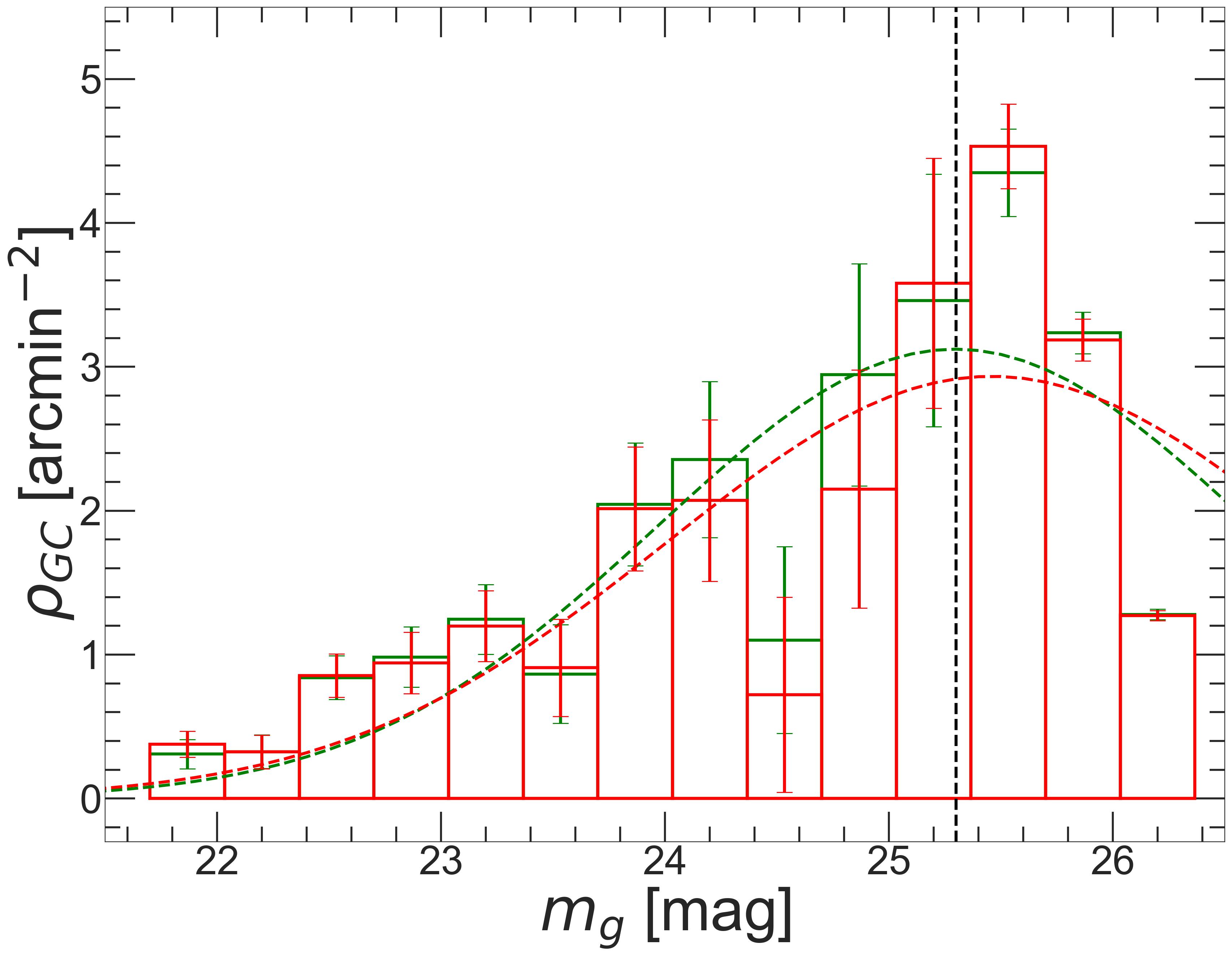}
    \end{minipage}
    \caption{Left: $g$-band completeness corrected luminosity function of GC candidates within $R_{e,GC}$ of NGC\,5018 (black dashed histogram) and for the \textit{annular background} (red dashed histogram) and \textit{random background} (green dashed histogram). Right: GCLF obtained after annular background subtraction (solid red histogram) and random background subtraction (solid green histogram), both fitted with Gaussian function (dashed red and green curves, respectively). The vertical dashed black line represents the expected TOM of the GCLF in this passband.}    \label{fig:gclf_cf_corr_5018_back_resid_g}
\end{figure*}

Figure \ref{fig:gclf_cf_corr_5018_back_resid_g} (left) presents the luminosity function of the GC candidates after applying the completeness correction within $R_{e,GC}$ of NGC\,5018 (black dashed histogram) and the \textit{annular} and \textit{random background}. Figure \ref{fig:gclf_cf_corr_5018_back_resid_g} (right) shows the residual luminosity functions obtained by subtracting the background luminosity functions from that of NGC\,5018. In both cases, we fit the residual with a Gaussian function for the annular and random background regions independently (red and green dashed curves, respectively). The fitting was performed using the curve\_fit\footnote{\url{https://docs.scipy.org/doc/scipy/reference/generated/scipy.optimize.curve_fit.html}} task from the SciPy python library. This was a two-step process: first, rough guesses for the Gaussian parameters were input, and the resulting fits were then used as new inputs to obtain optimised fits.

Table \ref{tab:gclf_fit_params_g} lists the various parameters obtained from the GCLF fitting along with the expected values (estimated in Sect. \ref{sec:phot_select}). The fitted peak of the GCLF is in good agreement with the expected values. By averaging the two $m^{\rm TOM}$ values and using our adopted $M_g^{TOM} = -7.5 \pm 0.2$ mag, we obtain a distance modulus of $32.90 \pm 0.45$ mag, corresponding to a distance of $38.02 \pm 7.89$ Mpc for NGC\,5018. This result agrees with the expected value ($36.0 \pm 1.7$ Mpc) reported in Table \ref{tab:gal_prop}. The observed dispersion of the GCLF in both cases appears slightly broader than the estimated value for both backgrounds.

\begin{table}[htb!]
    \centering
    \caption{Parameters derived from the GCLF fitting in the $g$-passband}
    \begin{tabular}{cccc}
    \hline
    \\[-2ex]
    Parameter & Expected$^a$ & Derived & Derived \\
     &  & \textit{annular back.}$^b$ & \textit{random back.}$^c$ \\
    \\[-2ex]
    \hline
    \\[-2ex]
     $m^{\rm TOM}$ (mag)$^d$ & $25.3 \pm 0.3$ & $25.5 \pm 0.5$ & $25.3 \pm 0.3$ \\ 
     $\sigma^{\rm GCLF}$ (mag)$^e$ & $1.2 \pm 0.2$ & $1.5 \pm 0.5$ & $1.3 \pm 0.3$ \\
     $N_{\rm GC}$$^f$ & - & $490 \pm 150$ & $480 \pm 150$ \\
     \\[-2ex]
      \hline
    \end{tabular}
    \tablefoot{$a$) Expected parameters estimated as explained in Sect. \ref{sec:phot_select}; $b$) Parameters extracted from fitting of \textit{annular background} subtracted GCLF (red curve in right panel in Fig. \ref{fig:gclf_cf_corr_5018_back_resid_g}); $c$) Parameters extracted from fitting of \textit{random background} subtracted GCLF (green curve in right panel in Fig. \ref{fig:gclf_cf_corr_5018_back_resid_g}); $d$) Position of the GCLF peak (TOM); $e$) Dispersion of the GCLF; $f$) Corrected total number of GCs in NGC\,5018.}
    \label{tab:gclf_fit_params_g}
\end{table}

As a test, we analysed the luminosity functions of the GC candidates in the elliptical overdensity region using the $gr$ matched catalogue. We also divided the full population of this region into blue ($g−r$ $\leq$ 0.65 mag) and red GC candidates ($g−r$ > 0.65 mag). Contrary to the single passband case, we did not make a more detailed analysis of the GCLF peak and width, as that would require a completeness correction on the combined $gr$ matched catalogue. For the full population, after background subtraction, we observed that the luminosity functions behaved as Gaussian peaking at $\sim$ 25.2 mag, as expected for GCs in the NGC\,5018 group. The peak of blue GCs residual was observed at a slightly brighter magnitude. The red GCs residual peak was instead observed at a slightly faint magnitude level ($\sim$ 25.5 mag). This is consistent with the findings in literature \citep[e.g.][]{brodie06,peng2009}.

\begin{figure*}
   \centering   
   \includegraphics[width=\textwidth]{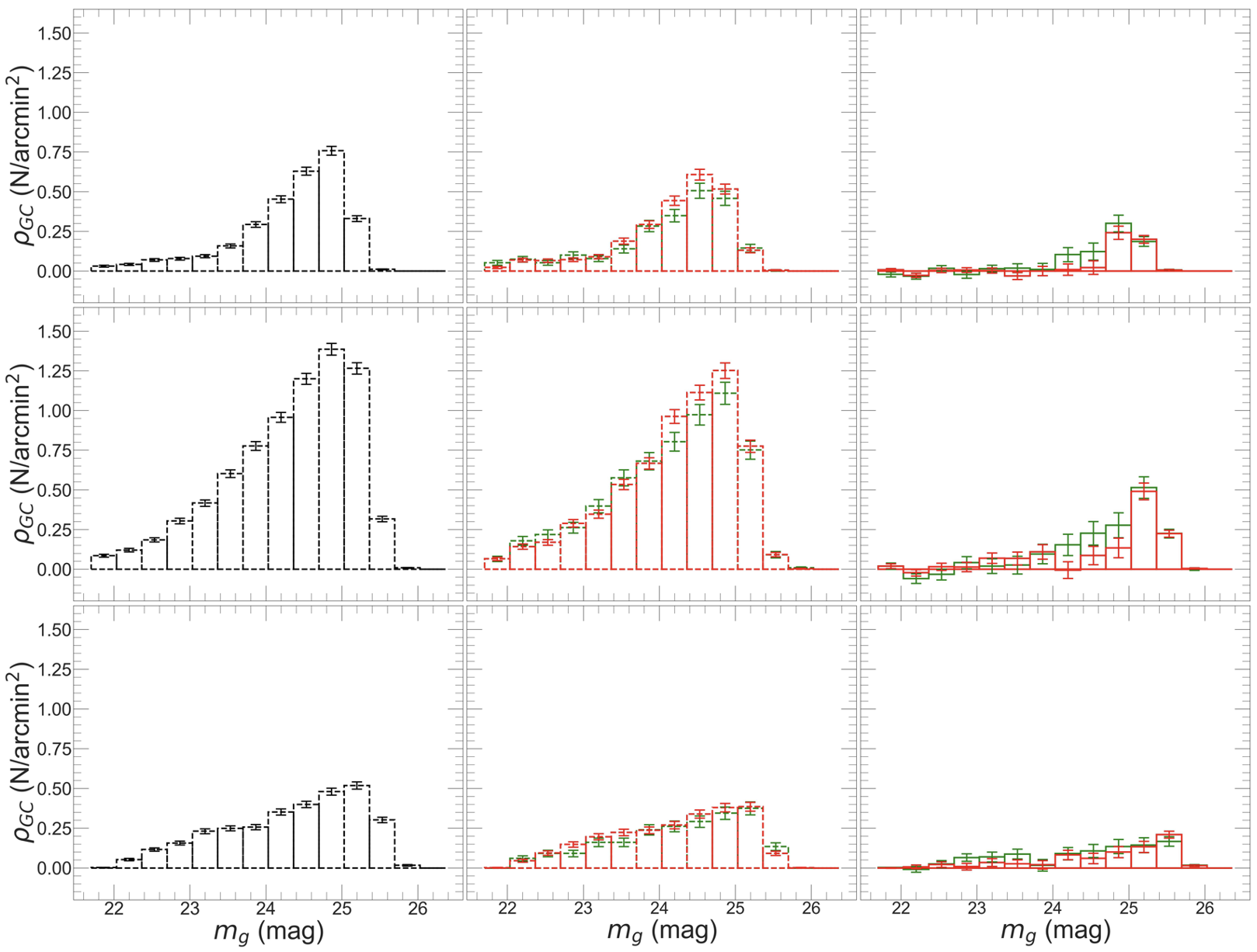}   \caption{Top row: Luminosity function of blue GC candidates (0.30 < $g-r$ $\leq$ 0.45 mag) from the $gr$ matched catalogue in the elliptical overdensity region (black dashed histogram in the left panel), \textit{annular} and \textit{random backgrounds} (red and green dashed histograms, respectively, in the middle panel) and residuals obtained after background subtractions (red and green solid histograms in the right panel). Middle and Bottom rows: Same as top row, but for intermediate GC candidates (0.45 < $g-r$ $\leq$ 0.85 mag) and red GC candidates (0.85 < $g-r$ $\leq$ 1.10 mag), respectively (see Sect. \ref{sec:gclf}).}    \label{fig:gclf_g_ellipse_blue_int_red_gcs_back_sub}
\end{figure*}

Although we do not observe ($g-r$) colour bimodality for NGC\,5018, we do observe it in the case of GC candidates in the elliptical overdensity region (Fig. \ref{fig:g_r_dist_5018_ell_back_sub}). It is worthy to note that the peak colours of the typical metal-poor and metal-rich GCs are ($g-r$) $\sim$ 0.60 and 0.75 mag, respectively, as shown in the FDS GC catalogue \citep[Figs. 8 and 15]{cantiello20}. So as a further test, we divide the GC sample of the elliptical overdensity region into three groups: blue GCs (0.30 < $g-r$ $\leq$ 0.45 mag), intermediate GCs (0.45 < $g-r$ $\leq$ 0.85 mag) and red GCs (0.85 < $g-r$ $\leq$ 1.10 mag); and inspect their luminosity functions shown in Fig. \ref{fig:gclf_g_ellipse_blue_int_red_gcs_back_sub}. The figure shows the luminosity functions of blue, intermediate and red GCs in top, middle and bottom rows, respectively, for the elliptical region (left panels), the two backgrounds (middle panels) and residual obtained after background subtractions (right panels). Like in the previous case, these luminosity functions are obtained using the $gr$ matched catalogue and, thus, are not completeness corrected. We observe a larger density of intermediate colour GCs in these luminosity functions. Also, we observe that the residuals for all three GC sub-populations appear roughly Gaussian. Although the sample size of blue and red GCs is relatively small,  we observe that the peak of luminosity functions of blue GCs is at a brighter magnitude ($\sim$ 24.9 mag) than the intermediate GCs ($\sim$ 25.2 mag). This is consistent with a possible brighter TOM for blue GCs or a residual contamination from young-ish bright star clusters. The peak of luminosity function of red GCs ($\sim$ 25.5 mag) appears to be fainter than the intermediate GCs consistent with the results in the literature \citep[e.g.][]{criscienzo2006}.

As a further test, we also inspected the radial density profiles (background subtracted) and 2D distribution maps of blue, intermediate and red GCs. Unsurprisingly, the intermediate GCs have a larger density of sources, compared to blue and red GCs, in the radial density profiles with the peaks lying between $R_{\rm ellipse}$ = 10\arcmin \ and 22\arcmin, corresponding to regions of NGC\,5018 and other GC overdensities surrounding the group (similar to Fig. \ref{fig:rad_gr_profs_ellipse_back_sub_linear}). On the other hand, the radial density profiles of blue and, especially of the red GCs appeared relatively flat. In the 2D distribution maps, we observed a clear overdensity of intermediate GCs on top of NGC\,5018 and in the red dashed ellipse shown in Fig. \ref{fig:2d_map_gr_regions_0.14}. Blue and red GCs instead appeared to have more patchy surface density. In particular, we observed blue/red GC overdensities aligning along the five bright galaxies of the group or with the tidal features.

\subsection{Total number of GCs ($N_{\rm GC}$) and Specific frequency ($S_{\!\rm N}$)}
\label{sec:ngc_sn}


To estimate the specific frequency for NGC\,5018, we use the mean of the two background corrected $N_{\rm GC}$ values ($N_{\rm GC}$ = $485 \pm 150$) obtained from the GCLF fitting described in the previous section. It is useful to note that the numbers reported in Table \ref{tab:gclf_fit_params_g}, obtained by integrating the GCLF up to the observed TOM from the fitting, are derived by doubling the results from integration twice: once for the GCLF part fainter than the TOM and once more to account for the spatial coverage due to the fact that we only consider the GC population within $R_{\rm e,GC}$ (see Sect. \ref{sec:morpho_select} for the definition).

For the $S_{\!\rm N}$ estimate, we use Eq. \ref{eqn:spec_freq} and derive an estimate for NGC\,5018 total $V$-passband magnitude using the transformation equation provided on the SDSS webpage (Lupton 2005) to convert $g$- and $r$-passband magnitudes to the $V$-passband magnitude:
\begin{linenomath}
\begin{equation}
\label{eqn:v_band_mag_lupton_eqn}
V = g - 0.5784 \times (g-r) - 0.0038. 
\end{equation}
\end{linenomath}
Using $m_g$ = 10.9 mag for NGC\,5018 from \citet{spavone18} and the ($g-r$) = $0.7 \pm 0.2$ mag from the same paper (as reported in their Table 6), we obtain $m_{V}$ = $10.52 \pm 0.12$ mag. At our adopted distance modulus this leads to $M_V$ = $-22.28 \pm 0.15$ mag. Our derived value for $M_V$ is consistent with the value reported in \citet{rampazzo07}. The resulting $S_{\!\rm N}$ = $0.59 \pm 0.27$ is consistent within the errors with the existing estimates from the literature reported in Table \ref{tab:sn_estimates}. 

\begin{table}[htb!]
    \centering
    \caption{$N_{\rm GC}$ and $S_{\!\rm N}$ estimates for NGC\,5018.}
    \begin{tabular}{ccc}
    \hline
    \\[-2ex]
    Source & $N_{\rm GC}$ & $S_{\!\rm N}$ \\
    \\[-2ex]
    \hline
    \\[-2ex]
      \citet{hilker96} & $1700 \pm 750$ & $1.10 \pm 0.60$ \\ 
       \citet{humphrey09} & $90 \pm 43$ & $0.46 \pm 0.22$ \\
        This work & $485 \pm 150$ & $0.59 \pm 0.27$ \\
        \\[-2ex]
      \hline
    \end{tabular}
    \label{tab:sn_estimates}
\end{table}



Considering the GC distribution over the elliptical overdensity region (red dashed ellipse in Fig. \ref{fig:2d_map_gr_regions_0.14}) and the total area of the assumed geometry, we derive an order-of-magnitude estimate of the total intra-group GC population. Unlike the approach used to estimate the total population of NGC\,5018, we use the $gr$ matched catalogue, which, as previously discussed, has different completeness properties compared to the single-band catalogues. For this, we assume the GCLF in the $g$- and $r$-bands of the intra-group GC candidates is roughly complete down to the TOM (see Fig. \ref{fig:comp_reg_funcs}). After background subtraction, the number obtained needs to be doubled to account for the fainter half of the GCLF and then multiplied by the total area of the ellipse. This gives a total GC population of $4800 \pm 470$ and $3100 \pm 370$ after \textit{annular} and \textit{random background} subtraction, respectively. However, since the $gr$ matched catalogue is less complete than the single-band catalogues, both total populations and their uncertainty should be considered lower limits.

We can use these numbers to estimate the specific frequency for the group: $S_{\!\rm {N,gr}}$. The area over which \citet[][see their Figure 5]{spavone18} estimates the total luminosity of the NGC\,5018 group ($L_g = 1.7 \times 10^{11} L_\odot$; Sect. \ref{sec:5018_literature}) has a similar geometry, though it is less extended, covering roughly half of our elliptical region. This total luminosity includes the contributions from the IGL component, NGC\,5018, NGC\,5022 and MCG-3-34-013. By doubling the total luminosity of the group\footnote{We converted the total $g$-band magnitude from \citet{spavone18} to the $V$-band using Eq. \ref{eqn:v_band_mag_lupton_eqn}.} and adding the $V$-band luminosities of NGC\,5006 \citep[using $M_B$ reported in Table \ref{tab:gal_prop} and assuming $B-V$ $\sim$ 0.6 mag for spiral galaxies;][]{tomita1996} and PGC\,140148 \citep[using $M_B$ reported in Table \ref{tab:gal_prop} and assuming $B-V$ $\sim$ 0.9 mag for lenticular galaxies;][]{barway2005}, we estimate the total $V$-band magnitude of the NGC\,5018 group to be $M_{V,gr}$ = \textminus 24.5 mag. Then using a mean intra-group GC population of $N_{GC,gr}\sim4000$, we obtain a specific frequency of $S_{\!\rm N,gr}\sim 0.6$. This should be considered a lower limit \citep{peng11,durrell14}, as the total IGL luminosity is likely overestimated (we assumed twice the estimate from \citet{spavone18}, neglecting the fact that the IGL fades at larger radii), while the number of GCs is underestimated for reasons explained above.

\section{Discussion}
\label{sec:discussion}

In this study, we used VEGAS survey data to investigate the GC system in the NGC\,5018 group. The quality and characteristics of the VEGAS imaging data allowed us to identify GC candidates across a field of 1.25 $\times$ 1.03 sq. degrees, allowing us to reveal the presence of an intra-group GC system in this galaxy group.

Our analysis revealed several key features of the GC system in the NGC 5018\,group. The 2D distribution maps show a notable elliptical overdensity of GC candidates, offset from the centre of NGC\,5018. This overdensity extends diagonally in the direction of the brightest galaxies in the group and includes a plume-like structure to the north-east of NGC\,5018 (Fig. \ref{fig:2d_map_gr_0.14_0.28}). The consistency of this feature across various tests (Figs. \ref{fig:2d_maps_narrow_sels_0.14} and \ref{fig:2d_maps_gr_TOM_elong_gr_2.0_0.14}), and its alignment with the geometry of the IGL, suggests that it represents a genuine intra-group GC population rather than an artefact of our selection criteria or background contamination.
It is important to recognise that using an elliptical geometry serves as a simplification  to make our analysis and presentation clearer, rather than accurately depicting the true geometry of the intra-group GCs component.

The radial density profiles support this finding, showing a relatively flat distribution of GC candidates accompanied by complex structures with several local peaks (Fig. \ref{fig:rad_gr_profs_ellipse_back_sub_linear}). Despite the relatively low efficiency of GC detection around NGC\,5018, we observed a significant peak in GC density on the galaxy, with a general decrease towards the outskirts (Fig. \ref{fig:rad_prof_5018_back_sub_log}). A plume-like feature in the 2D GCs density map points towards a local nucleated dwarf galaxy which is aligned with NGC\,5018. By inspecting the morphological and photometric parameters of this dwarf galaxy, NGC\,5018-LSB1, we find it appears to be a UDG candidate. The radial profiles of GC candidates around other bright galaxies in the group indicate less prominent GC systems, but the radial density profiles around the two spiral galaxies in the group, NGC\,5022 and NGC\,5006 (Fig. \ref{fig:rad_gr_profs_5006_back_sub_log}), suggest they blend into the intra-group GC population.

In terms of colour distribution, the GC candidates in NGC\,5018 display a broad range of colours without a clear bimodal distribution (left panel in Fig. \ref{fig:g_r_dist_5018_ell_back_sub}). This lack of bimodality may be influenced by photometric limitations, such as the relatively small separation between the $g$- and $r$-passbands. Nevertheless, it might also be due to the presence of a fraction of relatively young GCs \citep{hilker96}.

On the contrary, the $(g-r)$ colour profile of GC candidates in the elliptical overdensity region (right panel in Fig. \ref{fig:g_r_dist_5018_ell_back_sub}) reveals two peaks: a bluer peak at  $g-r\sim0.45$ and a red peak at  $\sim$0.80 mag. When inspecting the mean distance of blue ($g−r \leq 0.65$ mag) and red ($g−r > 0.65$ mag) GC populations from the centre of the elliptical region, we find that they have similar spatial distributions. However, with respect to the centre of NGC\,5018, the mean distance of the blue GC population is 16.3 arcminutes, while the red population is at 13.0 arcminutes. Thus, relative to NGC\,5018, the blue population of the intra-group GCs is more extended than the red population. This may suggest that a portion of this extended intra-group blue GC population originally belonged to NGC\,5018 and was later dispersed due to tidal interactions with neighbouring galaxies.

Our results from the GCLF analysis, based solely on the $g$-band data, indicate that the GCLF observed in NGC\,5018 has characteristics (peak and width) consistent with expectations (Fig. \ref{fig:gclf_cf_corr_5018_back_resid_g}). Using the results from the GCLF, we estimate the total population of GCs to be $N_{\rm GC} = 485 \pm 150$. This corresponds to a specific frequency of $S_{\!\rm N} = 0.59 \pm 0.27$, which agrees with previous estimates from the literature (based on smaller areas and/or shallower photometry). It also classifies NGC\,5018 as a relatively GC-poor galaxy, as elliptical galaxies similar to NGC\,5018 typically have $S_{\!\rm N} > 1$.

The GCLF analysis of intra-group GC candidates within the elliptical overdensity region (Fig. \ref{fig:gclf_g_ellipse_blue_int_red_gcs_back_sub}) show that peak of the luminosity function of blue GCs (0.30 < $g-r$ $\leq$ 0.45 mag) is observed at a slightly brighter magnitude compared to the intermediate GCs (0.45 < $g-r$ $\leq$ 0.85 mag). Also, the peak of luminosity function of red GCs (0.85 < $g-r$ $\leq$ 1.10 mag) is observed at a slightly faint magnitude level than the intermediate GCs, consistent with the results in the literature.

The low $S_{\!\rm N}$ value, combined with the blue GC population observed in the intra-group space, could be explained by the loss of GCs from NGC\,5018, which enriched the intra-group environment due to ongoing interactions between the galaxies in the group, together with the brighter magnitude due to the post-merger status of the group. For the intra-group GC population, we estimate a lower limit of $N_{\rm {GC,gr}} \sim 4000$, which translates to $S_{\!\rm {N,gr}} > 0.6$ when combined with the intra-group light and galaxy luminosity estimates from \citet{spavone18}.

We also examined the observed field for potential Ultra-Compact Dwarf (UCD) candidates \citep{hilker2006,evstigneeva2008}. In the $gr$ matched catalogue, which are complete at the magnitude levels expected for UCDs, we applied a cut two magnitudes brighter than the brightest limit for GCs in the $g$- and $r$-bands and used a stricter elongation cut ($\leq 1.5$ in both passbands). The $(g-r)$ colour and CI selection criteria remained unchanged. The resulting UCD candidates counts and 2D distribution map showed a flat, homogeneous spatial distribution. After background subtraction, the number of UCD candidates was consistent with zero. However, this does not necessarily imply the absence of UCDs, as they could be masked by statistical fluctuations, particularly from Milky Way stars in the field.

\section{Summary}
\label{sec:summary}

We presented the analysis of GCs in the NGC\,5018 group using the deep, multi-passband and wide field imaging from the VEGAS survey obtained using the VST telescope. A summary of the main results we obtained is as follows:

\begin{enumerate}

    \item The 2D distribution map shows an overdensity of GC candidates on the brightest member of the group, NGC\,5018. No significant GC overdensities are observed on the other galaxies in the group.
    
    \item We observe the presence of an intra-group GC population aligning along the five bright galaxies and surrounding the group itself. This observed geometry aligns with IGL detected in the group, but extended to larger group-centric distances.
    
    \item We report the discovery of a local nucleated LSB dwarf galaxy candidate, NGC\,5018-LSB1, which is possibly interacting with the nearby NGC\,5018 and is a candidate UDG galaxy in the group.
    
    \item The radial density profile of GC candidates in NGC\,5018 follows the surface brightness profile of the galaxy, although the central deviation of the profile ($R_{\rm gal}$ $\leq$ 1.2\arcmin) is due to dust and associated incompleteness in the GC detection and the bump observed in the profile at $R_{\rm gal}$ > 3.6\arcmin \ is due to observed plume of GC overdensity towards NGC\,5018 - LSB1.  
    
    \item The radial density profile of the spirals NGC\,5022 and NGC\,5006 reveals no central GC overdensity, but highlight the presence of an extended GC population that fades into the intra-group GC population. 
    
    \item The colour profile of the GC candidates in NGC\,5018 shows a major component with a peak at ($g-r$) $\sim$ 0.75 mag. The profile shows no hints of the well-known colour bimodality typically observed in such massive, bright ellipticals. This could be due to a combination of a part of GC population being young, short wavelength separation between $g$- and $r$-passbands and low S/N of GCs at faint magnitude levels.
    
    \item The colour profile of the intra-group GC candidates shows the presence of blue and red GC components with peaks at ($g-r$) $\sim$ 0.45 and $\sim$ 0.80 mag, respectively. The blue GC component is found to be more extended compared to the red GC component with respect to NGC\,5018. This, combined with previous results on the IGL of the system, might suggest that a part of blue GC population in the intra-group space originally  belonged to NGC\,5018 which was dispersed due to tidal interactions with the neighbouring galaxies. 
    
    \item By inspecting the GCLF, we obtain parameters consistent with the known distance to the galaxy and with the expected Gaussian parameters (width and peak). 
    
    \item Using the fitted GCLF we estimate $N_{\rm GC}$ = $485 \pm 150$, resulting in  $S_{\!\rm N}=0.59 \pm 0.27$ for NGC\,5018, in agreement with the previous works confirming that this galaxy harbours a relatively poor GCs system. For the intra-group GCs, we estimate $N_{\rm {GC,gr}}\sim 4000$ and a lower limit for specific frequency of the group $S_{\!\rm {N,gr}}> 0.6$.

    \item The GCLF analysis of intra-group GC candidates show that the peak of luminosity function of blue GCs is at a slightly brighter magnitude level than the intermediate GCs, while the same for red GCs is at a slightly faint magnitude level than the intermediate GCs which is consistent with the results in the literature.
    
\end{enumerate}

The results presented here relied mostly on deep $g$- and $r$- passband wide field images from the VEGAS survey. The work we presented over the dynamically interacting system of galaxies in the NGC\,5018 group shows the relevance of wide-area GC studies for constraining the properties of the host environment. In the future, thanks to the advent of large sky surveys, like the LSST from the Vera Rubin Observatory and Euclid, we will have wider areas covered with multiple filters and at  deep magnitude limits for the NGC\,5018 and other galaxy groups. This will greatly contribute in enhancing our understanding of GCs in galaxy groups and their role in formation and evolution of host galaxies and their environment.

\begin{acknowledgements}

This work is based on visitor mode observations taken at the European Southern Observatory (ESO) La Silla Paranal Observatory within the VST Guaranteed Time Observations; Programme IDs 096.B-0582(B), 097.B-0806(A), and 099.B-0560(A). M.S. wishes to thank the ESO staff of the Paranal Observatory for their support during the observations at VST. PL acknowledges financial support from the Astronomical Observatory of Abruzzo of the Italian National Institute for Astrophysics (INAF-OAAb) and University of Rome Tor Vergata. EI, MS and MC acknowledge the support by the Italian Ministry for Education University and Research (MIUR) grant PRIN 2022 2022383WFT “SUNRISE”, CUP C53D23000850006 and by  VST funds. EI, MS, MC and RH acknowledge funding from the INAF through the large grant PRIN 12-2022 "INAF-EDGE" (PI L. Hunt).

This research has made use of the NASA/IPAC Extragalactic Database (NED), which is funded by the National Aeronautics and Space Administration (NASA) and operated by the California Institute of Technology. We also acknowledge the usage of the HyperLeda database \citep[\url{http://leda.univ-lyon1.fr};][]{makarov14}) and the the Extragalactic Distance Database (EDD, \url{https://edd.ifa.hawaii.edu/}). We made extensive use of the softwares SExtractor \citep{bertin96} and Topcat \citep[\url{https://www.star.bris.ac.uk/~mbt/topcat/};][]{taylor05}). This research has made use of the AAVSO Photometric All Sky Survey (APASS) database, located at the AAVSO web site (\url{https://www.aavso.org}). Funding for APASS has been provided by the Robert Martin Ayers Sciences Fund. This research has made use of the VizieR catalogue access tool, Strasbourg Astronomical Data Center (CDS), Strasbourg, France \citep{10.26093/cds/vizier}. The original description of the VizieR service was published in \citet{vizier00}. This research has made use of the Sloan Digital Sky Survey (SDSS) database \url{www.sdss.org}. Funding for the SDSS V has been provided by the Alfred P. Sloan Foundation, the Heising-Simons Foundation, the National Science Foundation, and the Participating Institutions. SDSS acknowledges support and resources from the Center for High-Performance Computing at the University of Utah. SDSS telescopes are located at Apache Point Observatory, funded by the Astrophysical Research Consortium and operated by New Mexico State University, and at Las Campanas Observatory, operated by the Carnegie Institution for Science. This work made use of Astropy (\url{http://www.astropy.org}) a community-developed core Python package and an ecosystem of tools and resources for astronomy \citep{astropy:2013, astropy:2018, astropy:2022}. This research made use of Photutils, an Astropy package for
detection and photometry of astronomical sources \citep{bradley24}.

\end{acknowledgements}

\bibliography{biblio}{}
\bibliographystyle{aa}

\begin{appendix}
\label{sec:appendix}
\onecolumn 

\section{Testing the stability of the elliptical overdensity feature.}
\label{sec:ellipse_stab_tests}

As discussed in Sect. \ref{sec:2d_maps}, we detect the presence of an intra-group GC population roughly centred on the observed field using the 2D distribution map of GC candidates from the $gr$ matched catalogue (Fig. \ref{fig:2d_map_gr_regions_0.14}). This feature is aligned along the five bright galaxies in the group and shows a significant overlap with the IGL region studied by \citet{spavone18}, but extends to larger group-centric radii. To verify the persistence of this feature against our GC selection criteria, the quality of the observational dataset and the possibility that the feature is the result of non GC-contaminants remaining in our catalogue, we conducted several tests described below.

\begin{enumerate}
  
\item We narrowed down the GC selection criteria to study two cases: 
$i)$ \lq Case-1\rq, where we apply more stringent selections in the $g$-passband, using the reference magnitude selection for the $r$-band data; and 
$ii)$ \lq Case-2\rq, where more stringent selections are applied in both the $g$- and $r$-passbands. Additionally, we apply a narrower ($g-r$) colour range in both cases to obtain a less contaminated, although more incomplete, catalogue of GC candidates compared to our reference catalogue. The new criteria for the two cases are detailed in Tables \ref{tab:gr_gcsel_narrow_crit} and \ref{tab:gr_gcsel_very_narrow_crit}.

Figure \ref{fig:2d_maps_narrow_sels_0.14} shows the resulting 2D distribution maps for the new tests. Although the number of candidates decreased, both new maps confirm the overall appearance of our reference maps.  In these new tests though, the GC overdensity appears more concentrated in the regions covered by the five bright group members.

\begin{figure}[htb!]
    \centering    
    \begin{minipage}{0.495\textwidth}
        \centering        \includegraphics[width=0.995\textwidth]{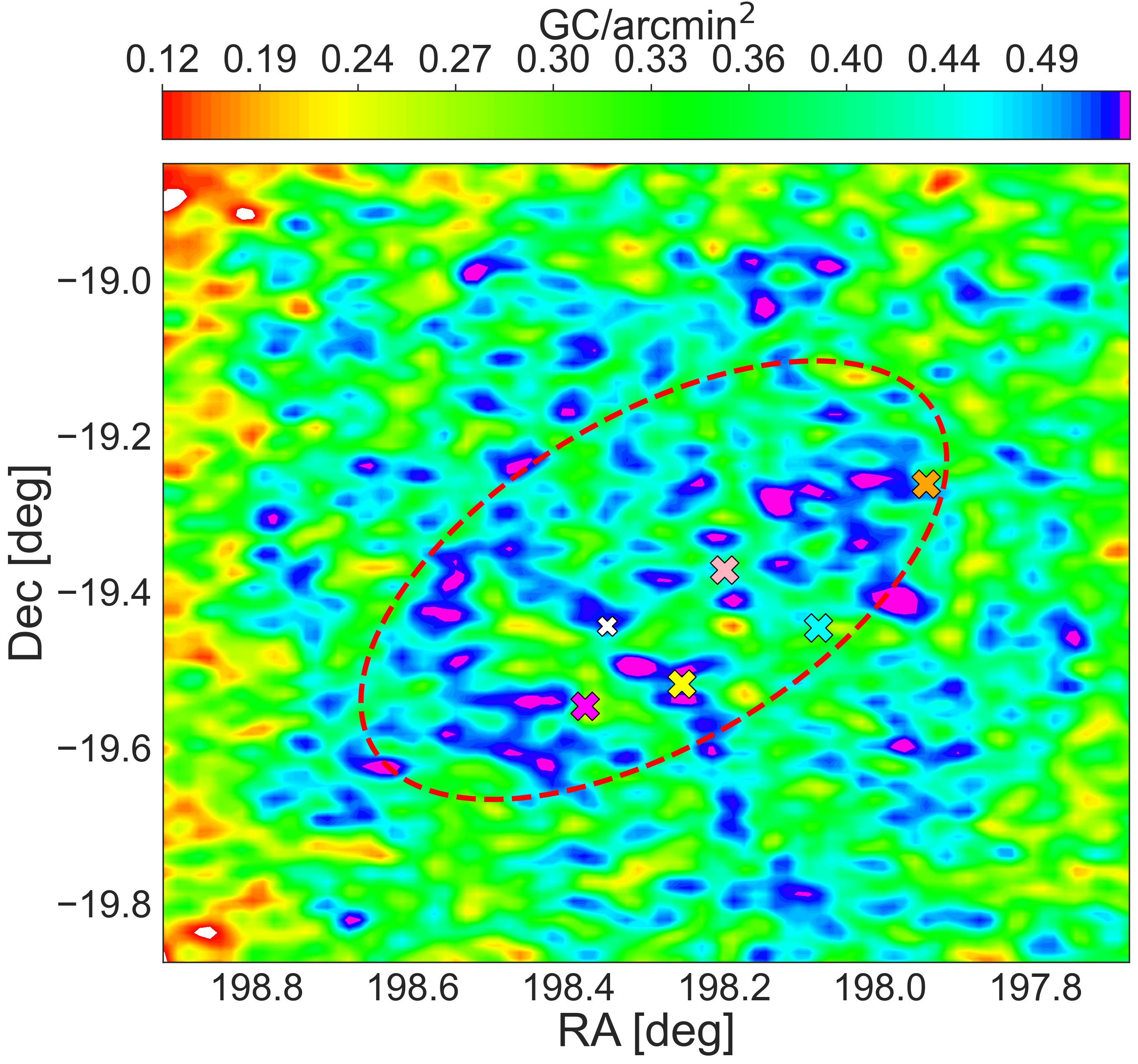}
    \end{minipage}
    \begin{minipage}{0.495\textwidth}
        \centering        \includegraphics[width=0.995\textwidth]{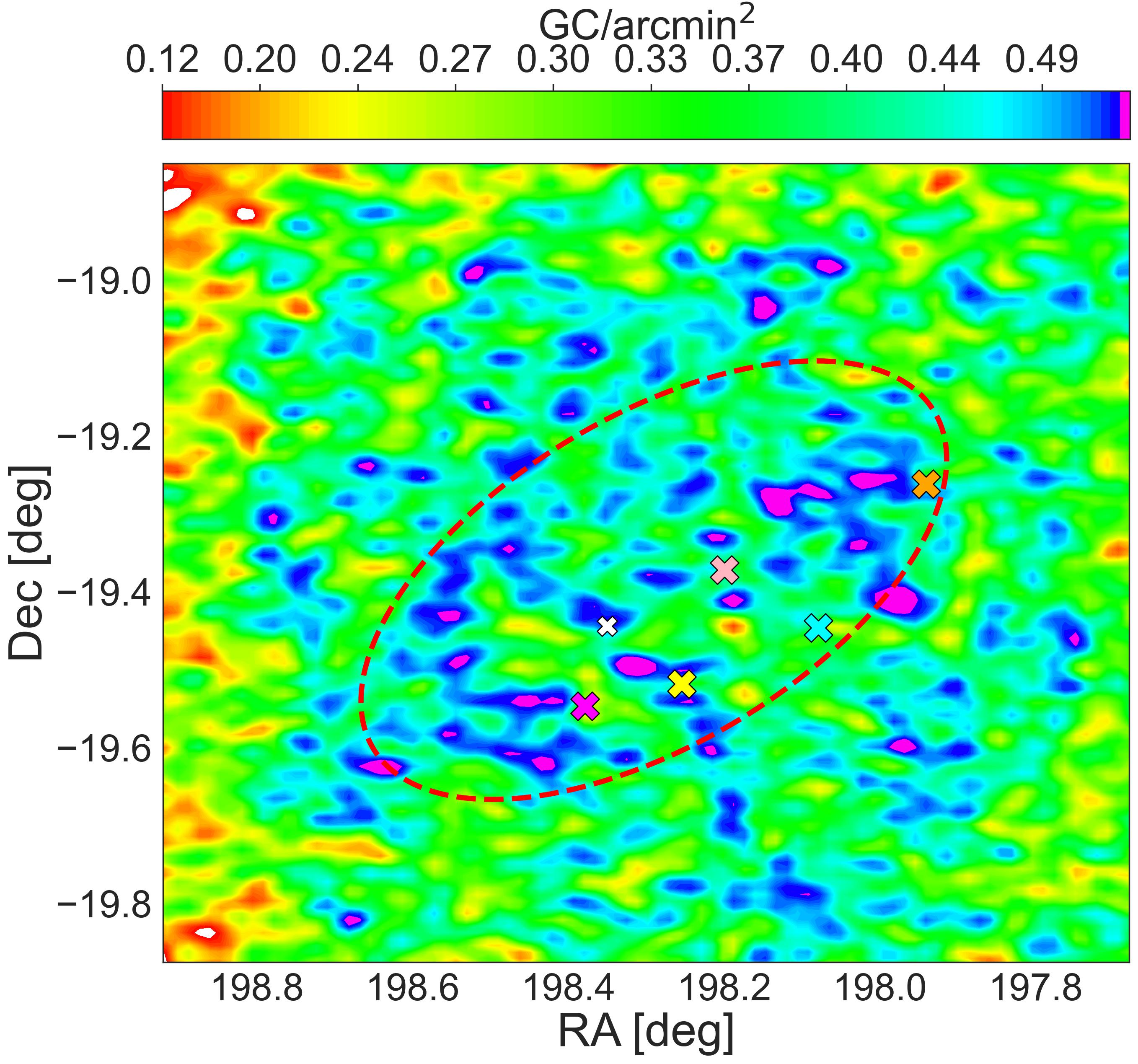}
    \end{minipage}
    \caption{2D distribution maps obtained for \lq Case-1\rq (left) and \lq Case-2\rq (right). The red dashed ellipse from Fig. \ref{fig:2d_map_gr_regions_0.14} and  the coloured crosses representing the five bright galaxies in the group are reproduced here for reference, including the dwarf galaxy (NGC\,5018 - LSB1; white cross).}    
    \label{fig:2d_maps_narrow_sels_0.14}
\end{figure}

\begin{table}[htb!]
\centering
\begin{minipage}[t]{0.485\linewidth}
\centering
\caption{Selection criteria adopted in \lq Case-1\rq.}
\label{tab:gr_gcsel_narrow_crit}
\begin{tabular}{ccc}
\hline
 Parameter & $g$-passband & $r$-passband \\
    \hline
      Mag. & $\geq 22.9, \leq$ 25.3 & $\geq 22.3, \leq$ 24.7 \\ 
      ELONG. & $\leq 2.0$ & $\leq 3.0$ \\
      CI & $\geq 0.5, \leq 1.3$ & $\geq 0.5, \leq 1.6$ \\
    \hline
    \noalign{\smallskip}
      Colour & \multicolumn{2}{c}{$0.4 \leq g-r \leq 0.9$} \\
      \hline    
\end{tabular}
\end{minipage}
\hfill%
\begin{minipage}[t]{0.485\linewidth}
\centering
\caption{Selection criteria adopted in \lq Case-2\rq.}
\label{tab:gr_gcsel_very_narrow_crit}
\begin{tabular}{ccc}
\hline
Parameter & $g$-passband & $r$-passband \\
    \hline
      Mag. & $\geq 22.9, \leq$ 25.3 & $\geq 22.3, \leq$ 24.7 \\ 
      ELONG. & $\leq 2.00$ & $\leq 2.25$ \\
      CI & $\geq 0.5, \leq 1.3$ & $\geq 0.5, \leq 1.4$ \\
    \hline
    \noalign{\smallskip}
      Colour & \multicolumn{2}{c}{$0.4 \leq g-r \leq 0.9$} \\
      \hline    
\end{tabular}
\end{minipage}
\end{table}

\item As a second check, we consider two additional cases: 
$i)$ \lq Case-3\rq, where we select bright point sources that should be foreground Milky Way stars; and 
$ii)$ \lq Case-4\rq, where we select candidate extended sources with GC-like magnitudes but non-GC colours, i.e, sources that are likely background galaxies in the observed field. The criteria adopted are shown in Tables \ref{tab:gr_star_sel_crit} and \ref{tab:gr_ext_sources_sel_crit}. If the ellipse were a contamination artefact, we would expect to see some residual evidence of the elliptical overdensity. 

The results shown in the two panels of Fig. \ref{fig:2d_maps_gr_stars_ext_sources_0.14} indicate that this is not the case, as we observe a relatively homogeneous distribution of sources in both cases.

\begin{figure*}[htb!]
    \centering    
    \begin{minipage}{0.495\textwidth}
        \centering        \includegraphics[width=0.995\textwidth]{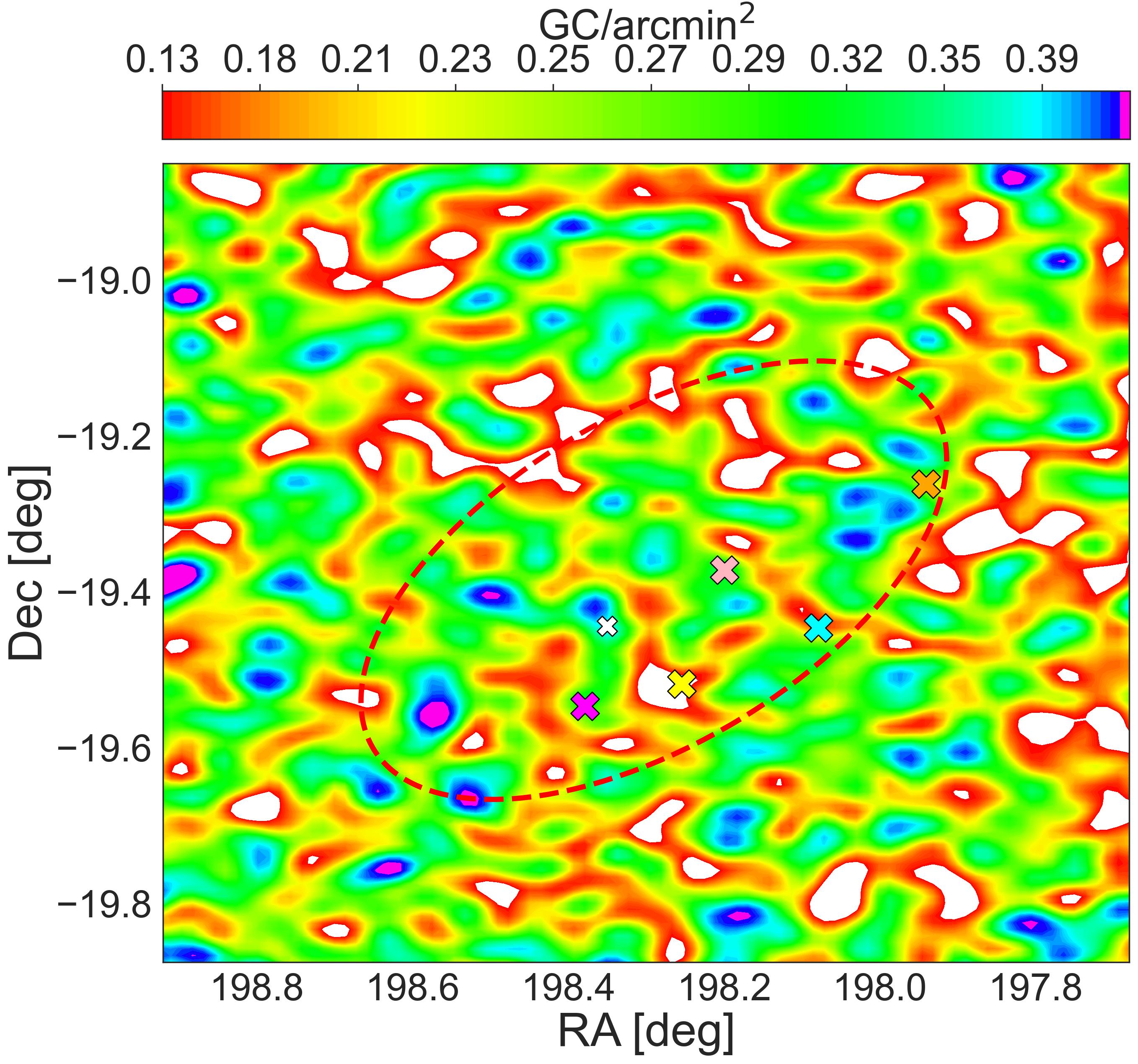}
    \end{minipage}
    \begin{minipage}{0.495\textwidth}
        \centering        \includegraphics[width=0.995\textwidth]{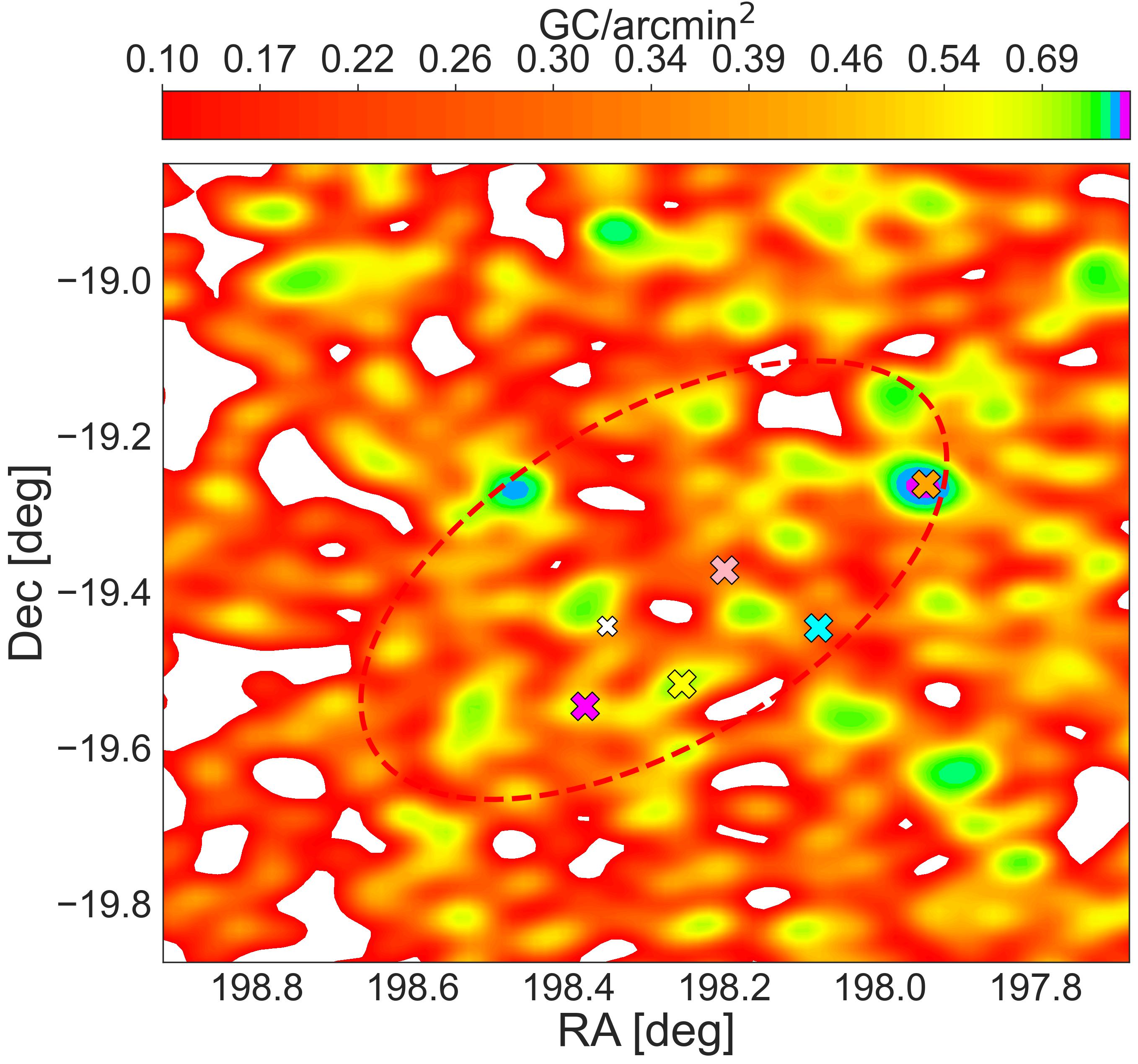}
    \end{minipage}
    \caption{2D distribution maps obtained for \lq Case-3\rq \ and \lq Case-4\rq. The red dashed ellipse from Fig. \ref{fig:2d_map_gr_regions_0.14} and the coloured crosses representing the five bright galaxies in the group are reproduced here for reference, including the dwarf galaxy (NGC\,5018 - LSB1; white cross).}
    \label{fig:2d_maps_gr_stars_ext_sources_0.14}
\end{figure*}

\begin{table}[htb!]
\centering
\begin{minipage}[t]{0.485\linewidth}
\centering
\caption{Selection criteria adopted in \lq Case-3\rq.}
\label{tab:gr_star_sel_crit}
\begin{tabular}{ccc}
\hline
    Parameter & $g$-passband & $r$-passband \\
    \hline
      Mag. & $\geq 18.0, \leq$ 22.0 & $\geq 18.0, \leq$ 22.5 \\ 
      ELONG. & $\leq 1.5$ & $\leq 2.0$ \\
      CI & $\geq 0.25, \leq 1.00$ & $\geq 0.25, \leq 1.50$ \\
    \hline
    \noalign{\smallskip}
      Colour & \multicolumn{2}{c}{$0.3 \leq g-r \leq 1.1$} \\
      \hline
\end{tabular}
\end{minipage}
\hfill%
\begin{minipage}[t]{0.485\linewidth}
\centering
\caption{Selection criteria adopted in \lq Case-4\rq.}
\label{tab:gr_ext_sources_sel_crit}
\begin{tabular}{ccc}
\hline
    Parameter & $g$-passband & $r$-passband \\
    \hline
      Mag. & $\geq 12.0, \leq$ 25.0 & $\geq 12.0, \leq$ 25.0 \\ 
      ELONG. & $\leq 50.0$ & $\leq 50.0$ \\
      CI & $\geq 1.30, \leq 100.00$ & $\geq 1.35, \leq 100.00$ \\
    \hline
    \noalign{\smallskip}
      Colour & \multicolumn{2}{c}{$g-r \leq 0.4, g-r \geq 0.9$} \\
      \hline
\end{tabular}
\end{minipage}
\end{table}

\item For a final check, we consider two additional cases: $i)$ \lq Case-5\rq, where we inspect the GC candidates brighter than TOM (as they are the most reliable candidates) while keeping the rest of the selection criteria unchanged; and $ii)$ \lq Case-6\rq, where we lower the elongation threshold compared to our reference. The criteria adopted are shown in Tables \ref{tab:gr_TOM_sel_crit} and \ref{tab:gr_elong_2.0_sel_crit}.

The results of these tests are shown in Fig. \ref{fig:2d_maps_gr_TOM_elong_gr_2.0_0.14}. In both cases, we observe that the elliptical overdensity feature persists; in these tests also the plume we associated with the LSB galaxy NGC\,5018-LSB1 appears.

\begin{figure*}[htb!]
    \centering    
    \begin{minipage}{0.495\textwidth}
        \centering        \includegraphics[width=0.995\textwidth]{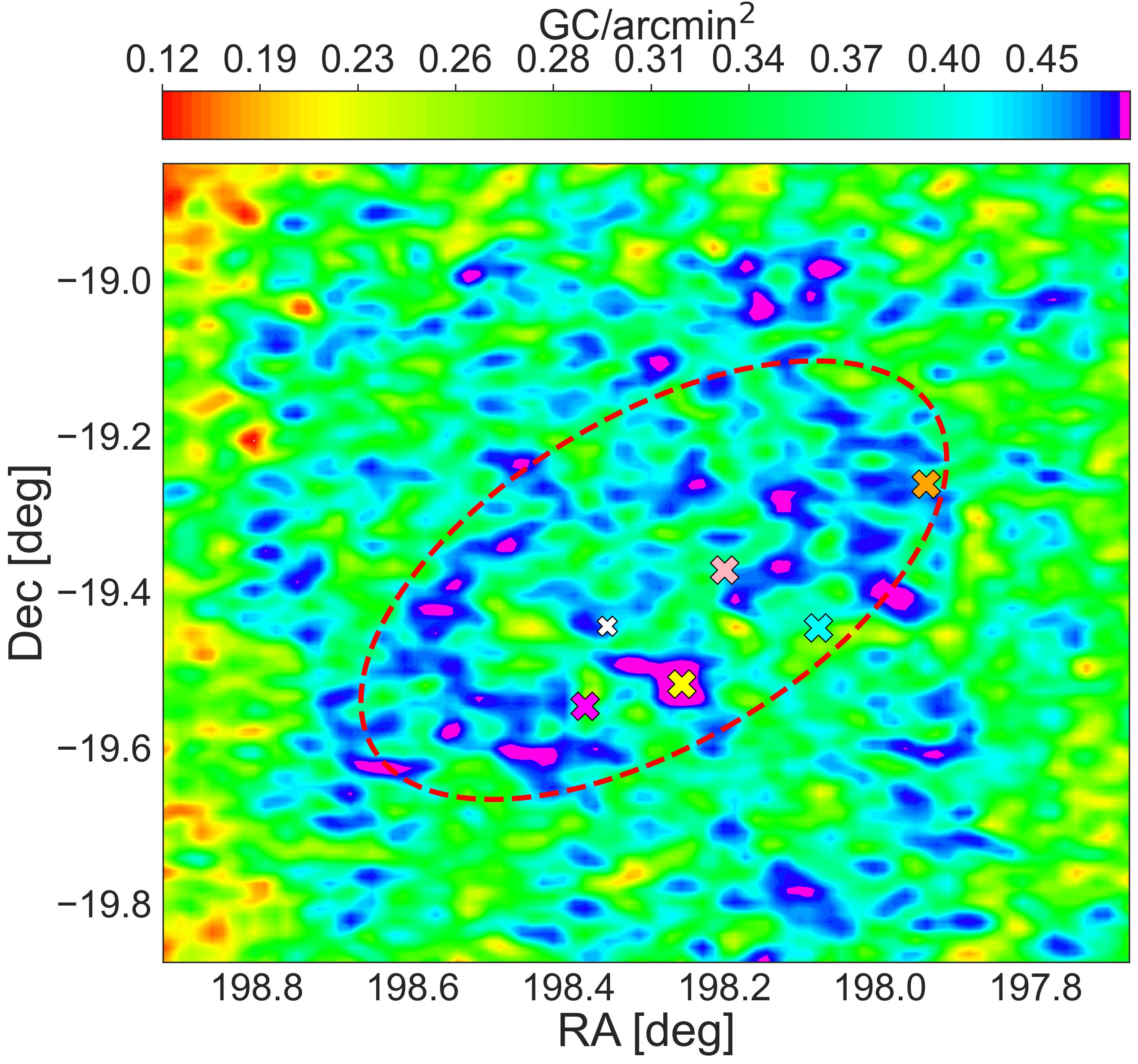}
    \end{minipage}
    \begin{minipage}{0.495\textwidth}
        \centering        \includegraphics[width=0.995\textwidth]{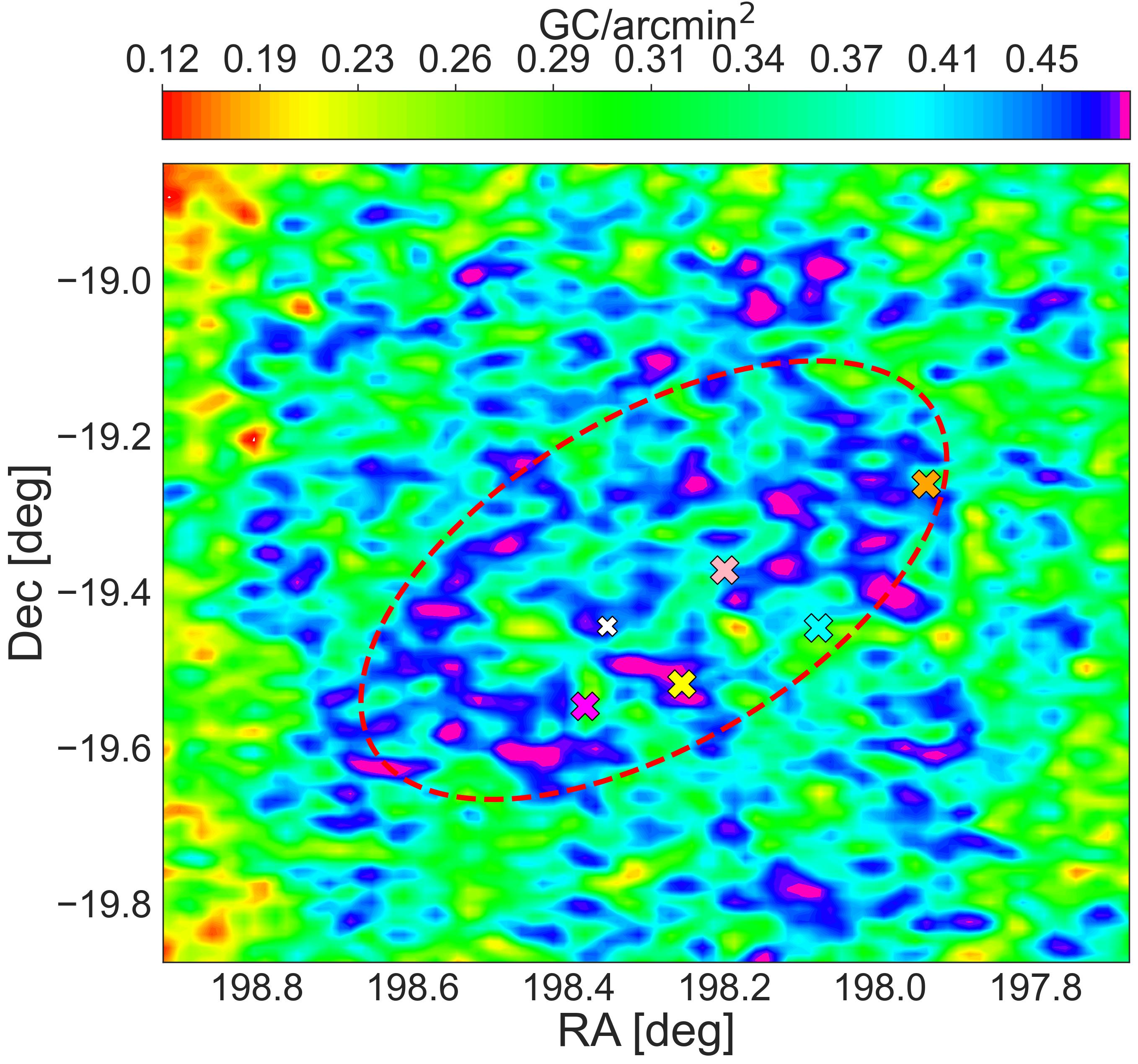}
    \end{minipage}
    \caption{2D distribution maps obtained for \lq Case-5\rq \ and \lq Case-6\rq. The red dashed ellipse from Fig. \ref{fig:2d_map_gr_regions_0.14} and the coloured crosses representing the five bright galaxies in the group are reproduced here for reference, including the dwarf galaxy (NGC\,5018 - LSB1; white cross).}
    \label{fig:2d_maps_gr_TOM_elong_gr_2.0_0.14}
\end{figure*}

\begin{table}[htb!]
\centering
\begin{minipage}[t]{0.485\linewidth}
\centering
\caption{Selection criteria adopted in \lq Case-5\rq.}
\label{tab:gr_TOM_sel_crit}
\begin{tabular}{ccc}
\hline
    Parameter & $g$-passband & $r$-passband \\
    \hline
      Mag. & $\geq 21.7, \leq$ 25.3 & $\geq 21.1, \leq$ 24.7 \\ 
      ELONG. & $\leq 3.0$ & $\leq 3.0$ \\
      CI & $\geq 0.5, \leq 1.6$ & $\geq 0.5, \leq 1.6$ \\
    \hline
    \noalign{\smallskip}
      Colour & \multicolumn{2}{c}{$0.3 \leq g-r \leq 1.1$} \\
      \hline
\end{tabular}
\end{minipage}
\hfill%
\begin{minipage}[t]{0.485\linewidth}
\centering
\caption{Selection criteria adopted in \lq Case-6\rq.}
\label{tab:gr_elong_2.0_sel_crit}
\begin{tabular}{ccc}
\hline
    Parameter & $g$-passband & $r$-passband \\
    \hline
      Mag. & $\geq 21.7, \leq$ 26.5 & $\geq 21.1, \leq$ 25.9 \\ 
      ELONG. & $\leq 2.0$ & $\leq 2.0$ \\
      CI & $\geq 0.5, \leq 1.6$ & $\geq 0.5, \leq 1.6$ \\
    \hline
    \noalign{\smallskip}
      Colour & \multicolumn{2}{c}{$0.3 \leq g-r \leq 1.1$} \\
      \hline
\end{tabular}
\end{minipage}
\end{table}

\end{enumerate}

To conclude, we emphasise that despite the changes possible in the selection criteria, the overdensity of GC candidates along the region of the five brightest group members persists. This result, combined with previous findings on the IGL distribution, supports the existence of an intra-group GC component aligned but more extended than the observed IGL. However, it is important to note that adopting an elliptical geometry is a simplification meant to facilitate a clearer analysis and presentation of our results, rather than representing the true geometry of the intra-group component.

\section{GC candidates catalogue}
\label{sec:gc_candidates_catalogue}

\begin{table*}[htb!]
  \centering
  \makebox[\linewidth][c]{%
    \rotatebox{90}{%
      \begin{minipage}{\textheight}
        \centering 
        \caption{Extract of the catalogue of GC candidates in the observed field.}
        \begin{tabular}{cccccccccccccc}
         \hline
         \\[-2ex]
            RA (J2000) & Dec (J2000) & $m_u$ & err $m_u$ & $m_g$ & err $m_g$ & $m_r$ & err $m_r$ & $CI_u$ & $CI_g$ & $CI_r$ & $ELONG_u$ & $ELONG_g$ & $ELONG_r$  \\
            
            [deg] & [deg] & [mag] & [mag] & [mag] & [mag] & [mag] & [mag] & [mag] & [mag] & [mag] & & & \\
            (1) & (2) & (3) & (4) & (5) & (6) & (7) & (8) & (9) & (10) & (11) & (12) & (13) & (14)\\
            \\[-2ex]
            \hline
            \\[-2ex]
            198.639064 & -19.870831 &	&  & 23.415 &	0.052	& 22.665 &	0.041 & &	1.221 &	1.201 &	&	1.350 &	1.134 \\
            
            197.768553 &	-19.872049 &	23.529 &	0.166 &	23.445 &	0.049 &	23.139 &	0.054 &	1.156 &	0.991 &	1.026 &	1.348 &	1.195 &	1.083 \\

            198.211328 &	-19.872226 &	23.303 &	0.134 &	22.792 &	0.026 &	22.203 &	0.023 &	1.619 &	1.218 &	1.046 &	2.041 &	1.455 &	1.272 \\

            198.907867 &	-19.868882 &	 &	&	24.382 &	0.203 &	23.505 &	0.161 &	& 	0.965 &	1.020 &	&	1.636 &	1.214 \\

            198.755168 &	-19.870046 &	 &	& 	22.521 &	0.023 &	22.178 &	0.027 &	& 	0.851 &	0.919 &	& 	1.087 &	1.289 \\

            198.451090 &	-19.870753 &	& 	&	24.721 &	0.156 &	24.401 &	0.180 &	&	0.989 &	1.155 &	& 	2.409 &	1.623 \\

            198.804428 &	-19.869229 &	23.323 &	0.137 &	23.602 &	0.083 &	22.865 &	0.071 &	0.943 &	0.904 &	1.021 &	1.176 &	1.115 &	1.102 \\

            198.235541 &	-19.871513 &	& 	& 	23.675 &	0.059 &	22.622 &	0.035 &	&	0.936 &	0.801 &	&	1.106 &	1.177 \\

            ... & ... & ... & ... & ... & ... & ... & ... & ... & ... & ... & ... & ... & ... \\

            ... & ... & ... & ... & ... & ... & ... & ... & ... & ... & ... & ... & ... & ... \\

            ... & ... & ... & ... & ... & ... & ... & ... & ... & ... & ... & ... & ... & ... \\

            ... & ... & ... & ... & ... & ... & ... & ... & ... & ... & ... & ... & ... & ... \\

            198.216974 &	-19.871168 &	&	&	23.153 &	0.036 &	22.566 &	0.032 & &	0.732 &	0.658 &	&	1.066 &	1.018 \\

            198.651816 &	-19.869746 &	23.381 &	0.142 &	22.917 &	0.032 &	22.512 &	0.035 &	1.076 &	0.857 &	0.917 &	1.608 &	1.177 &	1.050 \\
            
         \hline 
         \\[-1.5ex]

        \end{tabular}
        \tablefoot{Columns list: (1) Right Ascension; (2) Declination; (3-8) $ugr$-passband magnitudes (corrected for extinction) and their errors; (9-11) concentration index in $ugr$-passbands; (12-14) elongation (major-to-minor axis ratio) in $ugr$-passbands. All the listed parameters are derived from SExtractor (see Sect. \ref{sec:gc_select} for more details). Sources for which no $u$-passband data is available are left empty in the columns for this passband.}
        \end{minipage}
        
        \label{tab:gc_catalog}
    
    }%
  }%
\end{table*}

\end{appendix}

\end{document}